\documentclass[pdftex, twocolumn, epjc3]{svjour3}
\pdfoutput=1
\usepackage[T1]{fontenc}
\usepackage{setspace}
\usepackage{amsmath}
\usepackage[utf8]{inputenc}
\usepackage{graphicx}
\usepackage{placeins}
\usepackage{units}
\usepackage{xspace}
\usepackage{hyperref}
\usepackage{tabulary}
\usepackage{caption}
\usepackage{subcaption}
\usepackage{multirow}
\usepackage{amsmath}
\usepackage{amssymb}
\usepackage[numbers,sort&compress]{natbib}
\usepackage{color} 

\newcommand{\qq}{\ensuremath{Q^{2}}\xspace}
\newcommand{\mares}{\ensuremath{M_{\textrm{A}}^{\mbox{\scriptsize{RES}}}}\xspace}
\newcommand{\fa}{\ensuremath{F_{\mathsf{A}}(0)}\xspace}
\newcommand{\ndof}{\ensuremath{N_{\mathsf{DOF}}}\xspace}
\newcommand{\Ev}{\ensuremath{E_{\nu}}\xspace}
\newcommand{\ccppiplus}{\ensuremath{\nu_{\mu}p\rightarrow \mu^{-}p\pi^{+}}\xspace}
\newcommand{\ccnpizero}{\ensuremath{\nu_{\mu}n\rightarrow \mu^{-}p\pi^{0}}\xspace}
\newcommand{\ccnpiplus}{\ensuremath{\nu_{\mu}n\rightarrow \mu^{-}n\pi^{+}}\xspace}
\newcommand{\minerva}{MINER\ensuremath{\nu}A\xspace}
\newcommand{\nova}{NO\ensuremath{\nu}A\xspace}

\journalname{Eur. Phys. J. C}

\begin{document}

\title{Constraining the GENIE model of neutrino-induced single pion production using reanalyzed bubble chamber data}
\date{\today}

\author{Philip Rodrigues\thanksref{add1} \and 
  Callum Wilkinson\thanksref{add2} \and 
  Kevin McFarland\thanksref{add1}} 

\institute{University of Rochester, Department of Physics and Astronomy, Rochester 14627, New York, USA\label{add1}
  \and
  University of Bern, Albert Einstein Center for Fundamental Physics, Laboratory for High Energy Physics (LHEP), Bern 3012, Switzerland \label{add2}
}
\maketitle

\begin{abstract}
The longstanding discrepancy between bubble chamber measurements of $\nu_\mu$-induced single pion production channels has led to large uncertainties in pion production cross section parameters for many years. We extend the reanalysis of pion production data in deuterium bubble chambers where this discrepancy is solved (Wilkinson {\it et al.}, PRD 90 (2014) 112017) to include the \ccnpizero and \ccnpiplus channels, and use the resulting data to fit the parameters of the GENIE pion production model. We find a set of parameters that can describe the bubble chamber data better than the GENIE default parameters, and provide updated central values and reduced uncertainties for use in neutrino oscillation and cross section analyses which use the GENIE model. We find that GENIE's non-resonant background prediction has to be significantly reduced to fit the data, which may help to explain the recent discrepancies between simulation and data observed by the \minerva coherent pion and \nova oscillation analyses.
\end{abstract}

\section{Introduction}

A good understanding of single pion production by neutrinos with few-GeV energies is important for current and future oscillation experiments, where pion production is either a signal process, or a large background for analyses which select quasi-elastic events. At these energies the dominant production mechanism is via the production and subsequent decay of hadronic resonances.

Complete models of neutrino-nucleus single pion production interactions are usually factorized into three parts: the neutrino-nucleon cross section; additional nuclear effects which affect the initial interaction; and the ``final state interactions'' (FSI) of hadrons exiting the nucleus. Experimental data on nuclear targets presents a confusing picture, with recent data from the \minerva~\cite{minerva_ccpip_2014, minerva_anupi0_2015} and MiniBooNE~\cite{mb-cc1pplus-2010} experiments in poor agreement with each other in the framework of current theoretical models~\cite{zmuda_2014, mosel_2015}. An additional problem is the disagreement between measurements of the neutrino-nucleon single pion production cross section in the $\unit[100]{MeV}$ to few-\unit{GeV} energy range most relevant for current and planned neutrino oscillation experiments. The axial form factor for pion production on free nucleons cannot be constrained by electron scattering data, so relies upon data from the Argonne National Laboratory's \unit[12]{ft} bubble chamber (ANL) and the Brookhaven National Laboratory's \unit[7]{ft} bubble chamber (BNL). However, these datasets differ in normalization by 30--40\% for the leading pion production process \ccppiplus, which leads to large uncertainties in the predictions for oscillation experiments~\cite{sobczyk_2009, hernandez_spp_2007, leitner_spp_2008, hernandez_spp_2010, k2k_2003, t2k-nue-prd}, as well as in the interpretation of data taken on nuclear targets~\cite{gibuu_spp_tune}.

It has long been suspected that the discrepancy between ANL and BNL was due to an issue with the normalization of the flux prediction from one or both experiments\footnote{The ANL neutrino beam~\cite{ANL_Barish_1977} was produced by focusing 12.4 GeV protons onto a beryllium target. Two magnetic horns were used to focus the positive pions produced by the primary beam in the direction of the bubble chamber, these secondary particles decayed to produce a predominantly $\nu_{\mu}$ beam peaked at $\sim$0.5 GeV. The BNL neutrino beam~\cite{BNL_Baker_1982, BNL_Furuno_2003} was produced by focusing 29 GeV protons on a sapphire target, with a similar two horn design to focus the secondary particles. The BNL $\nu_{\mu}$ beam had a higher peak energy of $\sim$1.2 GeV, and was broader than the ANL beam.}, and it has been shown by other authors that their published results are consistent within the experimental uncertainties provided~\cite{sobczyk_2009, sobczyk_2014}. In Reference~\cite{anl_bnl_reanalysis}, we presented a method for removing flux normalization uncertainties from the ANL and BNL \ccppiplus measurements by taking ratios with charged-current quasielastic (CCQE) event rates in which the normalization cancels. Then we obtained a measurement of \ccppiplus by multiplying the ratio by an independent measurement of CCQE (which is well known for nucleon targets). Using this technique, we found good agreement between the ANL and BNL \ccppiplus datasets. In this work, we extend that method to include the subdominant \ccnpizero and \ccnpiplus channels, and use the resulting data, along with the \qq-spectra (where \qq is the four-momentum transfer) from the same experiments, to constrain the parameters of the GENIE single pion production model~\cite{genieMC}. While more sophisticated single pion production models exist~\cite{Morfin:2012kn, hayato_review_2014, Mosel:2016cwa}, the GENIE generator is widely used by current and planned neutrino oscillation experiments, so tuning the generator parameters represents a pragmatic approach to improving its description of available data. We find that the reanalyzed data, where the normalization discrepancy has been resolved, is able to significantly reduce the uncertainties on the pion production parameters. We also find that the non-resonant background prediction from GENIE needs to be significantly reduced to fit the data.

Reduced uncertainties on pion production parameters are vital for current and future neutrino oscillation experiments, which have very stringent systematic uncertainty requirements~\cite{lbne,t2k_sensitivity_2014}. We recommend that our new uncertainties should be used by experiments which use the GENIE neutrino interaction generator, and the reanalyzed ANL and BNL datasets presented here and in Reference~\cite{anl_bnl_reanalysis} should be used instead of the published ANL and BNL datasets for future model comparisons.

The paper is organized as follows. In Section~\ref{sec:datasets}, we describe the datasets used in this analysis. In Section~\ref{sec:genie_model} we describe the GENIE single pion production model and compare the nominal GENIE model and error bands with the data. The $\chi^{2}$ statistic which is minimized and the fit machinery are discussed in Section~\ref{sec:chi2}; the fit results are presented in Section~\ref{sec:results}; and there is a discussion of the goodness of fit in Section~\ref{sec:goodness_of_fit}. Finally, our conclusions are presented in Section~\ref{sec:conclusions}.

\section{Datasets used in this analysis}\label{sec:datasets}
In this work, we use the \qq and \Ev-spectra from ANL and BNL for all three charged-current single pion production modes (\ccppiplus, \ccnpizero and \ccnpiplus), giving a total of twelve datasets.

The \qq-dependent distributions used are presented as flux-integrated event rates without any invariant mass cut applied, and were digitized from References~\cite{ANL_Radecky_1982} (ANL) and~\cite{BNL_Kitagaki_1986} (BNL) for this work. To produce flux-integrated event rate predictions with GENIE, the flux was taken from References~\cite{ANL_Barish_1977} (ANL) and~\cite{BNL_Furuno_2003} (BNL).

The \Ev-dependent distributions for both ANL and BNL are taken from the reanalysis of \ccppiplus data presented in Reference~\cite{anl_bnl_reanalysis}, and the reanalysis of \ccnpizero and \ccnpiplus using the same technique and presented in~\ref{sec:reanalysis}. These datasets are neutrino-\emph{deuterium} cross sections, as no correction has been applied to account for deuterium nuclear effects. Additionally, in~\ref{sec:reanalysis_wcut} we present reanalyzed results for the three pion production channels in which an additional correction has been applied to include the effect of the invariant mass cut $W \leq 1.4$ GeV on the \Ev-dependent distributions. The reanalyzed results with  $W \leq 1.4$ GeV were not used in the present work, but we provide it for use in future model comparisons.

\begin{figure*}[p]
  \centering
  \begin{subfigure}{0.9\columnwidth}
    \includegraphics[width=\textwidth]{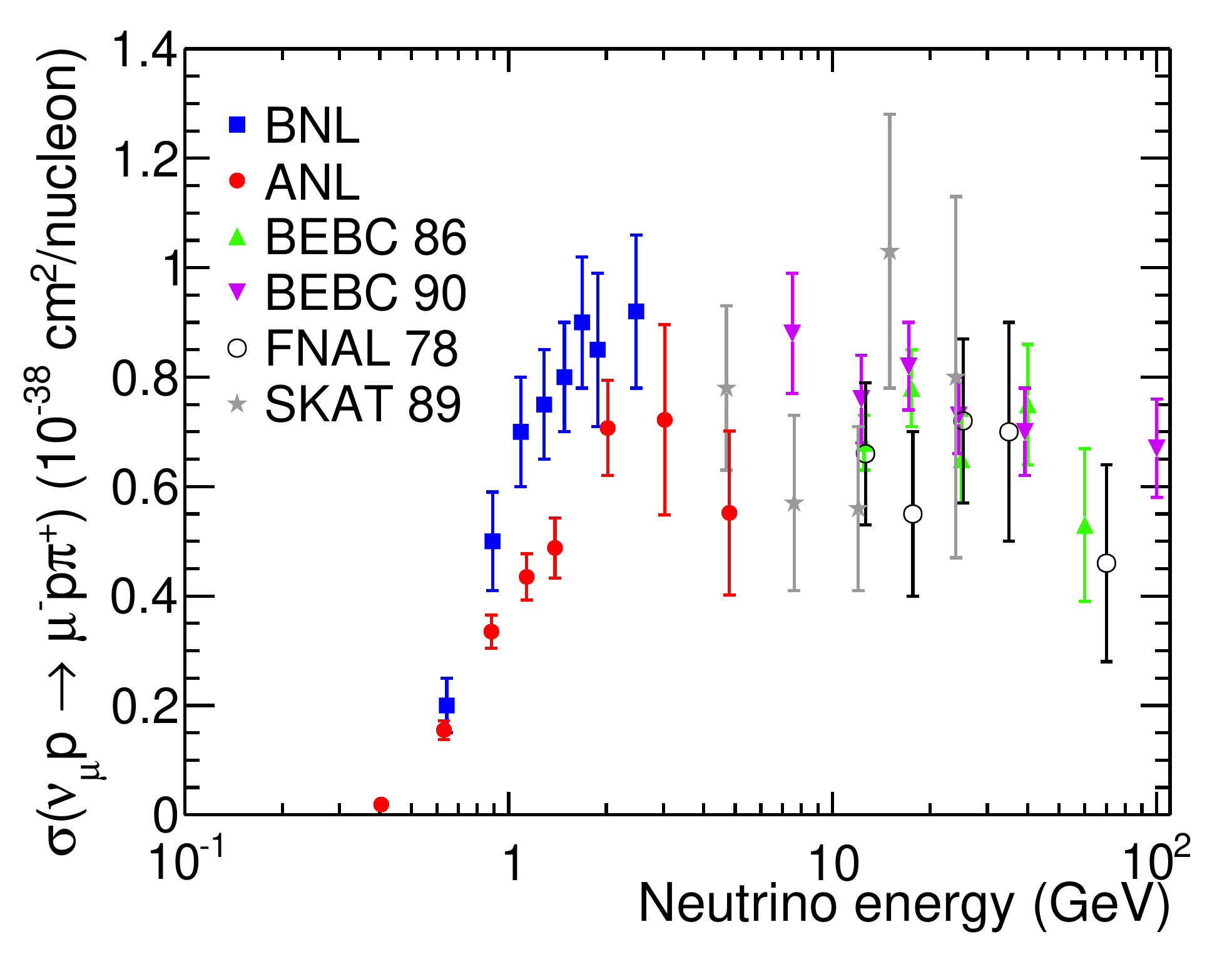}
    \caption{\ccppiplus published}
    \label{subfig:reanalysis_ccpiplus_pub}
  \end{subfigure}
  \begin{subfigure}{0.9\columnwidth}
    \includegraphics[width=\textwidth]{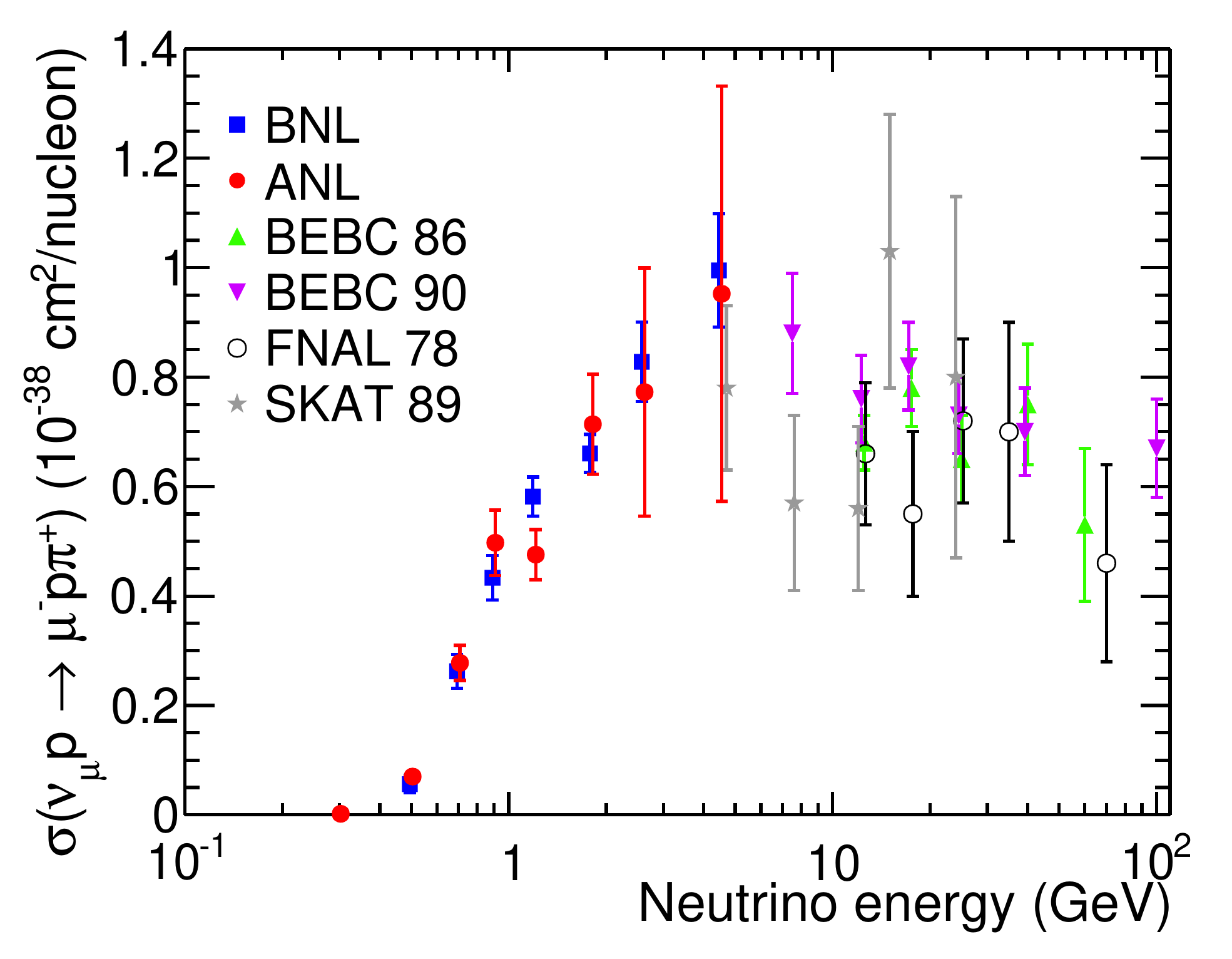}
    \caption{\ccppiplus reanalyzed}
    \label{subfig:reanalysis_ccpiplus_ext}
  \end{subfigure}
  \begin{subfigure}{0.9\columnwidth}
    \includegraphics[width=\textwidth]{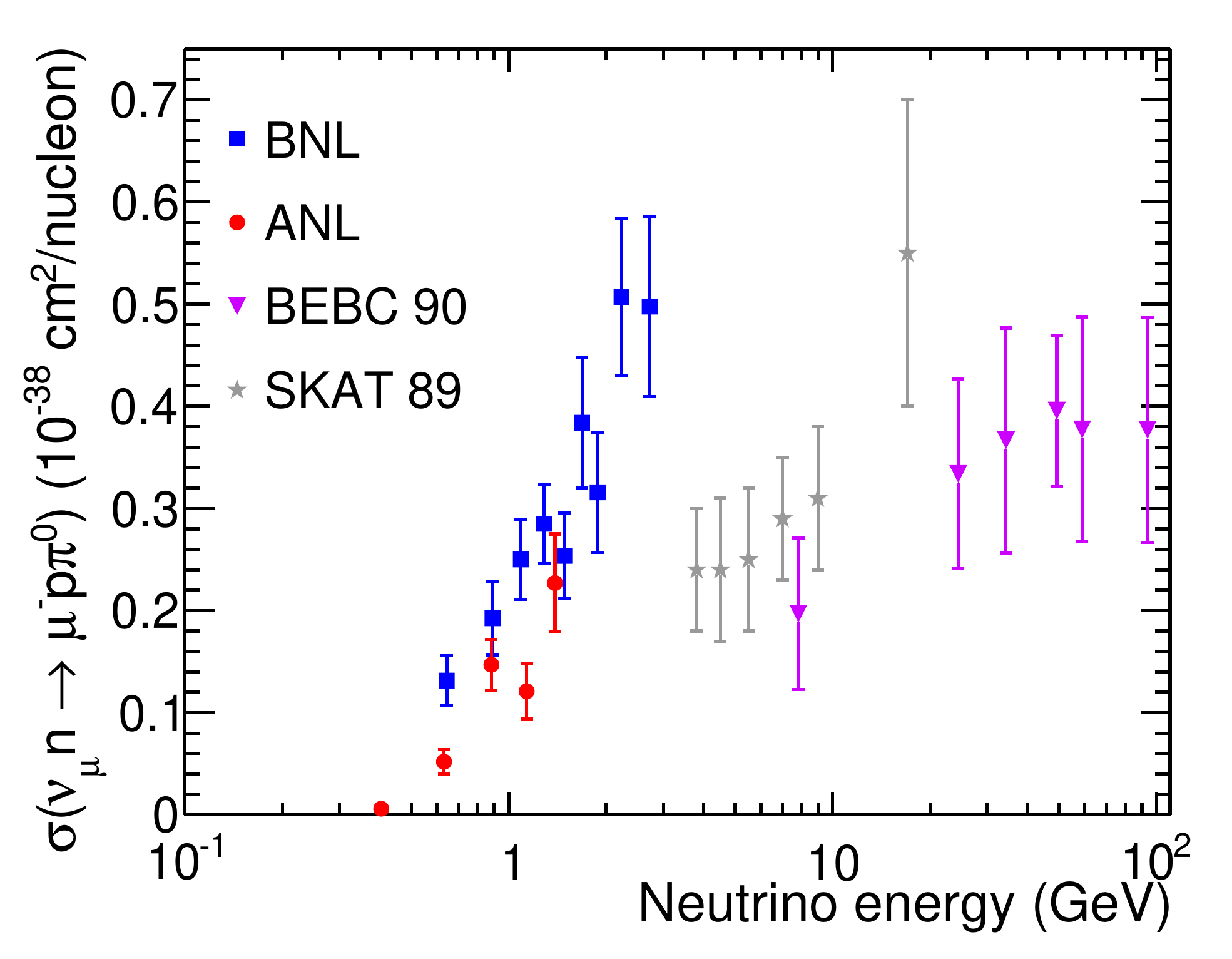}
    \caption{\ccnpizero published}
  \end{subfigure}
  \begin{subfigure}{0.9\columnwidth}
    \includegraphics[width=\textwidth]{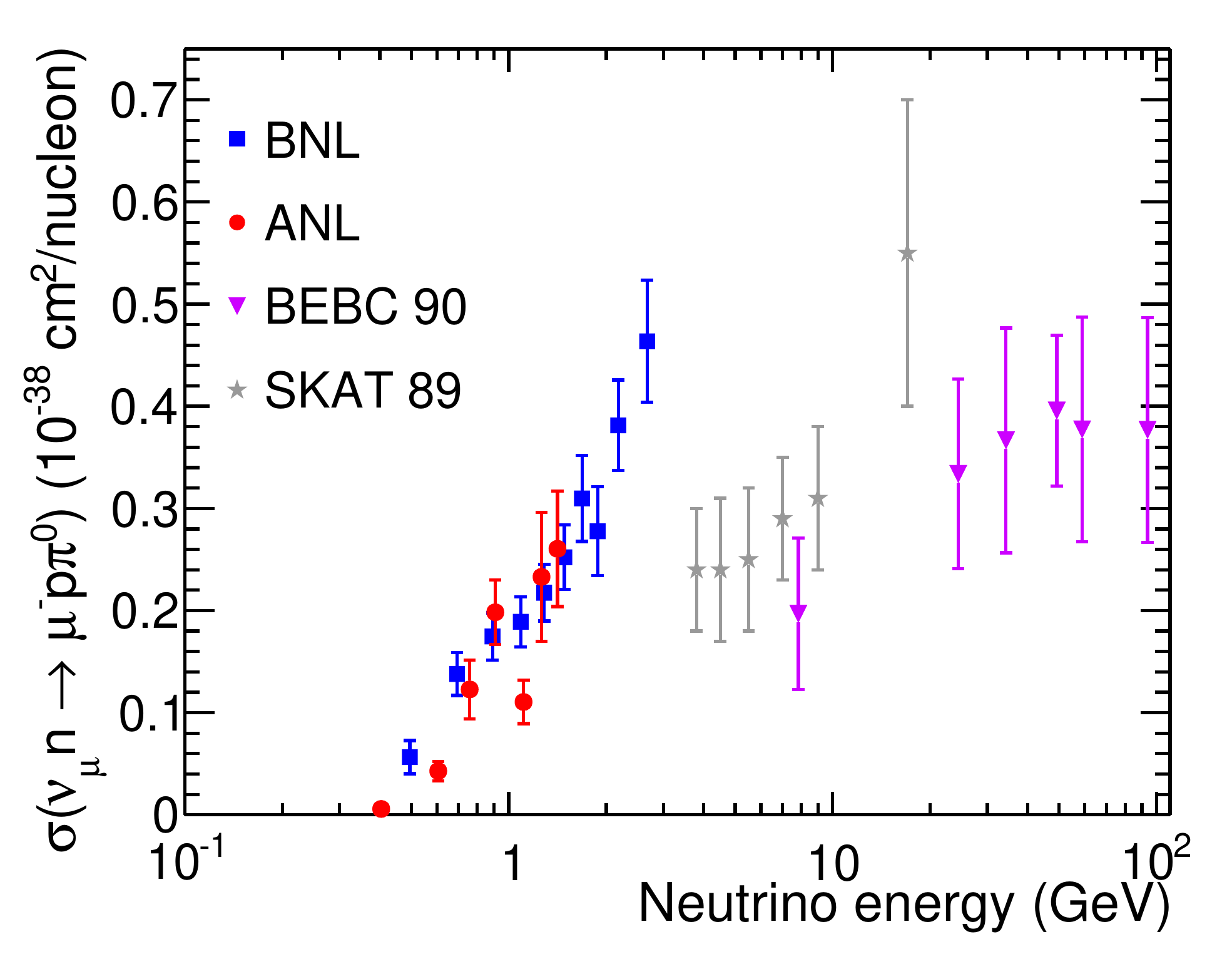}
    \caption{\ccnpizero reanalyzed}
  \end{subfigure}
  \begin{subfigure}{0.9\columnwidth}
    \includegraphics[width=\textwidth]{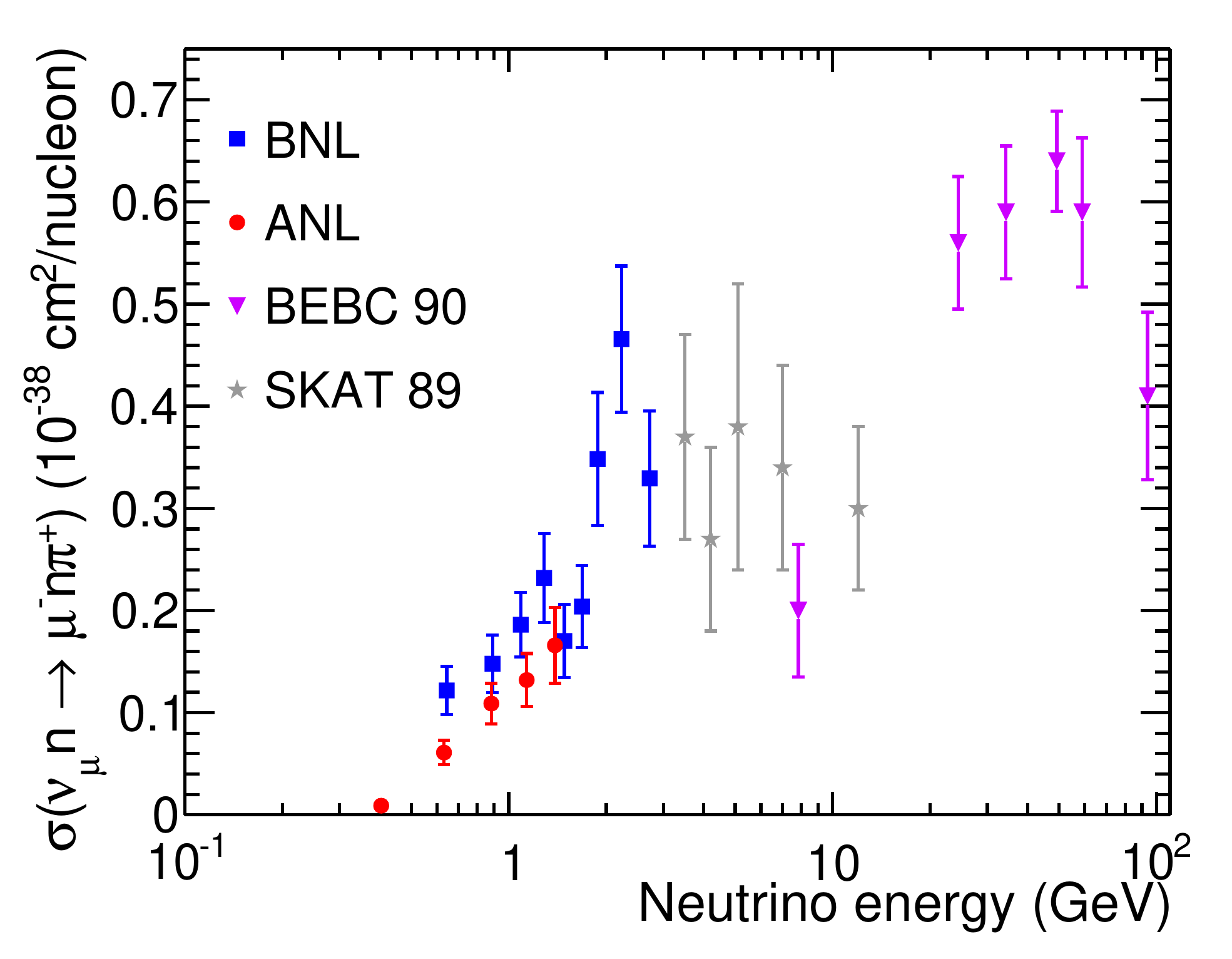}
    \caption{\ccnpiplus published}
  \end{subfigure}
  \begin{subfigure}{0.9\columnwidth}
    \includegraphics[width=\textwidth]{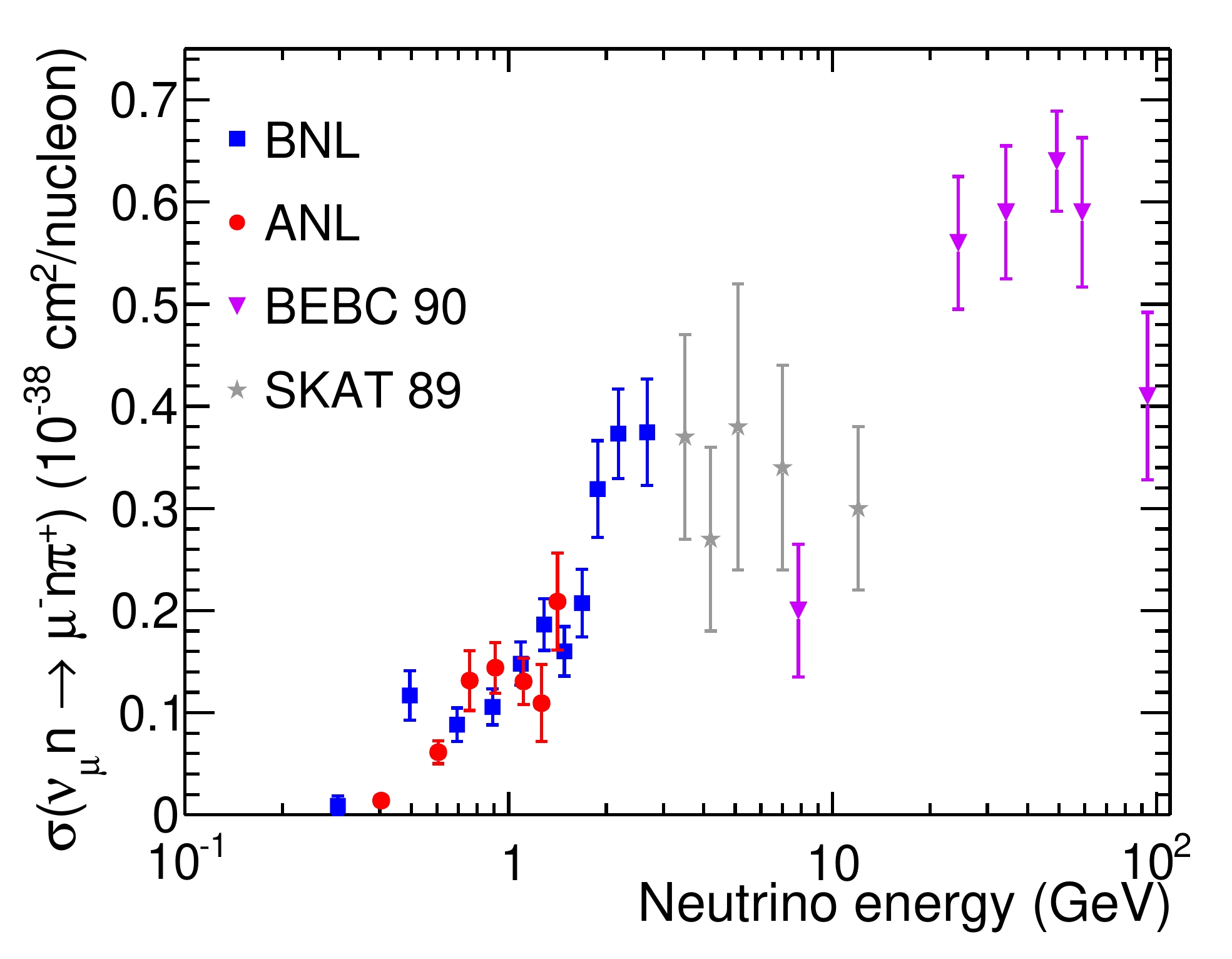}
    \caption{\ccnpiplus reanalyzed}
  \end{subfigure}
  \caption{The published and extracted ANL and BNL data are compared with other measurements of the three pion production channels~\cite{fnal_1978, skat_1989, bebc_1986, bebc_1990}. All data is on hydrogen and deuterium targets, except for SKAT 1989 data which was taken on heavy freon (CF$_{3}$Br). Note that both published and reanalyzed ANL and BNL data shown here have no invariant mass cut in the event selection, whereas the other datasets have an invariant mass cut of $W \leq \unit[2]{GeV}$ applied unless otherwise mentioned.}\label{fig:reanalysis_comparison}
\end{figure*}

The \Ev-dependent distributions for both ANL and BNL are shown for all three single pion production channels in Figure~\ref{fig:reanalysis_comparison} along with other bubble chamber measurements available for these channels. The original ANL and BNL results are also shown so the effect of the reanalysis can be seen. It is clear that the reanalysis affects all channels, although the effect is more pronounced for the dominant \ccppiplus channel, where the statistical errors are smaller and biases are easier to see. The ANL and BNL datasets agree well in all three channels after the reanalysis. In Figures~\ref{subfig:reanalysis_ccpiplus_pub} and~\ref{subfig:reanalysis_ccpiplus_ext}, the BEBC data on a hydrogen target has an invariant mass cut of $W \leq 2$ GeV~\cite{bebc_1986}, which removes contributions from diffractive processes. The FNAL data on a hydrogen target is selected with an invariant mass cut of $W \leq 1.4$ GeV, in order to isolate the $\Delta^{++}$ contribution to the cross section, which also cuts out any diffractive contributions from the cross section~\cite{fnal_1978}. Additionally, the FNAL result was scaled by 14\% to account for $\Delta^{++}$ contributions with $W > 1.4$ GeV.

Despite the caveats associated with the subdominant \ccnpizero and \ccnpiplus channels, we recommend that all three channels are used for future comparisons with ANL and BNL data. It should be stressed that most of the deficiencies in the reanalyzed results detailed here are also present in the original ANL and BNL results, and should be borne in mind when using reanalyzed or published ANL and BNL results.

A final note of caution regarding the use of these corrected datasets is that there is a hidden correlation between the three channels for each experiment. As the CCQE events used in the correction are common to all channels, statistical errors are correlated in a way which is difficult to quantify. An example of the problem can be seen by looking at the three ANL channels for $1.0 \leq E_{\nu} \leq 1.1$ GeV where an upward fluctuation of the CCQE event rate leads to a decrease in the reanalyzed cross section. It should be noted that this problem is also present in the published ANL cross section results because they used the measured CCQE event rate to correct their flux prediction. This problem is not dealt with in the fits presented in this work, we simply use the statistical errors from the reanalysis without considering correlations, but it will lead to an increase in the $\chi^{2}$ of the fits (because the $\chi^{2}$ penalty for a large statistical fluctuation is applied three times rather than once).

\section{GENIE single pion production model}\label{sec:genie_model}
The single pion production model in GENIE is described in Reference~\cite{genieMC} and does not change significantly between the GENIE major versions 2.6.X and 2.8.X (X denotes a the minor version number) investigated in this study\footnote{We include both versions as a sanity check which ensures that our results are consistent between the GENIE major versions used by currently running experiments.}. All processes simulated in GENIE use the Bodek-Ritchie RFG model to describe the initial state nucleon momentum distribution~\cite{BodekRitchie} for all nuclear targets, including deuterium. In GENIE, single pion production is separated into resonant and non-resonant terms, with interference terms between the two neglected. The resonant component (RES) is a modified version of the Rein-Sehgal (R-S) model~\cite{rein_sehgal}. In the original R-S model, the production and subsequent decay of 18 nucleon resonances with invariant masses $W \leq 2$ GeV are considered. In GENIE, only 16 resonances are included, based on the recommendation of the Particle Data Group~\cite{pdg2014}. The cross section calculation has not been modified to include lepton mass terms, but the effect of lepton mass terms on the phase space boundaries is taken into account. The cross section is cut off at a tunable invariant mass value, which is $W \leq 1.7$ GeV by default (and in this study). No in-medium modifications to resonances are considered, and interferences between resonances are neglected in the calculation. By default, the resonances decay isotropically in their center of mass frame. In general, there is an additional contribution to single pion production from coherent pion production processes, which are modeled (by default) in GENIE using the Rein-Sehgal coherent model~\cite{Rein:1982pf} with lepton mass corrections included~\cite{Rein:2006di}. However, for the ANL and BNL channels considered here, the selection criteria include requirements on the struck or spectator nucleon, which effectively excludes coherent pion production as the deuterium is no longer bound in the final state. As such, coherent contributions to single pion production are not considered further in this work\footnote{Additionally, diffractive processes~\cite{Rein:1986cd}, where the neutrino interacts coherently with a free nucleon (rather than an entire nucleus) can contribute to pion production. GENIE has an implementation of the diffractive pion production model described in Reference~\cite{Rein:1986cd}, but this is not included by default in the GENIE model. Diffractive processes do not affect the main body of this work because ANL and BNL have a deuterium, rather than free proton, target.}.

The original Rein-Sehgal model in Reference~\cite{rein_sehgal} includes non-resonant single pion production as an additional resonance amplitude, while in GENIE, the non-resonant component is implemented as an extension of the deep inelastic scattering model. The non-resonant (DIS) contribution to the GENIE single pion production model is calculated using the Bodek-Yang parametrization~\cite{Bodek:2002ps}, with other relevant parameters described in detail in Reference~\cite{genieMC}. Hadronization is described by the AKGY model~\cite{Yang:2009zx}, which uses KNO scaling~\cite{Koba:1972ng} for invariant masses of $W \leq 2.3$ GeV, and PYTHIA~\cite{Sjostrand:2006za} for invariant masses of $W \leq 3.0$ GeV, with a smooth transistion between. The low-$W$ KNO model is tuned to data from the Fermilab 15-foot bubble chamber experiment~\cite{Zieminska:1983bs}, and the high-$W$ PYTHIA model is tuned to BEBC data~\cite{Allen:1981vh}. We note that retuning of the PYTHIA model has been discussed elsewhere~\cite{Katori:2014fxa}, although is not considered here as it affects larger $W$ values than are relevant for this study.

A major difference between the GENIE versions 2.6.X and 2.8.X is the change to the default Final State Interaction (FSI) model, which is applied to all outgoing particles produced at the vertex for both the resonant and non-resonant contributions to single pion production. The default FSI model for both versions of GENIE is the {\it hA} intranuclear cascade model~\cite{genieMC}, which is tuned to $\pi^{+}$-$^{56}Fe$ and $p$-$^{56}Fe$ data, then extrapolated to other targets based on $A^{2/3}$ scaling (where $A$ is the atomic number). In GENIE v2.6.X, FSI effects are negligible for deuterium, but the {\it hA} model was retuned for GENIE v2.8.X~\cite{genieWeb}, which leads the deuterium FSI to reduce the total cross section predictions for all channels relevant for this work by 20--30\%. In this work we have ignored this difference and make the assumption that interactions on deuterium can be treated as interactions on quasi-free nucleons which are only loosely bound together, and so neglect FSI effects. Low-\qq bins (\qq $<$ 0.1 GeV$^{2}$) are not included in the fit to avoid the region where FSI effects are expected to have a significant effect in deuterium. We note that in Reference~\cite{sato_2014} a careful study of FSI effects for pion production interactions on deuterium was carried out. This work found that interactions between the final state nucleons significantly modifies the cross section for the \ccppiplus and \ccnpiplus channels, the most notable feature being a suppression of the cross section for very forward pions. A more careful treatment of FSI based on the work presented in Reference~\cite{sato_2014} would be an improvement to future iterations of this work, but the calculation does not currently predict the entire final state of the interaction so is not ready to be implemented in a generator.

In GENIE, there are a number of systematic parameters which can be varied to change the single pion production model~\cite{genie_manual}. These parameters are summarized in Table~\ref{tab:genie_parameters}. Note that although GENIE allows the normalization of charged-current non-resonant single pion production on protons and neutrons separately, we have grouped them here into a single category (labelled ``DIS'') in the absence of any reason to treat them differently\footnote{Note also that when tuning GENIE based on this work, it would be reasonable to consider these dials to be fully correlated with the normalization of the neutral current non-resonant single pion production prediction for interactions on both a target neutron and target proton, and as fully correlated with the corresponding antineutrino dials.}.

The axial form factor used for resonant pion production in GENIE is given by 
\begin{equation}
        F_{\mathrm{A}}(\qq) = \frac{\fa}{\left(1 + \frac{Q^{2}}{\left(\mares\right)^{2}}\right)^{2}},
\end{equation}
\noindent where $\fa \equiv \frac{5}{3}Z \equiv g_{\mathrm{A}} = 1.267$, $Z$ is a renomalization factor for the axial-vector coupling constant of a quark obtained from data in Reference~\cite{fkr_1971} (also considered in Reference~\cite{kuzmin_2006}) and $\mares$ is the resonant axial mass, available as a parameter in GENIE's reweighting framework.
The normalization of the axial form factor is not available in the GENIE reweighting framework (changing its value requires events to be regenerated with a modified value of the ``RS-Zeta'' parameter), but a similar parameter for the overall resonant normalization is available. Modifying \fa is the more physically-motivated alternative, but as modifying the resonant normalization is more convenient for users, we perform two fits, each with one of these parameters modified.

\begin{table*}
  \centering      
      {\renewcommand{\arraystretch}{1.2}
        \begin{tabular}{>{\centering}p{5cm}|>{\centering}p{5cm}|p{5cm}<{\centering}}
          \hline
          Parameter & GENIE value & Parameter name \\
          \hline
          Resonant axial mass (\mares) & 1.12 $\pm$ 0.22 GeV~\cite{kuzmin_2006} & GXSec\_MaCCRES \\
          Resonant normalization & \multirow{2}{*}{100 $\pm$ 20\%} & \multirow{2}{*}{GXSec\_NormCCRES} \\
          (RES norm.) & & \\
          Non-resonant normalization & \multirow{2}{*}{100 $\pm$ 50\%} & GXSec\_RvnCC1pi \\ 
          (DIS norm.) & & GXSec\_RvpCC1pi \\
          Normalization of the axial & \multirow{2}{*}{100\% (no GENIE uncertainty)} & \multirow{2}{*}{N/A} \\
          form factor (\fa) & & \\
          \hline
      \end{tabular}}
\caption{Variable parameters in the GENIE single pion production model~\cite{genie_manual}. The normalization of the axial form factor is not a variable parameter in GENIE currently, but is varied in this work as described in the text.}\label{tab:genie_parameters}
\end{table*}

\begin{figure}[htbp]
  \centering
  \begin{subfigure}{0.9\columnwidth}
    \includegraphics[width=\textwidth]{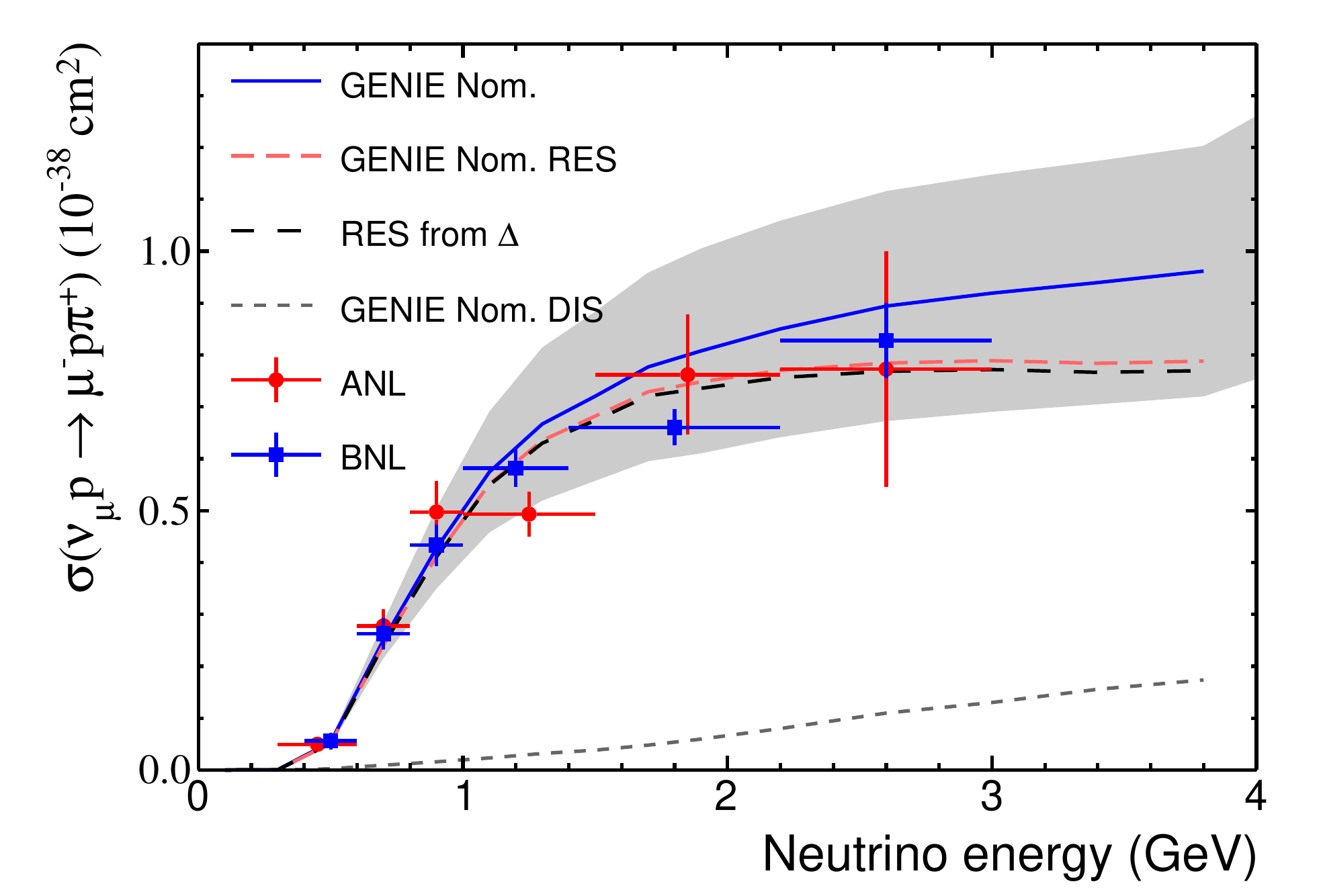}
    \caption{\ccppiplus}
  \end{subfigure}
  \begin{subfigure}{0.9\columnwidth}
    \includegraphics[width=\textwidth]{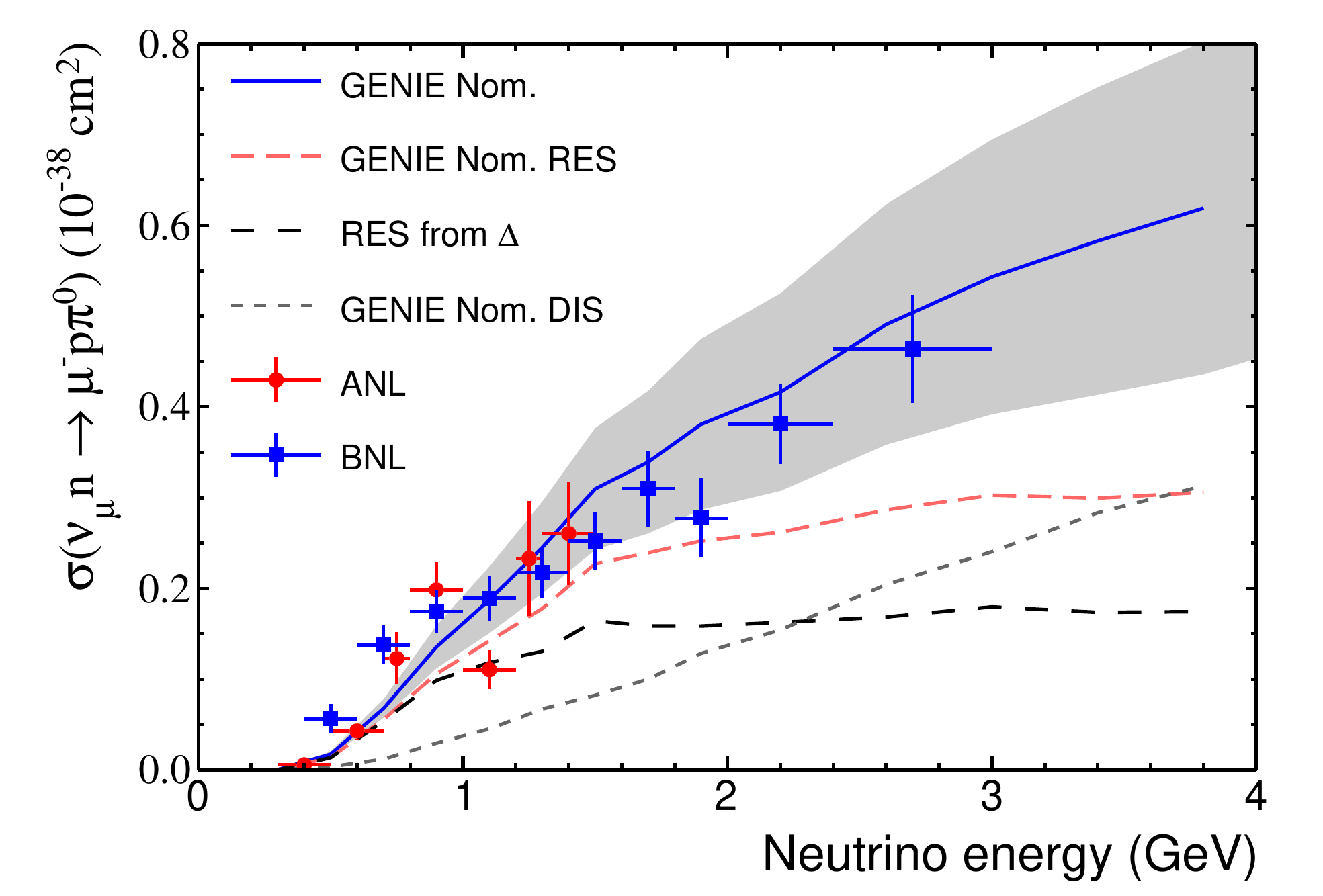}
    \caption{\ccnpizero}
  \end{subfigure}
  \begin{subfigure}{0.9\columnwidth}
    \includegraphics[width=\textwidth]{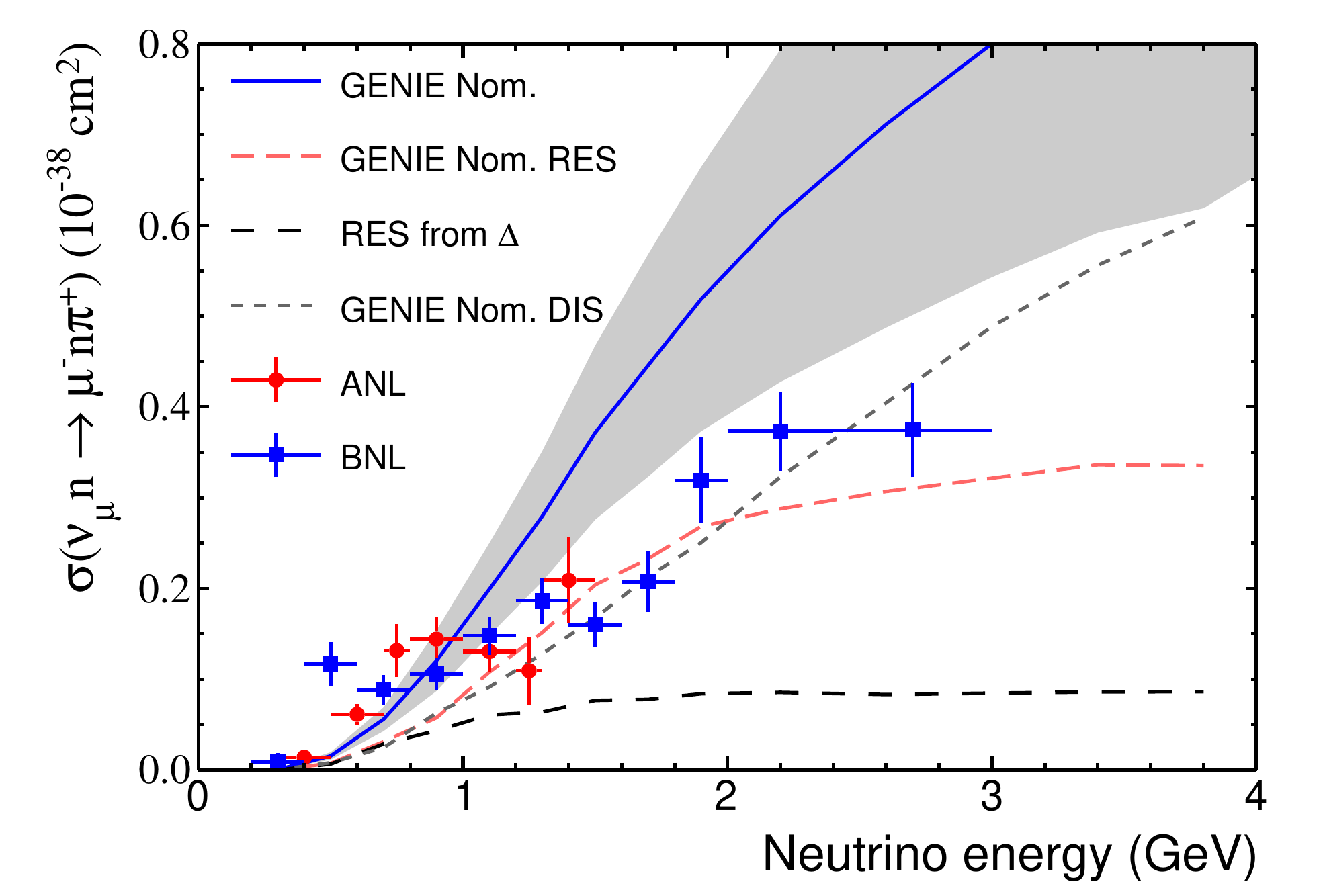}
    \caption{\ccnpiplus}
  \end{subfigure}
  \caption{The nominal GENIE prediction is shown as a function of \Ev for the three single pion production channels of interest, and is compared to the corrected ANL and BNL data. The total prediction is broken down into the resonant (RES) and non-resonant (DIS) contributions, and additionally, the $\Delta$-contribution to the RES component is shown. The error band shown on the total GENIE prediction shows the 1$\sigma$ error bands for all default GENIE parameters given in Table~\ref{tab:genie_parameters} combined in quadrature (note that \fa is not a default GENIE parameter so is not included in the error band).}\label{fig:nominal_Enu_comparisons}
\end{figure}

\begin{figure*}[htbp]
  \centering
  \begin{subfigure}{0.9\columnwidth}
    \includegraphics[width=\textwidth]{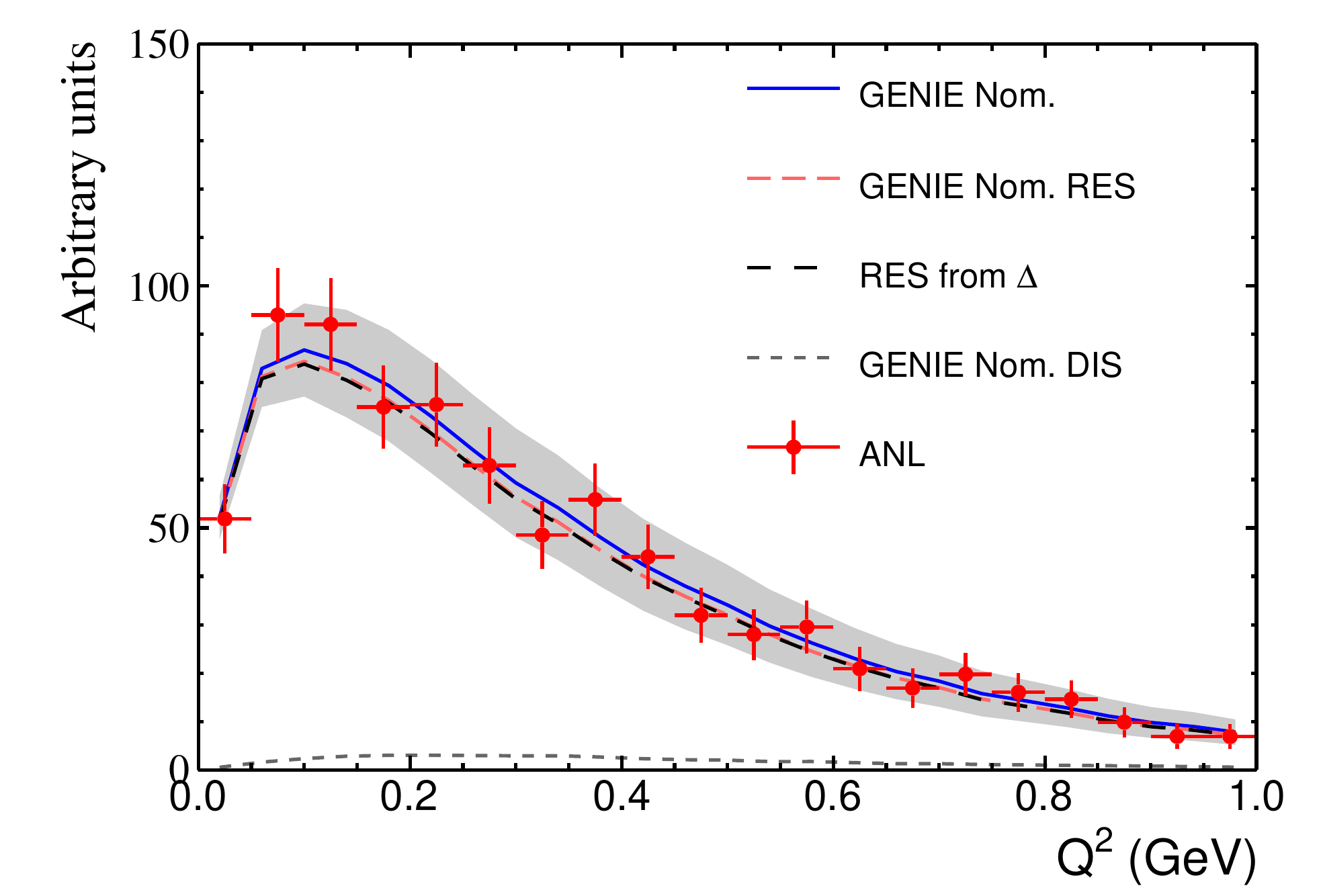}
    \caption{ANL \ccppiplus}
  \end{subfigure}
  \begin{subfigure}{0.9\columnwidth}
    \includegraphics[width=\textwidth]{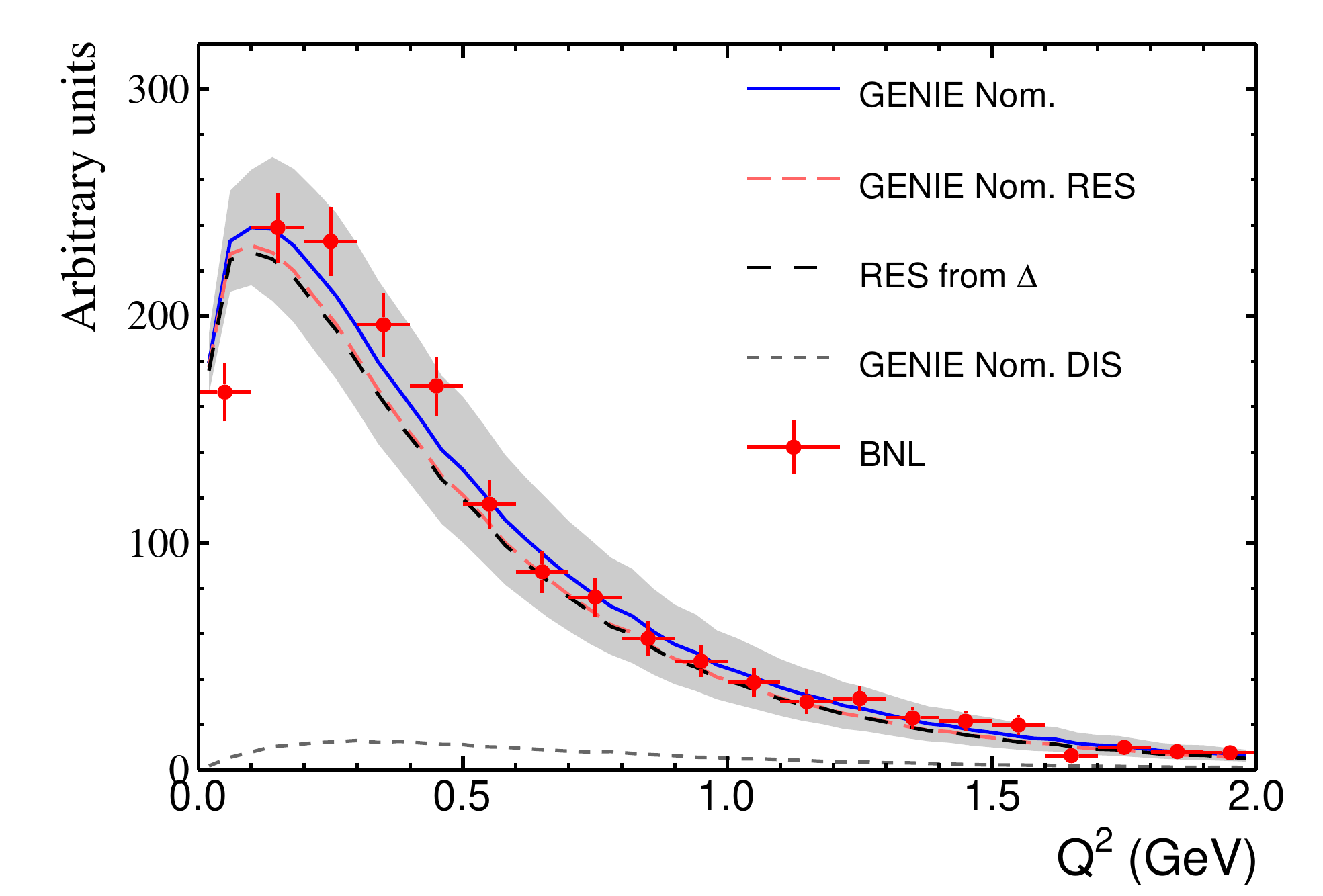}
    \caption{BNL \ccppiplus}
  \end{subfigure}
  \begin{subfigure}{0.9\columnwidth}
    \includegraphics[width=\textwidth]{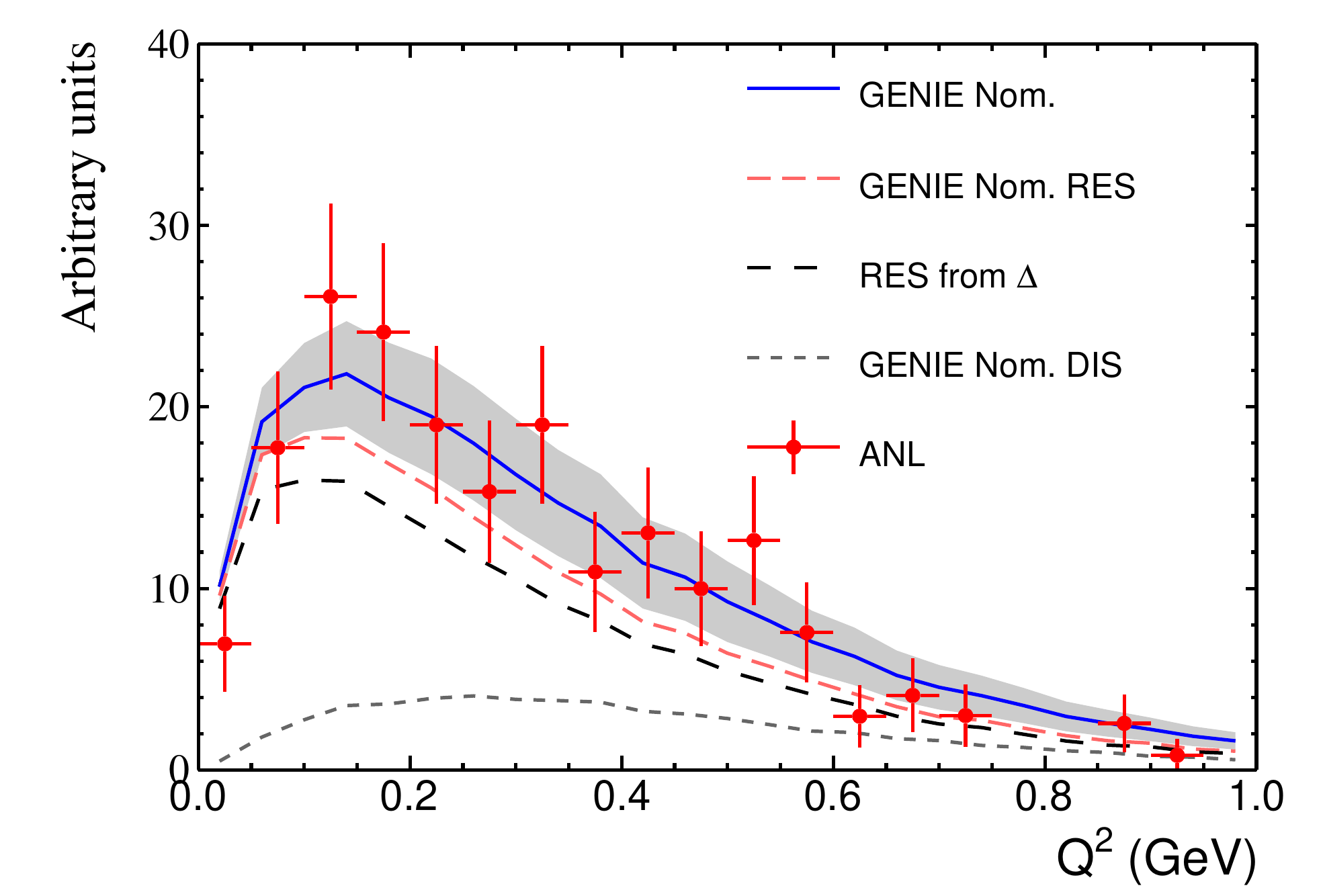}
    \caption{ANL \ccnpizero}
  \end{subfigure}
  \begin{subfigure}{0.9\columnwidth}
    \includegraphics[width=\textwidth]{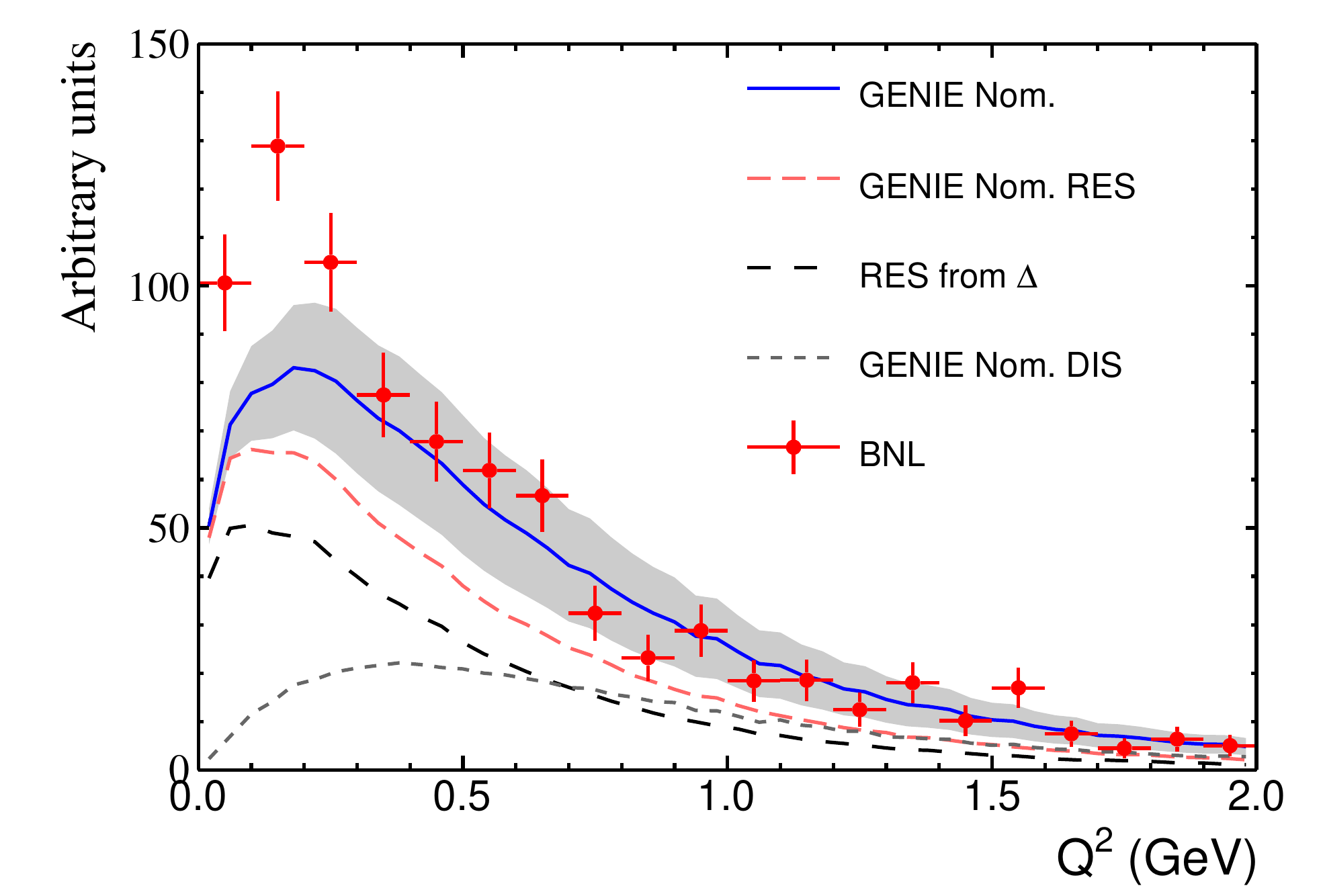}
    \caption{BNL \ccnpizero}
  \end{subfigure}
  \begin{subfigure}{0.9\columnwidth}
    \includegraphics[width=\textwidth]{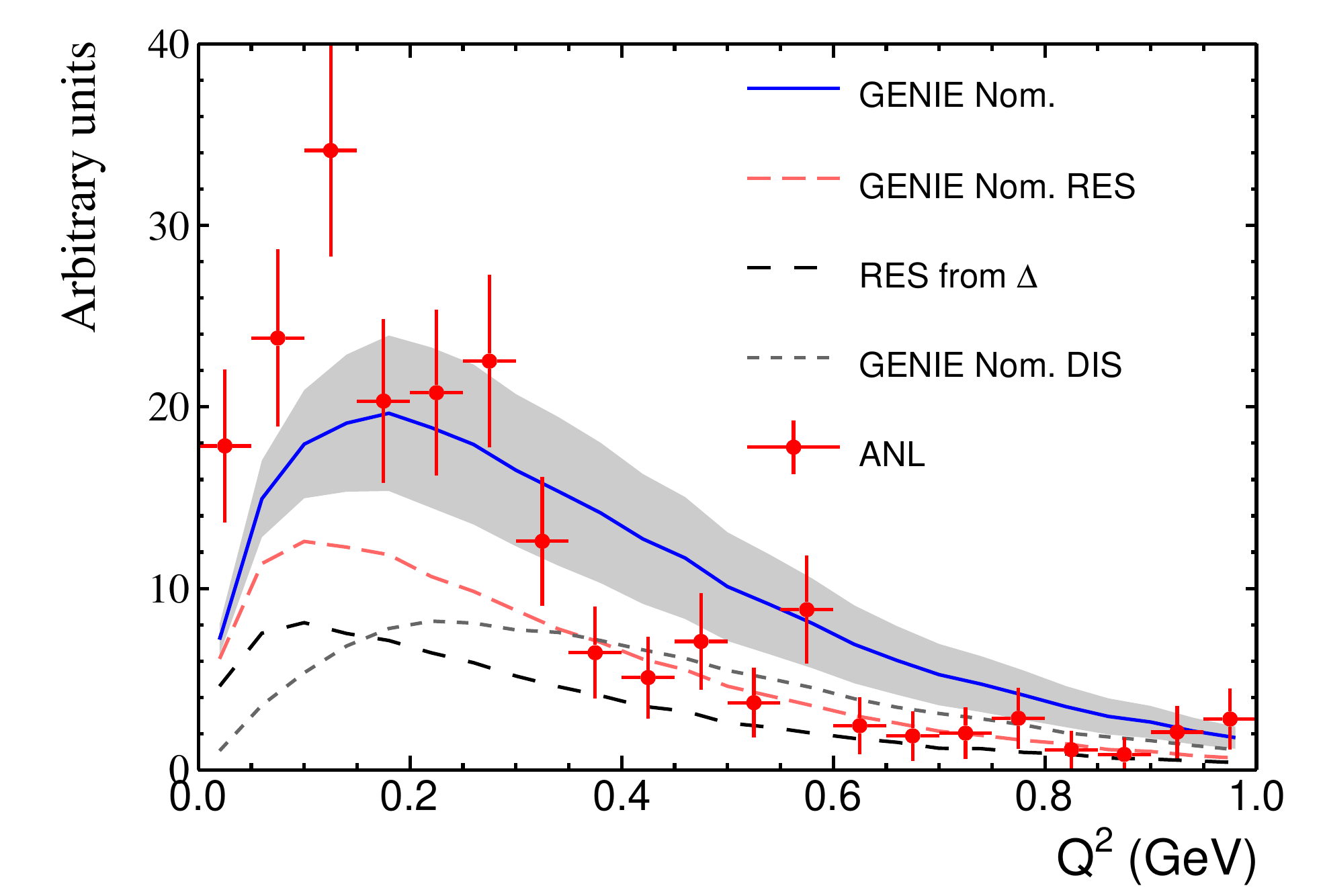}
    \caption{ANL \ccnpiplus}
  \end{subfigure}
  \begin{subfigure}{0.9\columnwidth}
    \includegraphics[width=\textwidth]{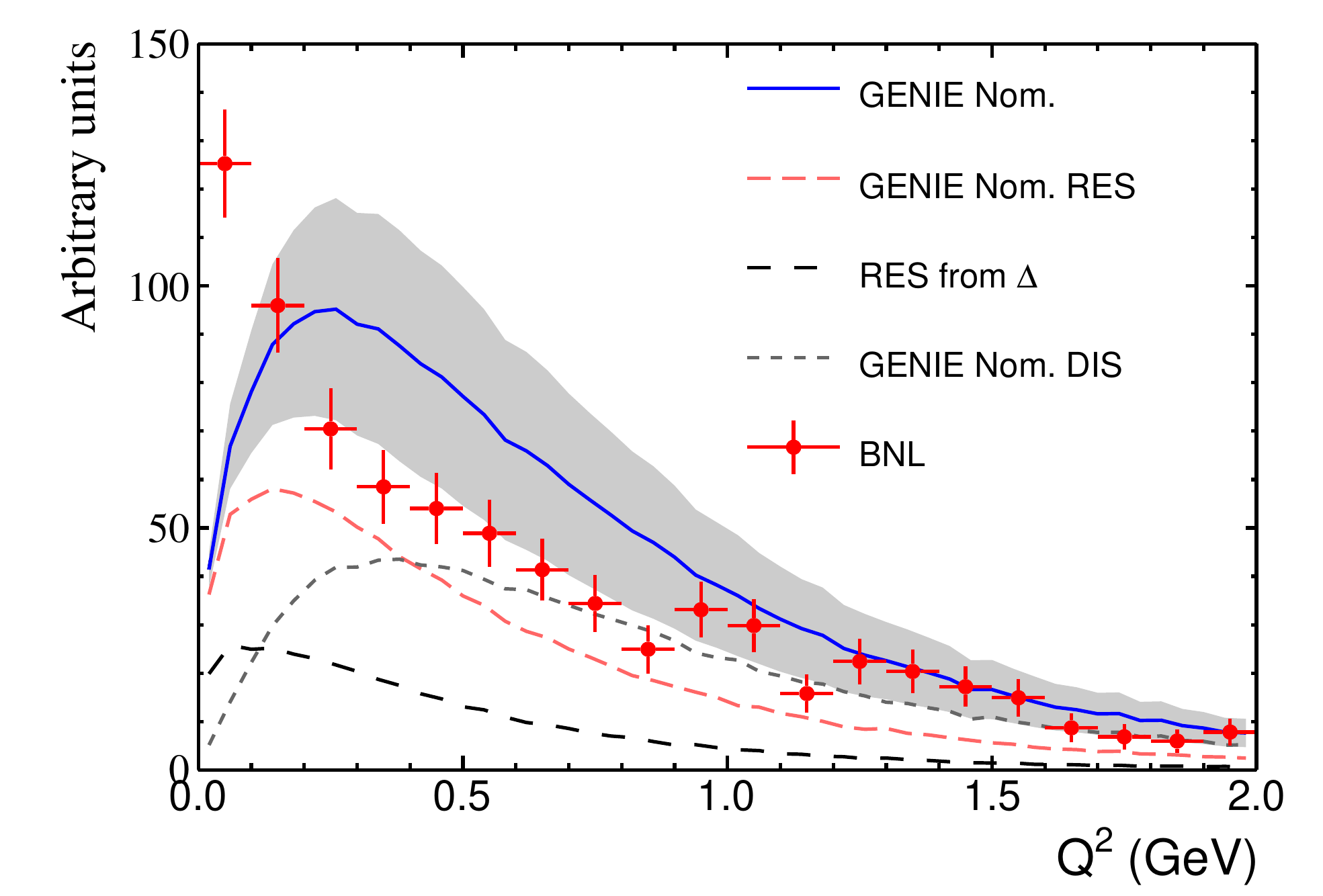}
    \caption{BNL \ccnpiplus}
  \end{subfigure}  
  \caption{The nominal GENIE prediction is shown as a function of \qq for the three single pion production channels of interest, and is compared separately to the ANL and BNL data. The total prediction is broken down into the resonant (RES) and non-resonant (DIS) contributions, and additionally, the $\Delta$-contribution to the RES component is shown. The error band shown on the total GENIE prediction shows the 1$\sigma$ error bands for all GENIE default parameters given in Table~\ref{tab:genie_parameters} combined in quadrature (note that \fa is not a default GENIE parameter so is not included in the error band).}\label{fig:nominal_Q2_comparisons}
\end{figure*}

The nominal GENIE v2.8.2 cross sections for the three single pion production channels considered in this work (\ccppiplus, \ccnpizero and \ccnpiplus) are shown as a function of the neutrino energy \Ev in Figure~\ref{fig:nominal_Enu_comparisons}, and compared with ANL and BNL data. The nominal GENIE v2.8.2 event rate predictions as a function of \qq, produced using the ANL and BNL fluxes, are compared separately to ANL and BNL data in Figure~\ref{fig:nominal_Q2_comparisons}. The GENIE prediction for the \qq distributions is normalized separately for ANL and BNL such that the total prediction is equal to the measured rate summed over all three channels. The \qq predictions shown are therefore shape-only, but with the relative normalization between the different pion production channels preserved\footnote{It is not possible to make a correctly normalized event rate prediction because neither experiment gives sufficient information on the number of target nucleons in the bubble chamber.}. All data considered have no invariant mass cuts applied and the event selection in GENIE is based on the particles produced at the initial interaction vertex, not those surviving GENIE's FSI model as previously discussed. In Figures~\ref{fig:nominal_Enu_comparisons} and~\ref{fig:nominal_Q2_comparisons}, the GENIE prediction is also shown broken down into resonant (RES) and non-resonant (DIS) contributions. Additionally, the dominant $\Delta$ contribution to the RES component is shown separately for reference. The total GENIE prediction is the incoherent sum of the RES and DIS contributions, where interference terms have been neglected. On each plot the 1$\sigma$ error band produced by combining the nominal GENIE uncertainties on \mares, RES normalization and DIS normalization is also shown for comparison.

It is clear from Figures~\ref{fig:nominal_Enu_comparisons} and~\ref{fig:nominal_Q2_comparisons} that the nominal GENIE prediction cannot describe all of the pion production channels well for the reanalyzed datasets. In Figure~\ref{fig:nominal_Enu_comparisons}, it is noticeable that, while the measured cross sections for the subdominant \ccnpizero and \ccnpiplus channels are similar, there are large differences between the nominal GENIE predictions for these channels. The non-resonant component of the GENIE prediction, which contributes strongly to these channels, appears to be too large. It can be seen from Figure~\ref{fig:nominal_Q2_comparisons} that the nominal GENIE prediction fails to describe the low-\qq data well for some channels. We also note that the GENIE uncertainties are larger than the data suggests, and may be reduced by tuning the GENIE model to the ANL and BNL data. These observations motivate this work. 

\section{Fitting the GENIE model}\label{sec:fit}
In this section, the datasets described in Section~\ref{sec:datasets} are used to constrain the GENIE model introduced in Section~\ref{sec:genie_model}. The $\chi^{2}$ statistic which is minimized is given in Section~\ref{sec:chi2}, and results are given in Section~\ref{sec:results}. A discussion of the goodness of fit is given in Section~\ref{sec:goodness_of_fit}. The MINUIT package~\cite{minuit} as implemented in the ROOT library~\cite{root} is used to perform all fits.

The fits are performed separately for four GENIE configurations. Both GENIE v2.6.2 and GENIE v2.8.2 predictions are fit using all parameters available in the standard version of GENIE (as described in Table~\ref{tab:genie_parameters}). A fit was performed to GENIE v2.8.2 where the normalization of the resonant axial form factor, \fa, is used as a fit parameter instead of the resonant normalization, as motivated in Section~\ref{sec:genie_model}. Finally, a fit was performed to GENIE v2.8.2 without the resonant normalization or \fa to investigate the effect that correlations between the axial mass (which has a strong effect on the normalization of the cross section) and the normalization parameters have on the results.

\subsection{\texorpdfstring{\boldmath$\chi^{2}$}{chi^2} definition}\label{sec:chi2}
No information about systematic uncertainties, correlations within datasets or correlations between datasets is available, so only statistical errors are considered for all datasets, and the function to be minimized can be expressed as a sum over the datasets included in the fit. This is reasonable as the statistical uncertainties are large. Additionally, the datasets used are efficiency corrected by the experiments, but are not unfolded, so the treatment of detector effects is likely to be inadequate\footnote{Unfortunately, insufficient information has been published to do a more sophisticated analysis, so this is a caveat which applies to all analyses which use ANL or BNL data.}. For these reasons, any measure of goodness of fit should be treated as approximate.

A Poisson-likelihood statistic is used for the datasets as a function of \qq because many of the higher \qq bins have low event rates. Note that for \qq datasets, the sum is over the $N$ bins with \qq $\geq$ 0.1 GeV$^{2}$.
\begin{align}
  \chi^{2} &= \sum_{\qq\;\mathrm{datasets}} \left\{2\sum^{N}_{i=1}\left[ \mu_{i}(\vec{\mathbf{x}}) - n_{i} + n_{i}\ln\frac{n_{i}}{\mu_{i}(\vec{\mathbf{x}})}\right] \right\}\notag\\
  &+ \sum_{E_{\nu}\;\mathrm{datasets}}\left\{\sum^{N}_{i=1}\frac{\left[n_{i}-\mu_{i}(\vec{\mathbf{x}})\right]^{2}}{\sigma_{i}^{2}}\right\}
  \label{eq:chi2}
\end{align}
\noindent where $n_{i}$ and $\mu_{i}(\vec{\mathbf{x}})$ are the measured and predicted number of events in the $i$th bin, $\sigma_{i}$ is the statistical error on the $i$th bin, $\vec{\mathbf{x}}$ are the model parameters varied in the fit and the inner summations are over the $N$ bins of each dataset. $\vec{\mathbf{x}}$ also contains normalization terms for ANL and BNL which affect the \qq datasets only. As previously remarked, the \qq datasets are shape-only in the fit, but the relative normalization between the three pion production modes is preserved separately for ANL and BNL.

Note that this statistic is appropriate for minimization, but $\chi^{2}/\ndof$ is not strictly correct as measure of the goodness of fit because the Poisson-likelihood terms contribute constant terms to the $\chi^{2}$. A more rigorous measure of the goodness of fit is discussed in Section~\ref{sec:goodness_of_fit}.

\FloatBarrier
\subsection{Results}\label{sec:results}
Two fake data studies were performed to validate the fitter. Firstly, Asimov~\cite{asimov} fake data fits were produced for all four GENIE configurations considered. These provide basic validation that the fitter found the correct minimum and give the expected size of the parameter uncertainties, which can be used to validate the fit results. Secondly, pull studies were performed for all GENIE configurations to check that the test statistic is an unbiased estimator of central values and uncertainties for the parameters varied in the fits. No biases were observed.

\begin{figure}[htbp]
  \centering
  \begin{subfigure}{0.9\columnwidth}
    \includegraphics[width=\textwidth]{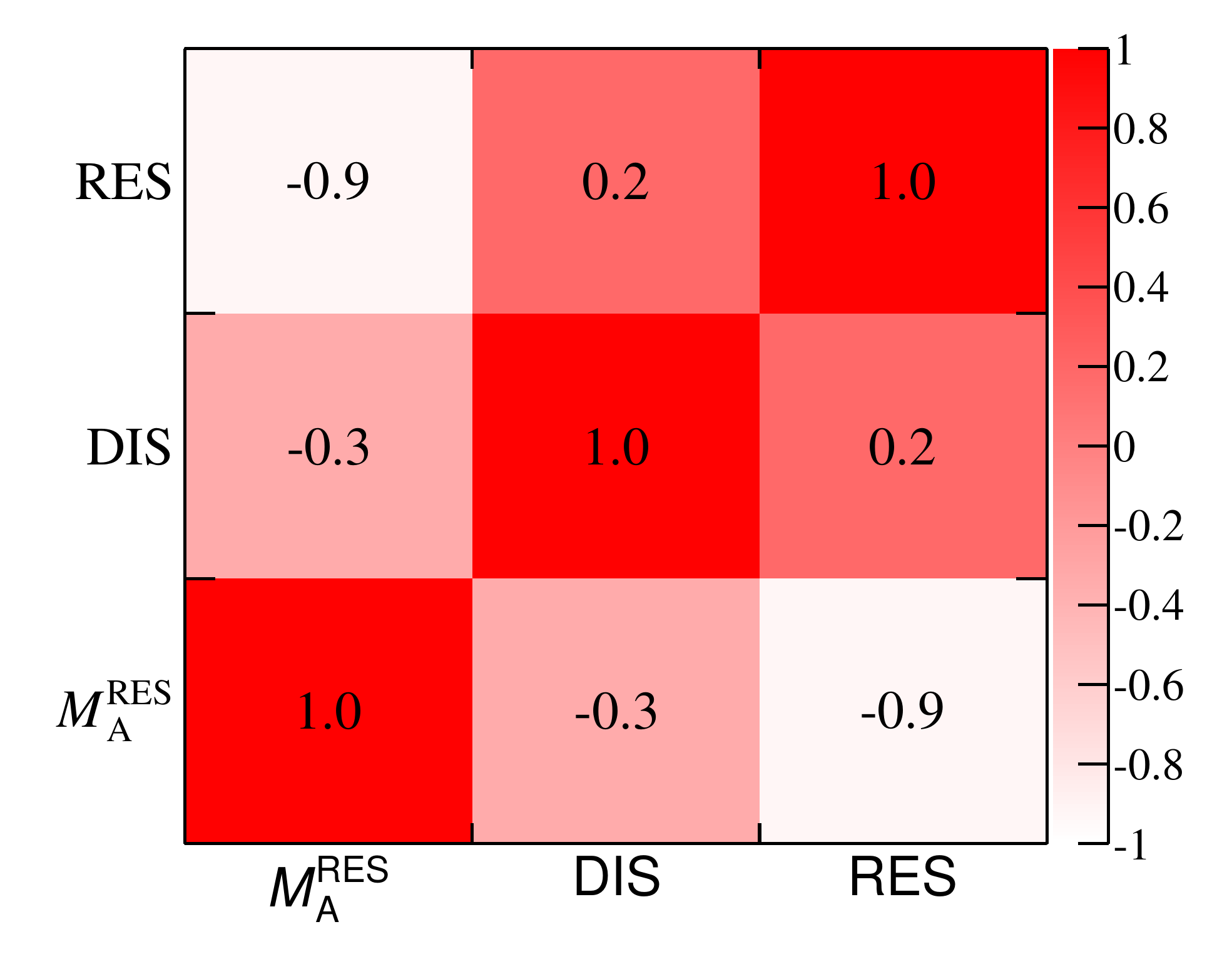}
    \caption{GENIE v2.8.2 (RES)}
  \end{subfigure}
  \begin{subfigure}{0.9\columnwidth}
    \includegraphics[width=\textwidth]{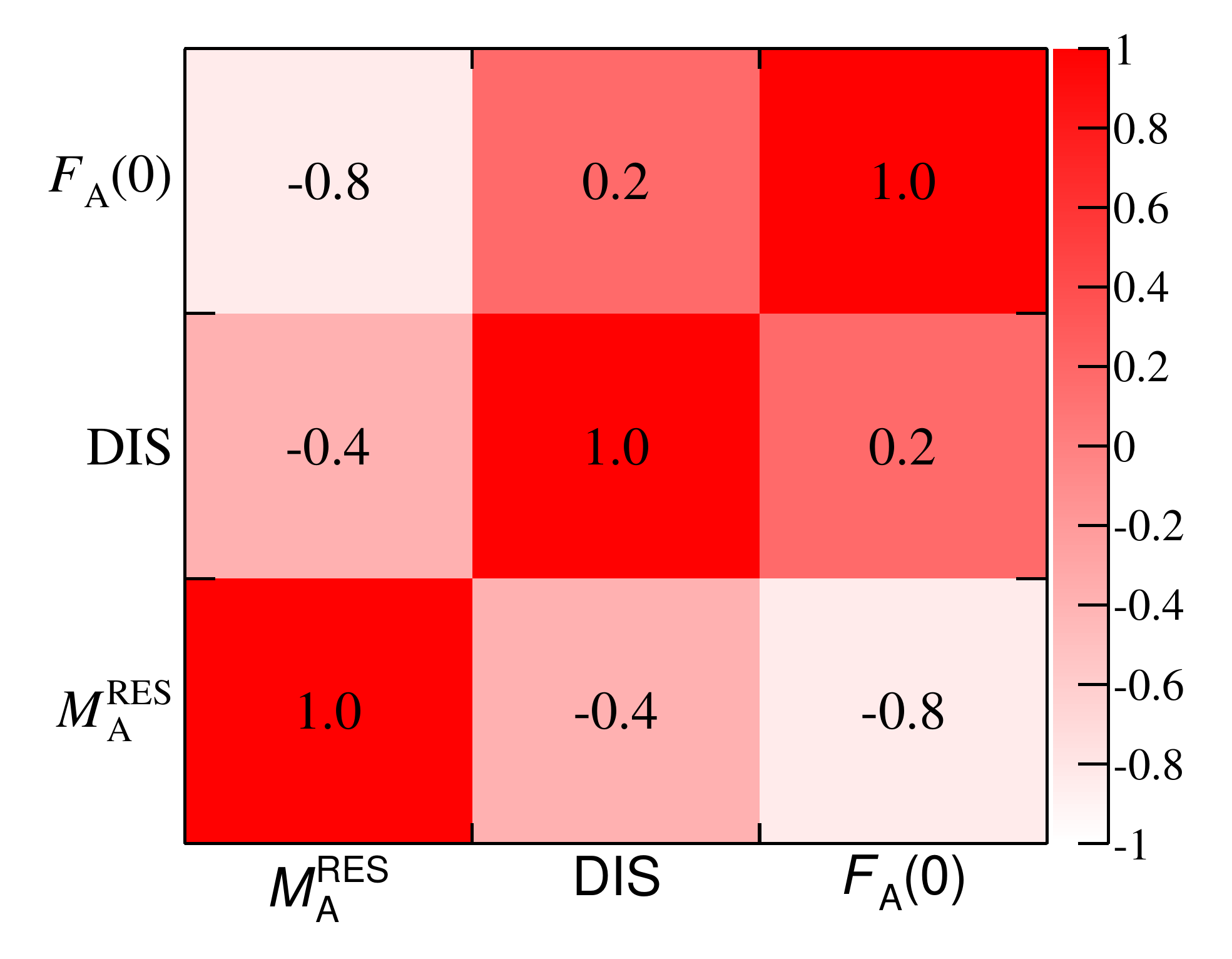}
    \caption{GENIE v2.8.2 (\fa)}
  \end{subfigure}
  \caption{Correlation matrices from the GENIE v2.8.2 (RES) and GENIE v2.8.2 (\fa) fits. The GENIE v2.6.2 correlation matrix is very similar to the GENIE v2.8.2 (RES) matrix, and the GENIE v2.8.2 (no norm.) matrix is very similar to the relevant bins of both of the matrices shown.}\label{fig:covariances}
\end{figure}

The best fit results to all twelve datasets are shown in Table~\ref{tab:best_fit_results} for the four GENIE configurations considered in this work. The parameter uncertainties given by the fits are consistent with those predicted by the Asimov fake data study. The normalization of the non-resonant background was reduced significantly in all fits, as was expected given the nominal model comparisons shown in Section~\ref{sec:genie_model}. The resonant axial mass, \mares, was also reduced from the GENIE nominal value of \mares = 1.12 GeV in all fits, although we note that there is a  strong anticorrelation between \mares and the RES normalization, and between \mares and \fa (as can be seen in Figure~\ref{fig:covariances}). The GENIE v2.8.2 fits were also repeated using GENIE's free nucleon cross sections in order to ensure that the results are not biased by GENIE's initial state nuclear model. It was found that the results were consistent to within 1$\sigma$ for all parameters in all fits.

The GENIE v2.6.2 and GENIE v2.8.2 (RES) results, with the same parameters available in the standard version of GENIE (as described in Table~\ref{tab:genie_parameters}) give consistent results. The fit which uses the normalization of the axial form factor, \fa, as a fit parameter is consistent with the other fits, although as \fa has a small \qq dependence, the correlation with \mares is different and the value for \mares is less suppressed than in fits with RES normalization free (which has no \qq dependence).
\begin{table*}
  \centering      
      {\renewcommand{\arraystretch}{1.2}
        \begin{tabular}{c|cccc}
          \hline
          GENIE & v2.6.2 (RES) & v2.8.2 (RES) & v2.8.2 (\fa) & v2.8.2 (no norm.)\\
          \hline
          $\chi^{2}_{nom}$ & 558.8 & 615.3 & 615.3 & 615.3 \\
          $\chi^{2}_{min}$ & 311.2 & 324.4 & 327.3 & 330.3 \\
          \ndof & 157 & 157 & 157 & 158 \\
          \mares (GeV) & 0.94 $\pm$ 0.05 & 0.94 $\pm$ 0.05 & 1.00 $\pm$ 0.04 & 1.04 $\pm$ 0.02 \\
          DIS norm. (\%) & 46 $\pm$ 4 & 43 $\pm$ 4 & 43 $\pm$ 4 & 42 $\pm$ 4 \\
          RES norm. (\%) & 115 $\pm$ 7 & 115 $\pm$ 7 & -- & -- \\
          \fa norm. (\%) & -- & -- & 107 $\pm$ 4 & -- \\
          \hline
      \end{tabular}}
\caption{Best fit results for the four GENIE configurations used in this analysis.}\label{tab:best_fit_results}
\end{table*}

\begin{figure*}[htbp]
  \centering
  \begin{subfigure}{0.9\columnwidth}
    \includegraphics[width=\textwidth]{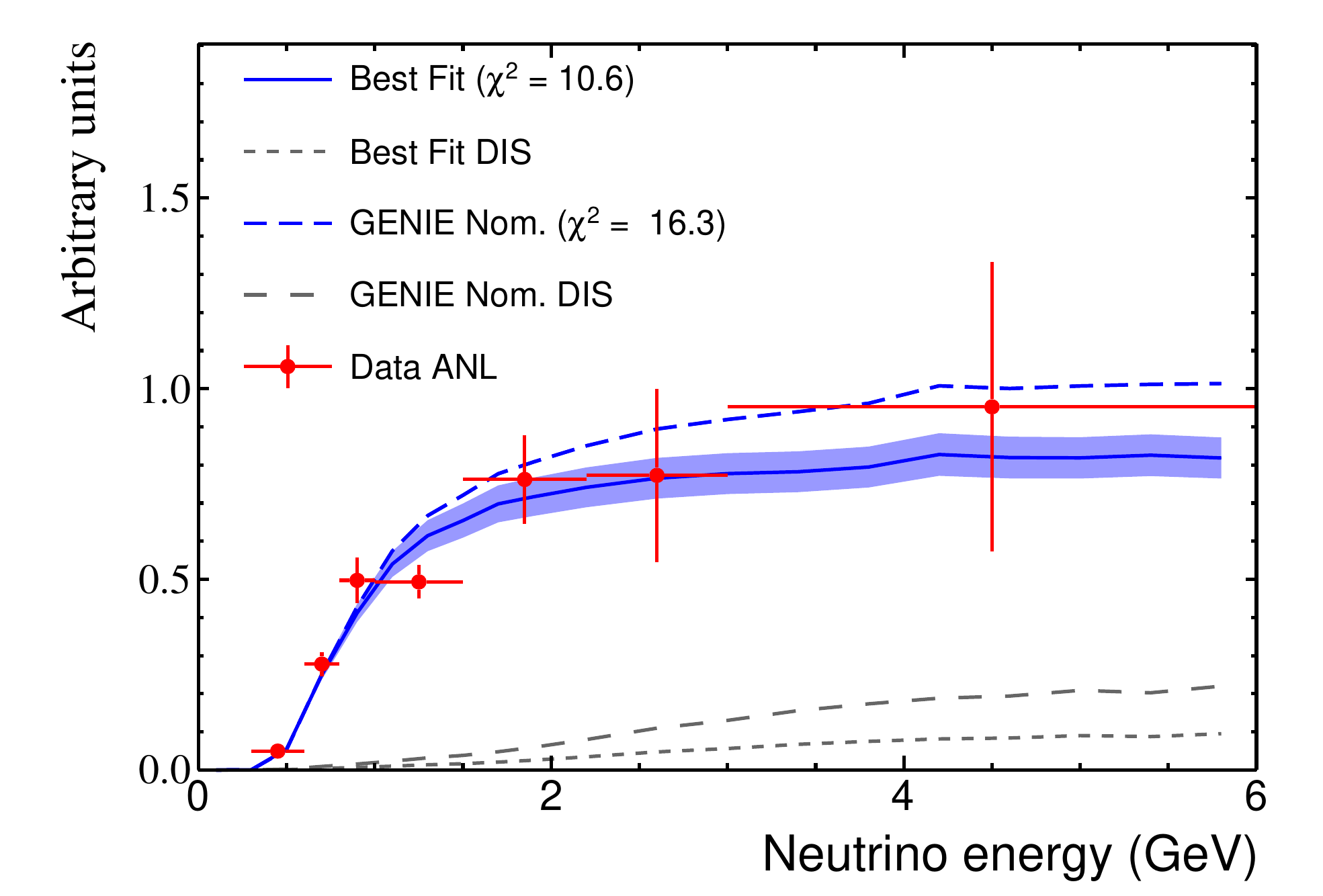}
    \caption{ANL \Ev}
  \end{subfigure}
  \begin{subfigure}{0.9\columnwidth}
    \includegraphics[width=\textwidth]{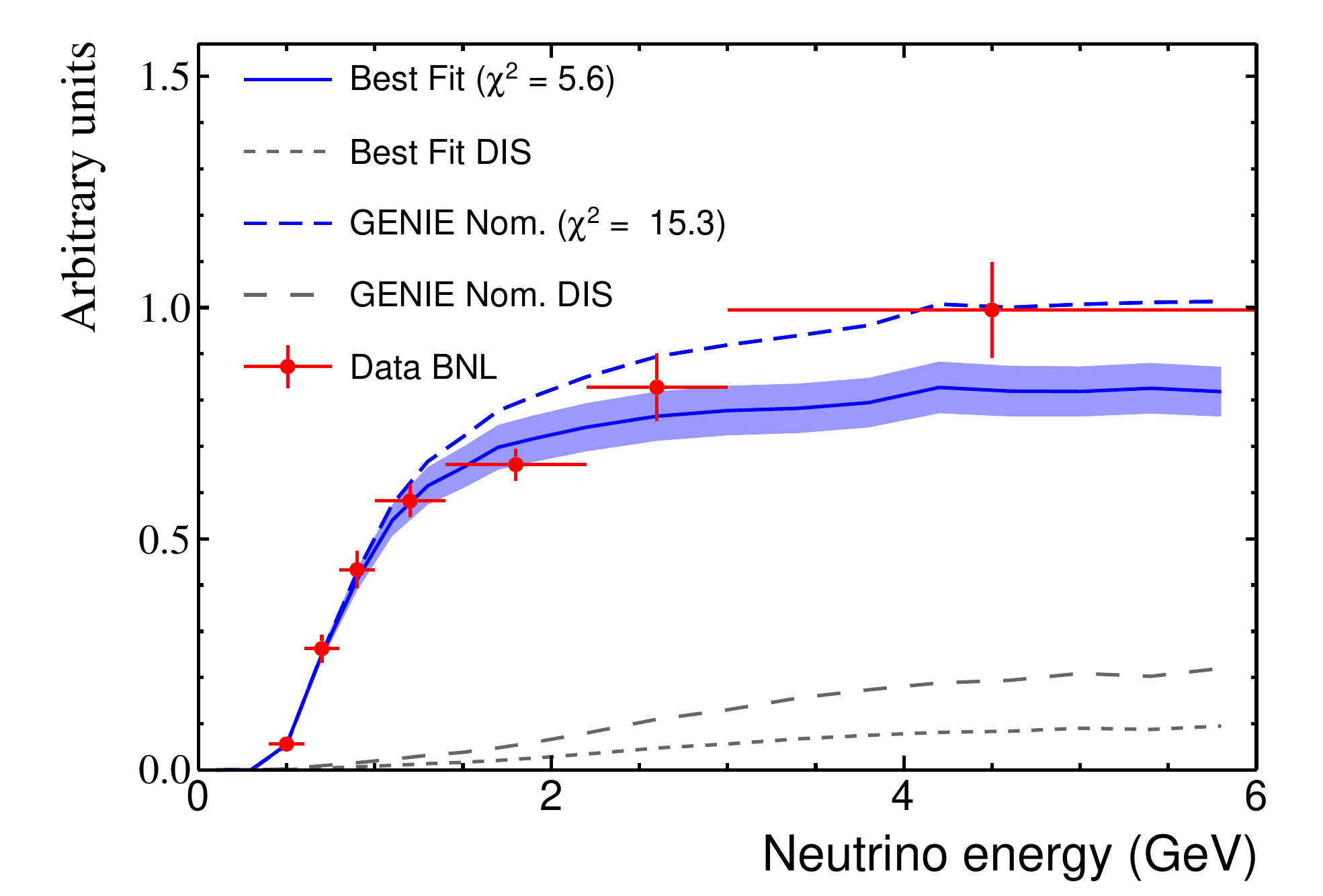}
    \caption{BNL \Ev}
  \end{subfigure}
  \begin{subfigure}{0.9\columnwidth}
    \includegraphics[width=\textwidth]{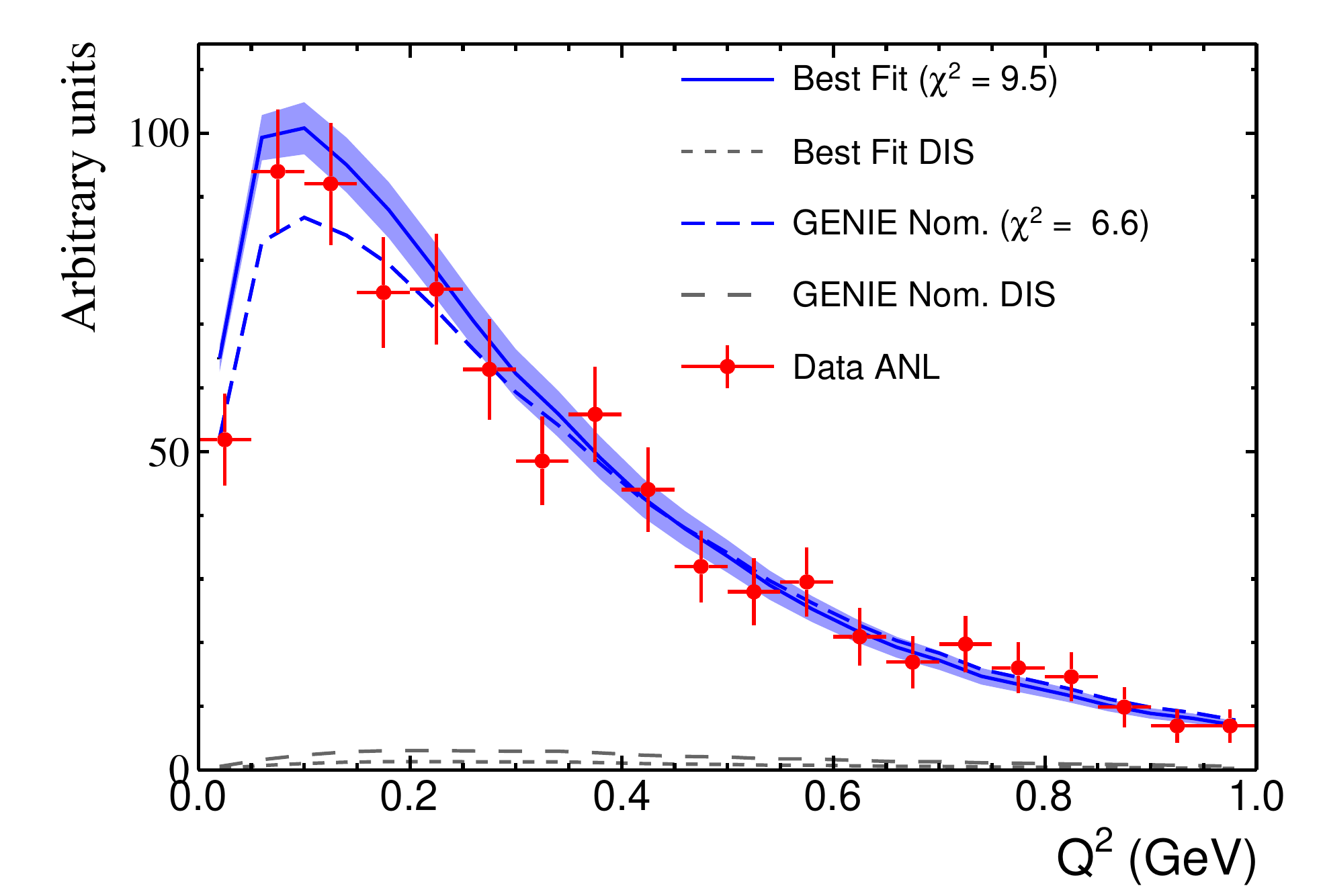}
    \caption{ANL \qq}
  \end{subfigure}
  \begin{subfigure}{0.9\columnwidth}
    \includegraphics[width=\textwidth]{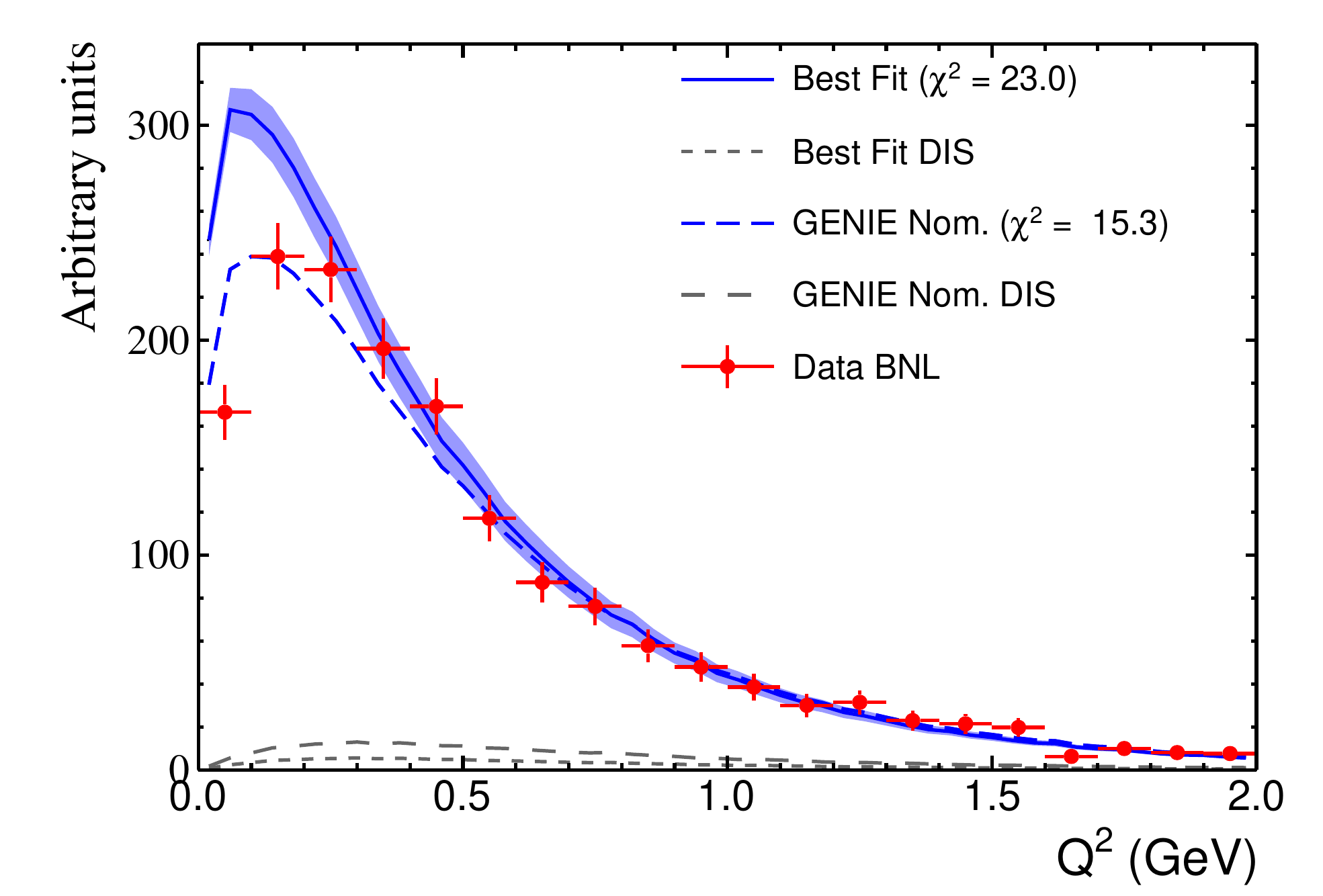}
    \caption{BNL \qq}
  \end{subfigure}
  \caption{Best fit distributions and post-fit uncertainties for the four \ccppiplus datasets included in the GENIE v2.8.2 (RES) fit. The nominal prediction is shown for reference, and the $\chi^{2}$ contribution from each dataset is given in the legend for both the nominal and best fit distributions. The nominal and best fit DIS contribution to the total GENIE prediction (RES+DIS) are also shown for reference.}\label{fig:best_fit_channel0}
\end{figure*}

\begin{figure*}[htbp]
  \centering
  \begin{subfigure}{0.9\columnwidth}
    \includegraphics[width=\textwidth]{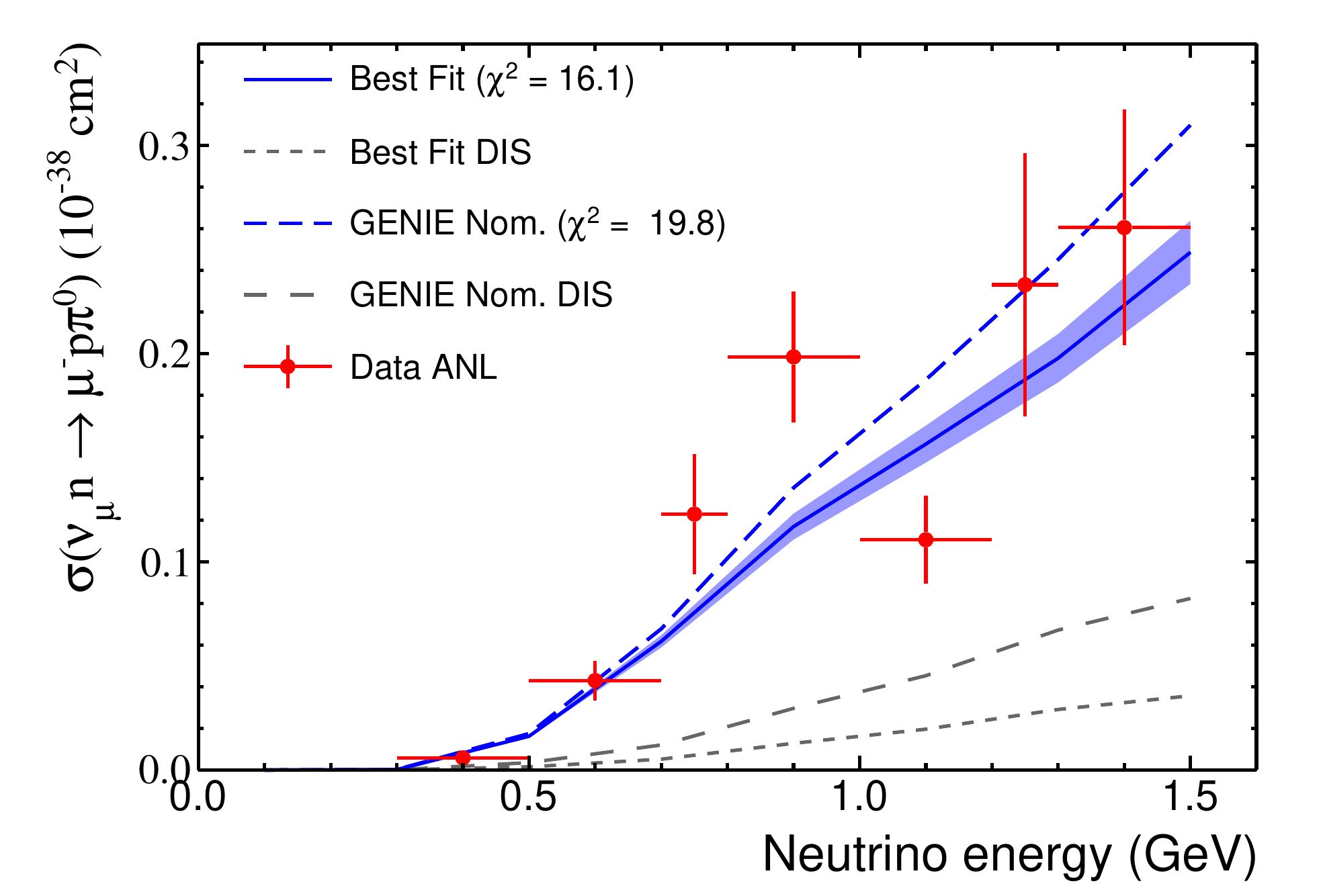}
    \caption{ANL \Ev}
  \end{subfigure}
  \begin{subfigure}{0.9\columnwidth}
    \includegraphics[width=\textwidth]{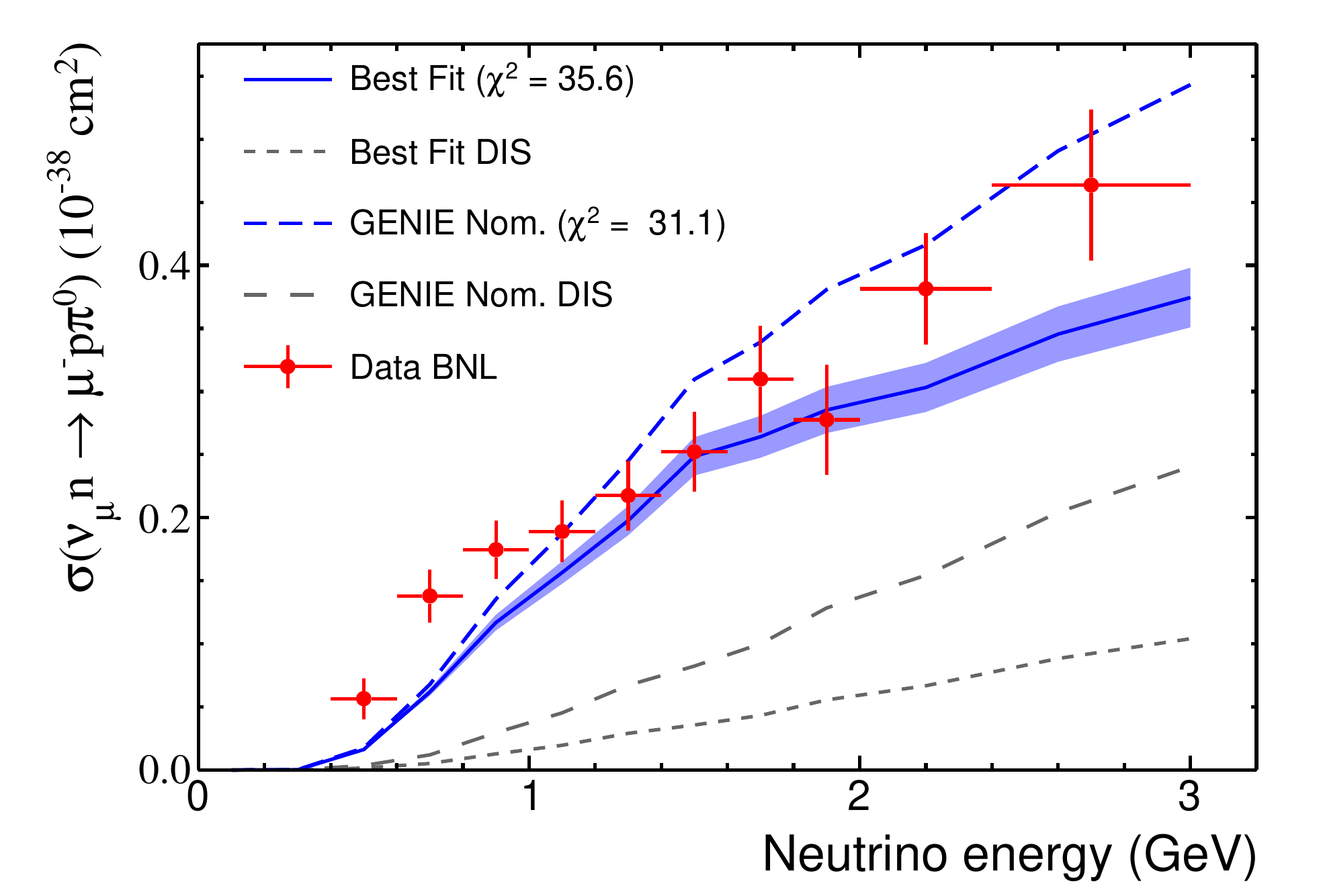}
    \caption{BNL \Ev}
  \end{subfigure}
  \begin{subfigure}{0.9\columnwidth}
    \includegraphics[width=\textwidth]{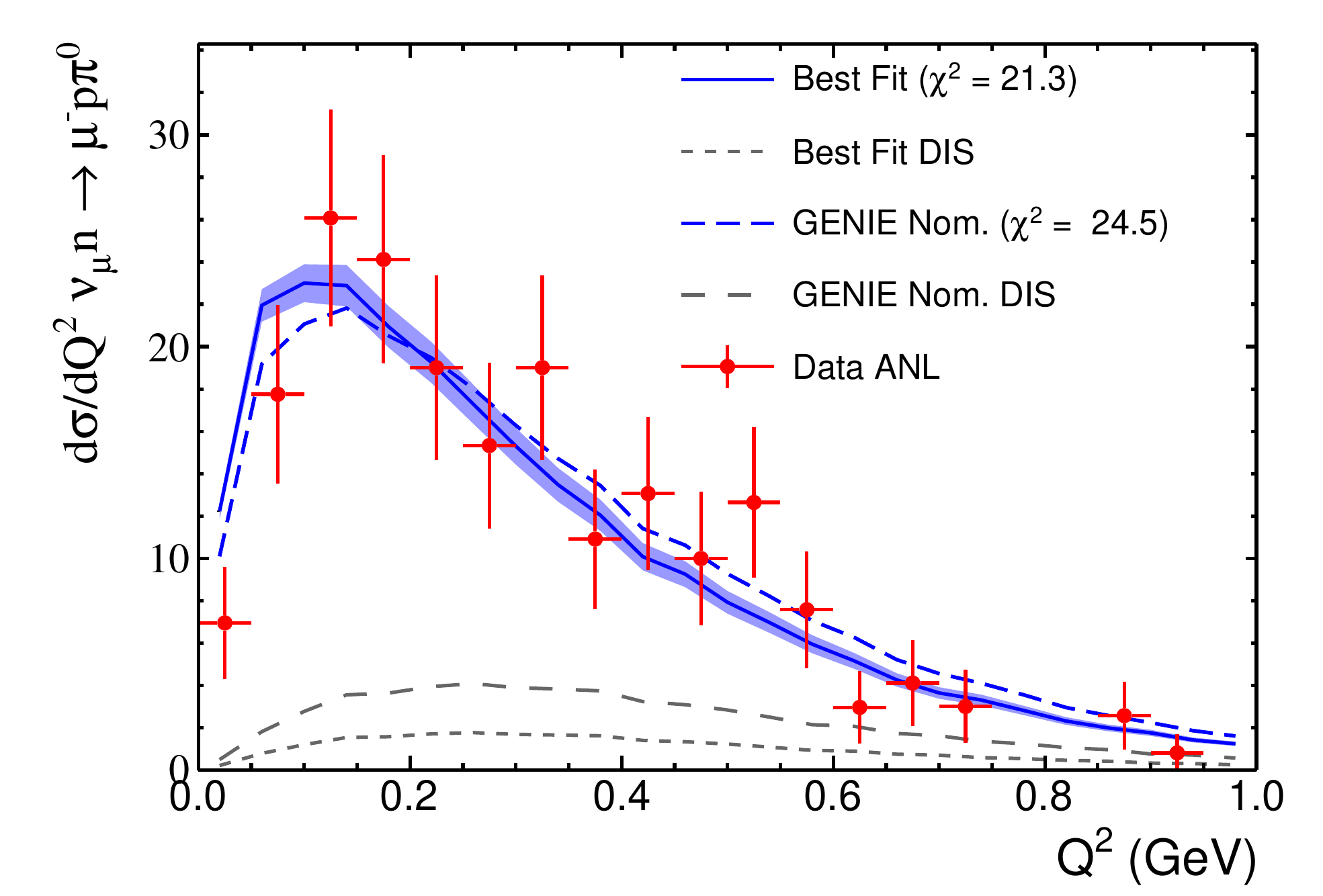}
    \caption{ANL \qq}
  \end{subfigure}
  \begin{subfigure}{0.9\columnwidth}
    \includegraphics[width=\textwidth]{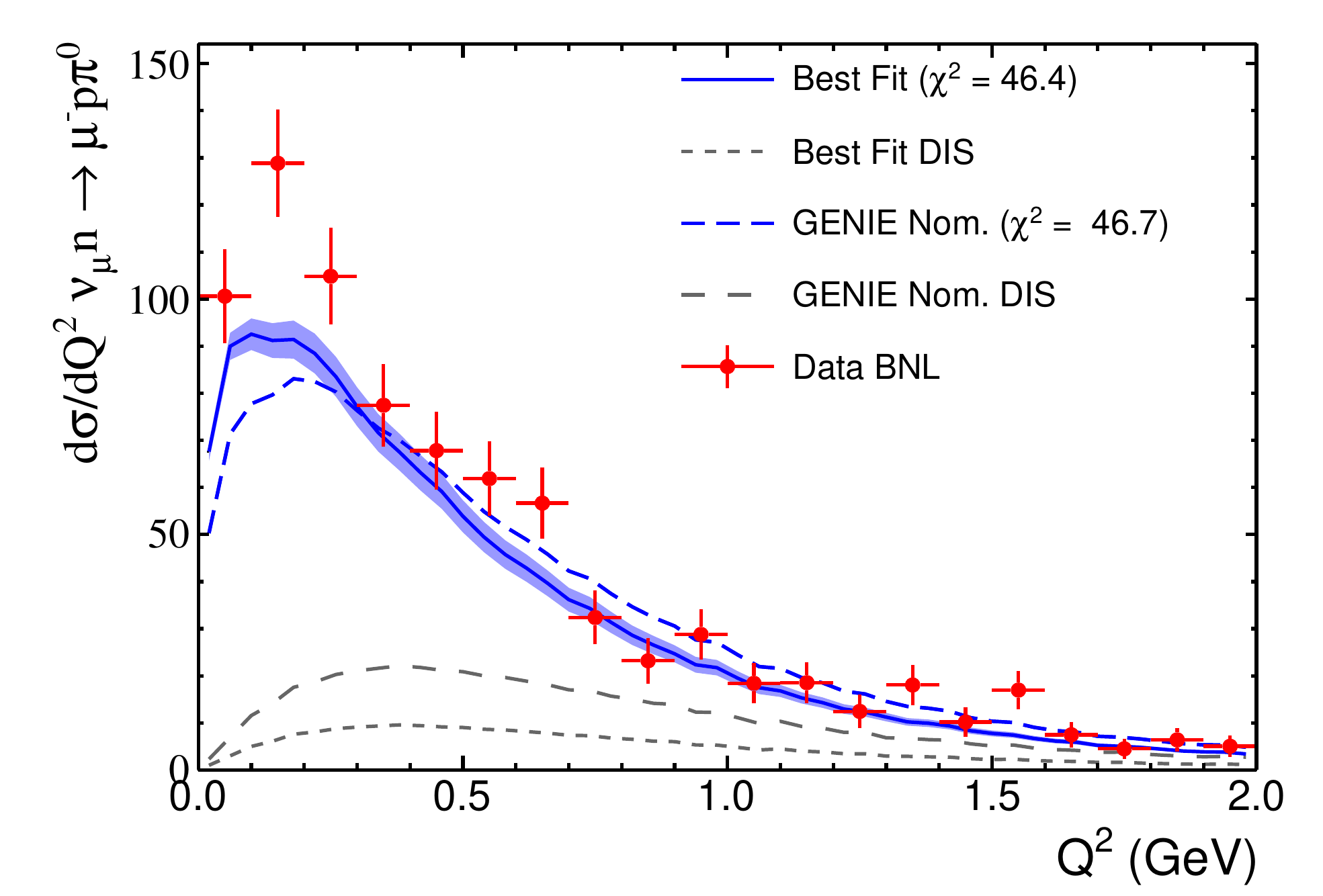}
    \caption{BNL \qq}
  \end{subfigure}
  \caption{Best fit results and post-fit uncertainties for the four \ccnpizero datasets included in the GENIE v2.8.2 (RES) fit. The nominal prediction is shown for reference, and the $\chi^{2}$ contribution from each dataset is given in the legend for both the nominal and best fit distributions. The nominal and best fit DIS contribution to the total GENIE prediction (RES+DIS) are also shown for reference.}\label{fig:best_fit_channel1}
\end{figure*}

\begin{figure*}[htbp]
  \centering
  \begin{subfigure}{0.9\columnwidth}
    \includegraphics[width=\textwidth]{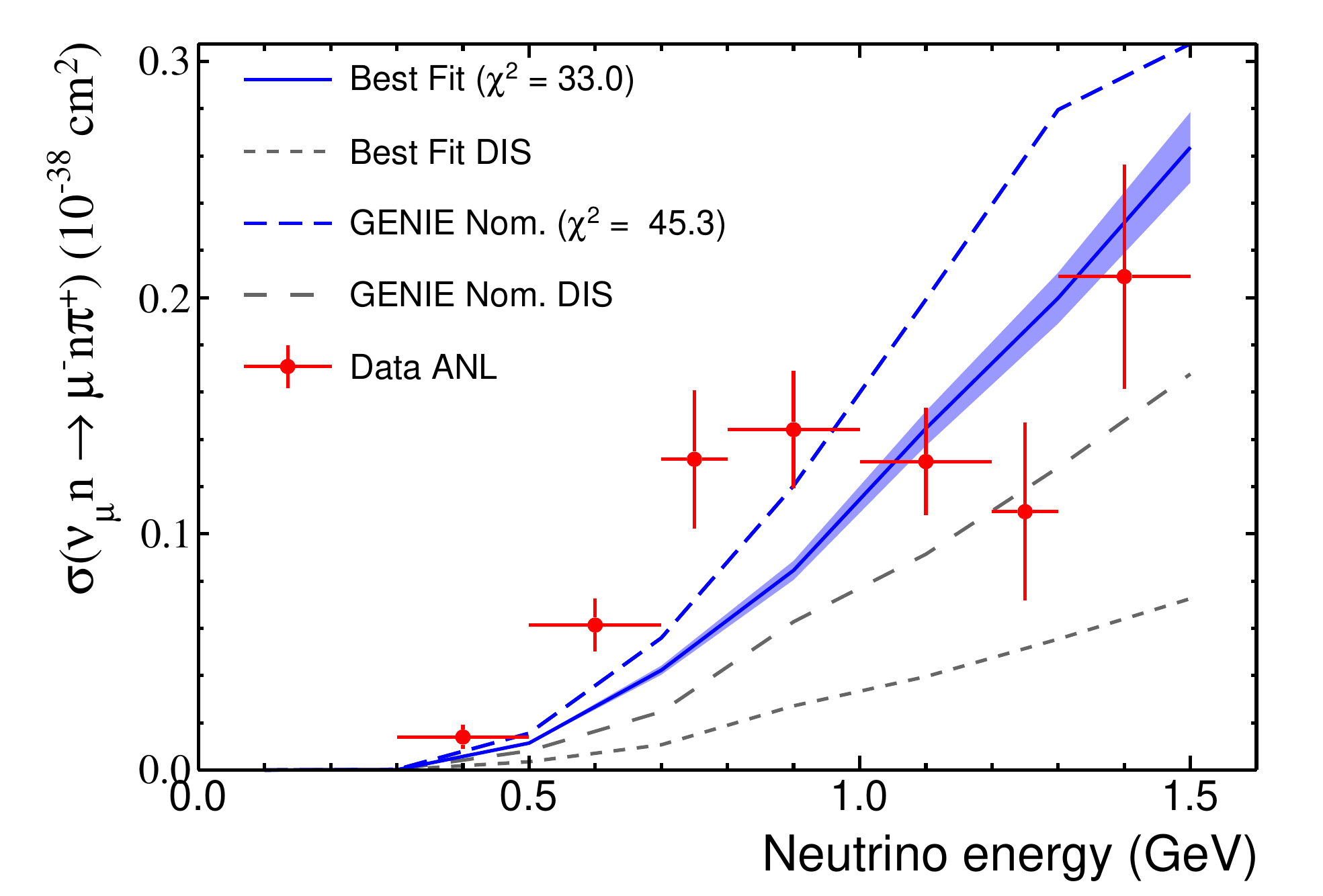}
    \caption{ANL \Ev}
  \end{subfigure}
  \begin{subfigure}{0.9\columnwidth}
    \includegraphics[width=\textwidth]{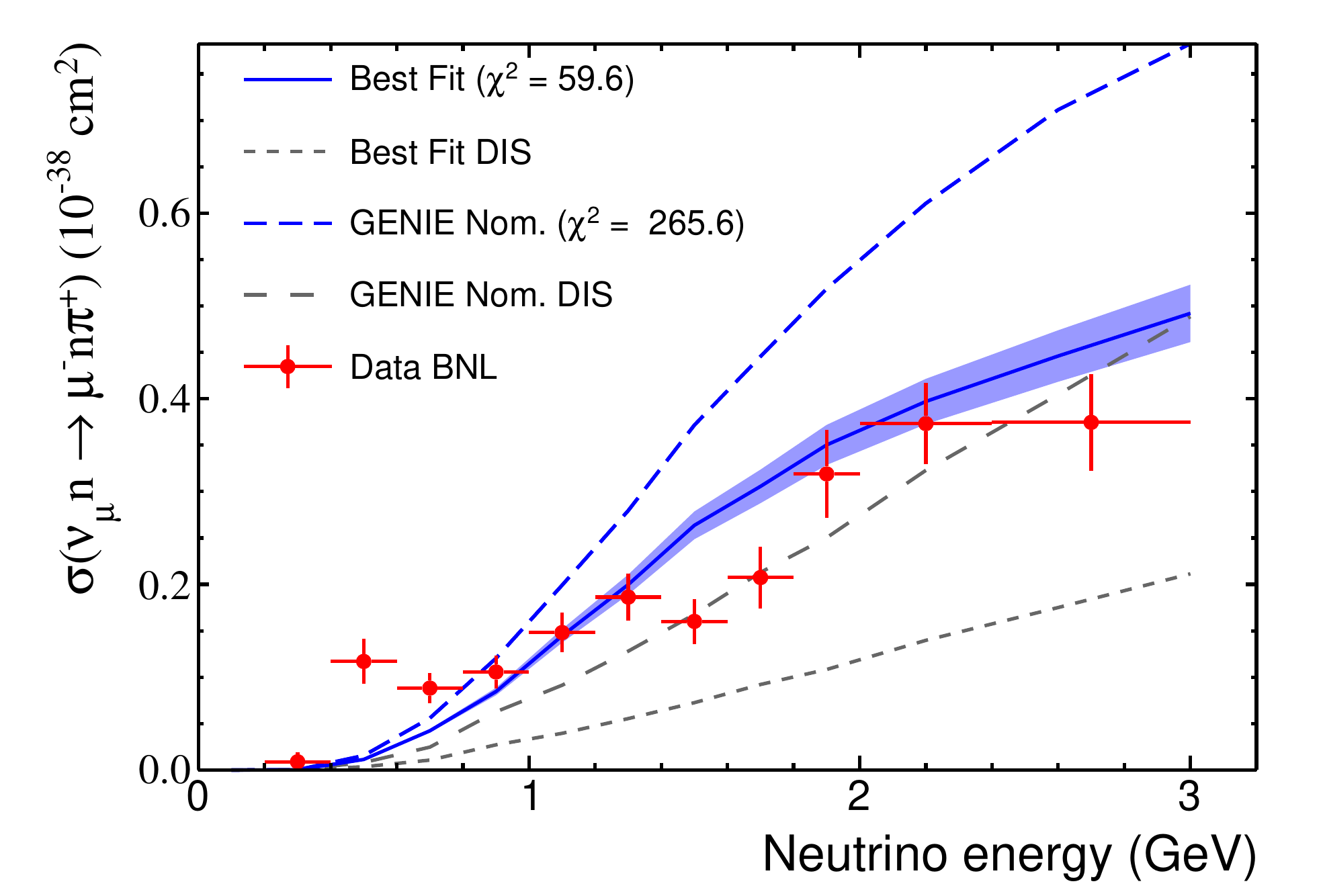}
    \caption{BNL \Ev}
  \end{subfigure}
  \begin{subfigure}{0.9\columnwidth}
    \includegraphics[width=\textwidth]{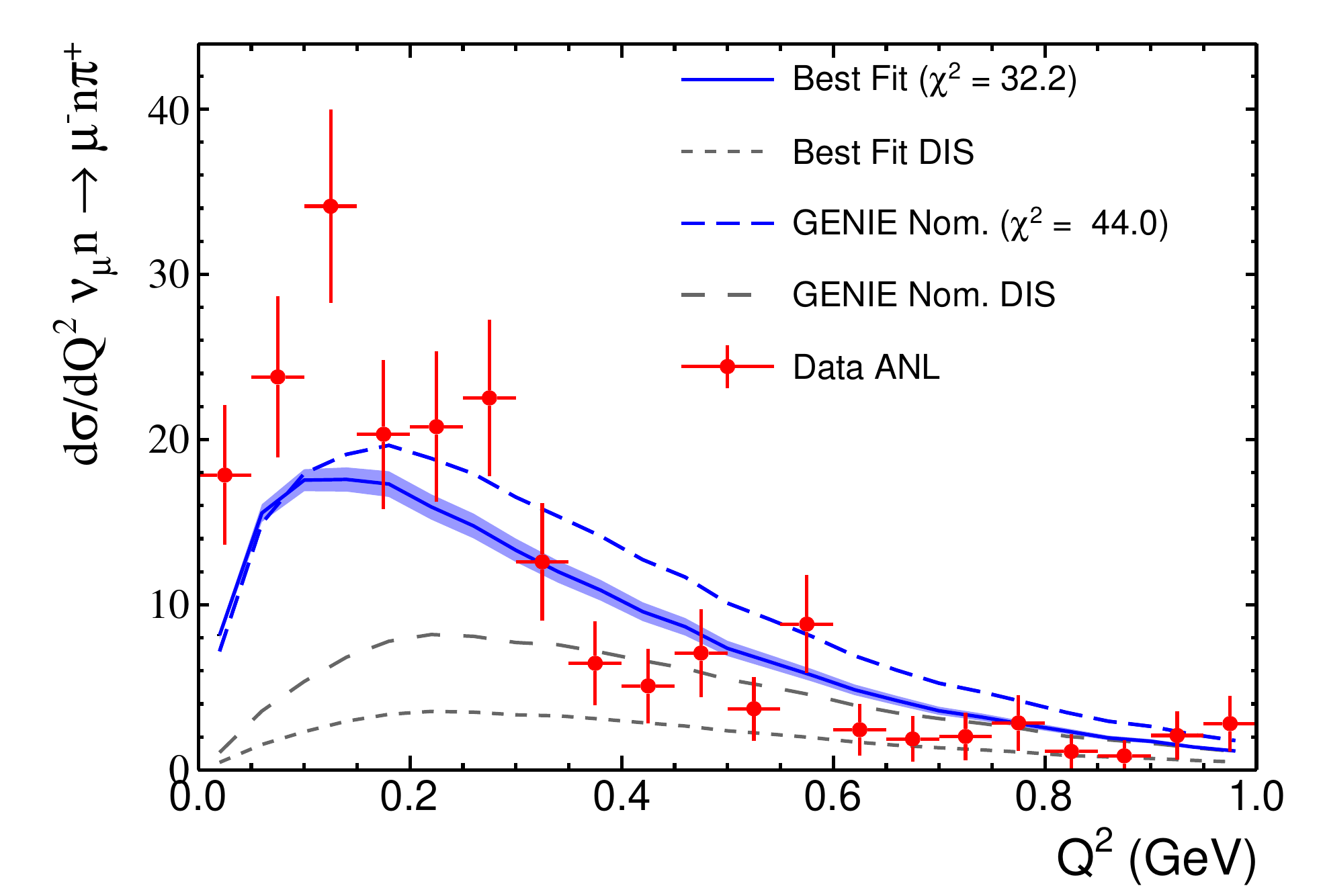}
    \caption{ANL \qq}
  \end{subfigure}
  \begin{subfigure}{0.9\columnwidth}
    \includegraphics[width=\textwidth]{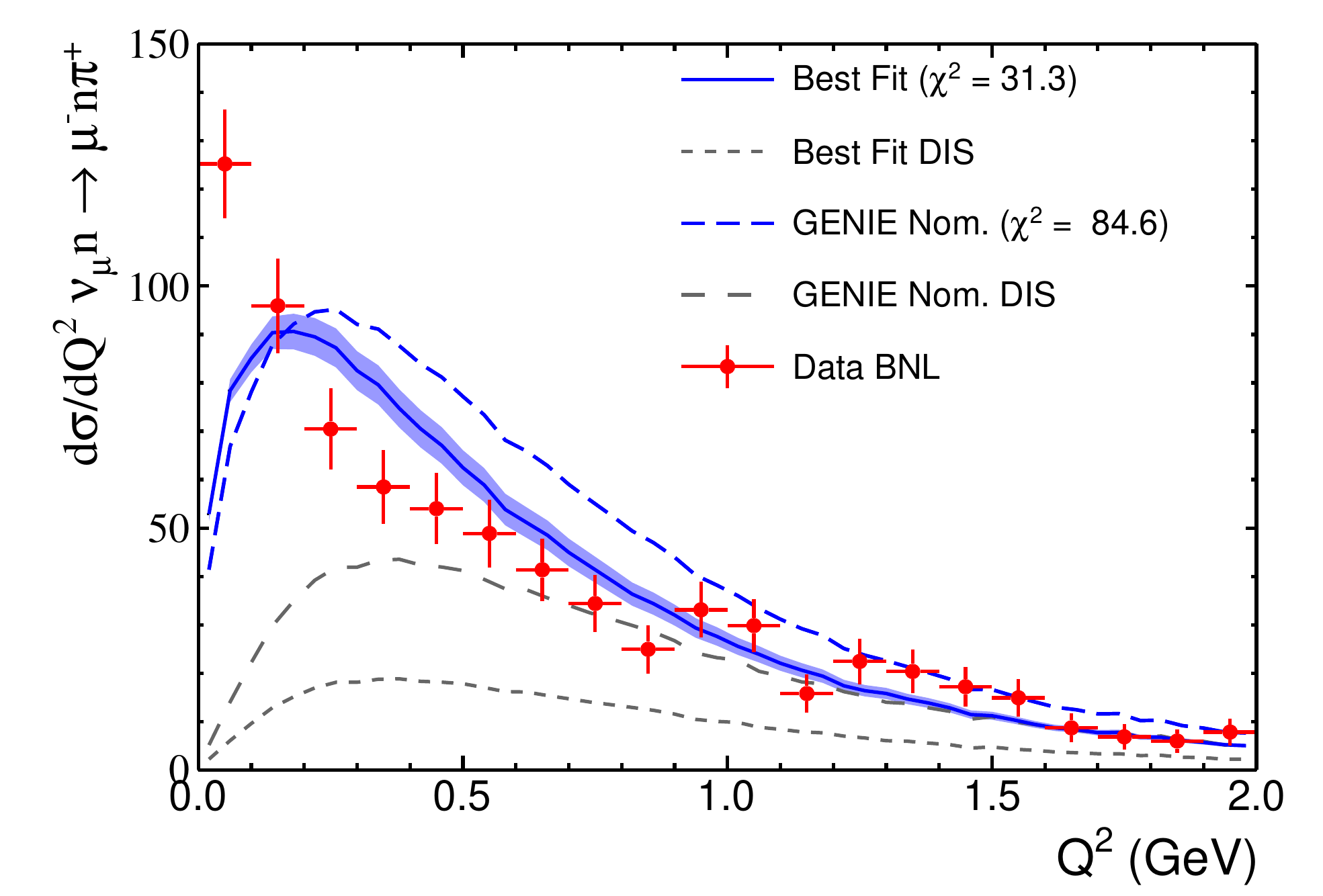}
    \caption{BNL \qq}
  \end{subfigure}
  \caption{Best fit results and post-fit uncertainties for the four \ccnpiplus datasets included in the GENIE v2.8.2 (RES) fit. The nominal prediction is shown for reference, and the $\chi^{2}$ contribution from each dataset is given in the legend for both the nominal and best fit distributions. The nominal and best fit DIS contribution to the total GENIE prediction (RES+DIS) are also shown for reference.}\label{fig:best_fit_channel2}
\end{figure*}

The best fit distributions for the GENIE v2.8.2 (RES) fit are shown for the \ccppiplus distributions in Figure~\ref{fig:best_fit_channel0}, for the \ccnpizero distributions in Figure~\ref{fig:best_fit_channel1}, and for the \ccnpiplus distributions in Figure~\ref{fig:best_fit_channel2}. Data points with \qq $\leq 0.1$ GeV$^{2}$ are shown but not included in the $\chi^{2}$ calculation.

\subsection{Goodness of fit}\label{sec:goodness_of_fit}
\begin{figure}[htbp]
  \centering
  \includegraphics[width=0.9\columnwidth]{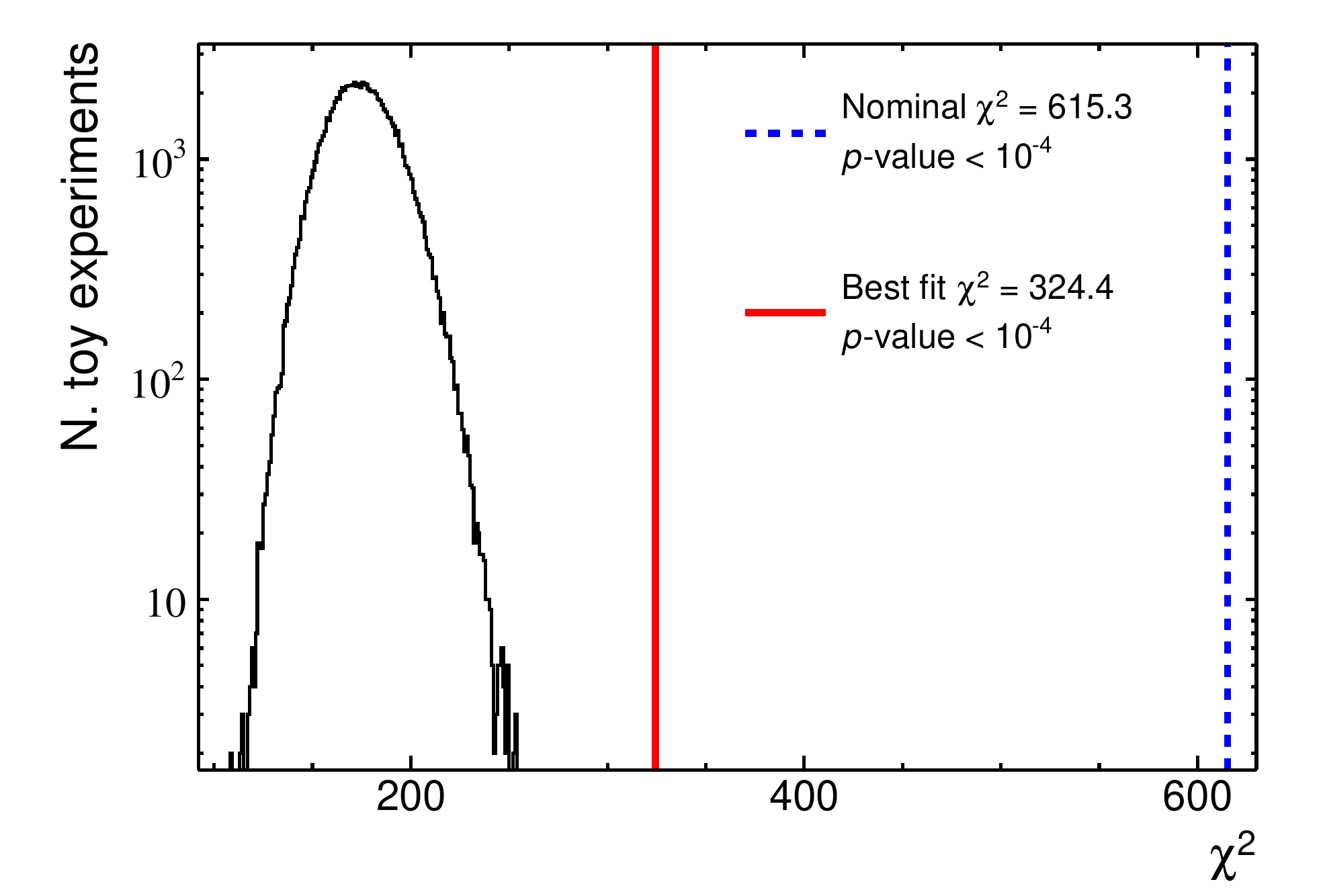}
  \caption{Expected $\chi^{2}$ distribution for the test statistic defined in Equation~\ref{eq:chi2} produced using 100,000 toy experiments. The nominal and best fit $\chi^{2}$ values for twelve datasets fit in Section~\ref{sec:results} are shown for comparison. The nominal $\chi^{2}$ for GENIE v2.8.2 is shown, as well as the best fit $\chi^{2}_{min}$ from the GENIE v2.8.2 (RES) fit.}\label{fig:goodness_of_fit}
\end{figure}

As has been previously remarked, the $\chi^{2}$/\ndof is not an appropriate measure of the goodness of fit for Equation~\ref{eq:chi2} as it involves Poisson-likelihood terms with low-statistics bins. In Figure~\ref{fig:goodness_of_fit}, the expected $\chi^2$ distribution has been produced by making 100,000 toy experiments in which a fake dataset has been produced with statistical errors thrown for all twelve datasets included in the fit. The $\chi^2$ is calculated between each toy experiment and the nominal data (without thrown statistical errors). The $p$-value of any fit can be calculated by integrating the distribution to the right of any given $\chi^{2}_{min}$ fit value as the $\chi^{2}$ distribution is independent of the fit type. Figure~\ref{fig:goodness_of_fit} shows the sampling distribution from this method, with the actual fit value for GENIE v2.8.2 (\fa), which has a $p$-value $\ll 10^{-4}$, indicating a poor fit between data and the model, even after the fit procedure.
\begin{table*}[htb]
  \centering
  \setlength{\tabcolsep}{.5em}
  {\renewcommand{\arraystretch}{1.4}
    \begin{tabular}{c|c|c|c|c|c|c|c|c|c|c|c|c}
      \hline
      \multirow{3}{*}{Dataset} & \multicolumn{4}{c|}{\ccppiplus} & \multicolumn{4}{c|}{\ccnpizero} & \multicolumn{4}{c}{\ccnpiplus} \\
      \cline{2-13}
      & \multicolumn{2}{c|}{ANL} & \multicolumn{2}{c|}{BNL} & \multicolumn{2}{c|}{ANL} & \multicolumn{2}{c|}{BNL} & \multicolumn{2}{c|}{ANL} & \multicolumn{2}{c}{BNL} \\
      \cline{2-13}
      & \Ev & \qq & \Ev & \qq & \Ev & \qq & \Ev & \qq & \Ev & \qq & \Ev & \qq \\
      \hline
      Nominal $\chi^{2}$  & 16.3 & 6.6 & 15.3 & 15.3 & 19.8 & 24.5 & 31.1 & 46.7 & 45.3 & 44.0 & 265.6 & 84.6 \\
      Best fit $\chi^{2}$ & 10.6 & 9.5 &  5.6 & 23.0 & 16.1 & 21.3 & 35.6 & 46.4 & 33.0 & 32.2 &  59.6 & 31.3 \\
      \ndof & 7 & 18 & 7 & 19 & 7 & 18 & 10 & 19 & 7 & 18 & 11 & 19 \\
      \hline
  \end{tabular}}
  \caption{Contributions to the nominal $\chi^{2}$ for GENIE v2.8.2 and to the best fit $\chi^2_{min}$ for the GENIE v2.8.2 (RES) fit from each of the twelve datasets included in the fit.}
  \label{tab:results_breakdown}
\end{table*}

In Table~\ref{tab:results_breakdown}, the contribution that each of the twelve datasets makes to the nominal and best fit $\chi^{2}$ values is given for the GENIE v2.8.2 (RES) fit. The \ndof contributed by each dataset is also shown for comparison. It is clear that there is a disproportionate contribution from the reanalyzed \Ev-dependent datasets for the subdominant channels (\ccnpizero and \ccnpiplus). The uncertainty on these distributions only include statistical errors, which are dominant, but there are significant normalization uncertainties due to detector corrections and background subtractions which are not included. These corrections are also likely to have an effect on the shape of the distributions, but it is not possible to calculate meaningful shape-uncertainties for these effects (nor are they included in the published results from ANL or BNL). It should also be noted that there is a correlation between the three \Ev-dependent datasets for both ANL and separately for BNL, introduced by the procedure for reanalyzing the datasets, although this is unlikely to be a significant issue. These issues are discussed further in~\ref{sec:reanalysis}.

To ensure that including the four subdominant \Ev-dependent datasets (\ccnpizero and \ccnpiplus for both ANL and BNL) does not badly bias the results, the fit was repeated without these four problematic datasets included. The fit results are within 1$\sigma$ of the values in Table~\ref{tab:best_fit_results}, which indicates that these datasets do not bias the fit strongly. Indeed, the \qq-dependent and \Ev-dependent distributions for each channel and experiment agree reasonably well with each other. When these four datasets are excluded, the $p$-value returned at the best fit point is more reasonable ($\sim$0.02), indicating that the poor quality of fit seen in Figure~\ref{fig:goodness_of_fit} can be mostly attributed to these four datasets. As there is no reason to suspect that these datasets are biasing the fit, we prefer to leave these datasets in the fit and present the fit with all twelve datasets included as the main result of this work. Using both \qq-dependent and \Ev-dependent datasets helps break the degeneracy between \mares and normalization parameters, so there is a strong reason for including all datasets if possible.

\section{Conclusions}\label{sec:conclusions}
ANL and BNL provide the only neutrino-nucleon data for the energies in the few-GeV region most relevant for current and future oscillation experiments. The large normalization discrepancy between them has led to large uncertainties in pion production parameters, which presents a problem for meeting the stringent error budgets required by current and future oscillation analyses. In this work, we use the reanalyzed ANL and BNL \ccppiplus datasets from Reference~\cite{anl_bnl_reanalysis}, where this normalization discrepancy has been solved, to constrain the GENIE single pion production model parameters. The reanalysis method from Reference~\cite{anl_bnl_reanalysis} is applied to the subdominant pion production channels \ccnpizero and \ccnpiplus, and \qq-dependent distributions for all three channels for both ANL and BNL are also used in the fits. Although the GENIE single pion model is not state of the art, it is widely used by many currently running experiments, so improvements to the parametrization are of interest to the community. Additionally, the fits described here provide a blueprint for their use in constraining other models.
\begin{figure}[htbp]
  \centering
  \begin{subfigure}{0.9\columnwidth}
    \includegraphics[width=\textwidth]{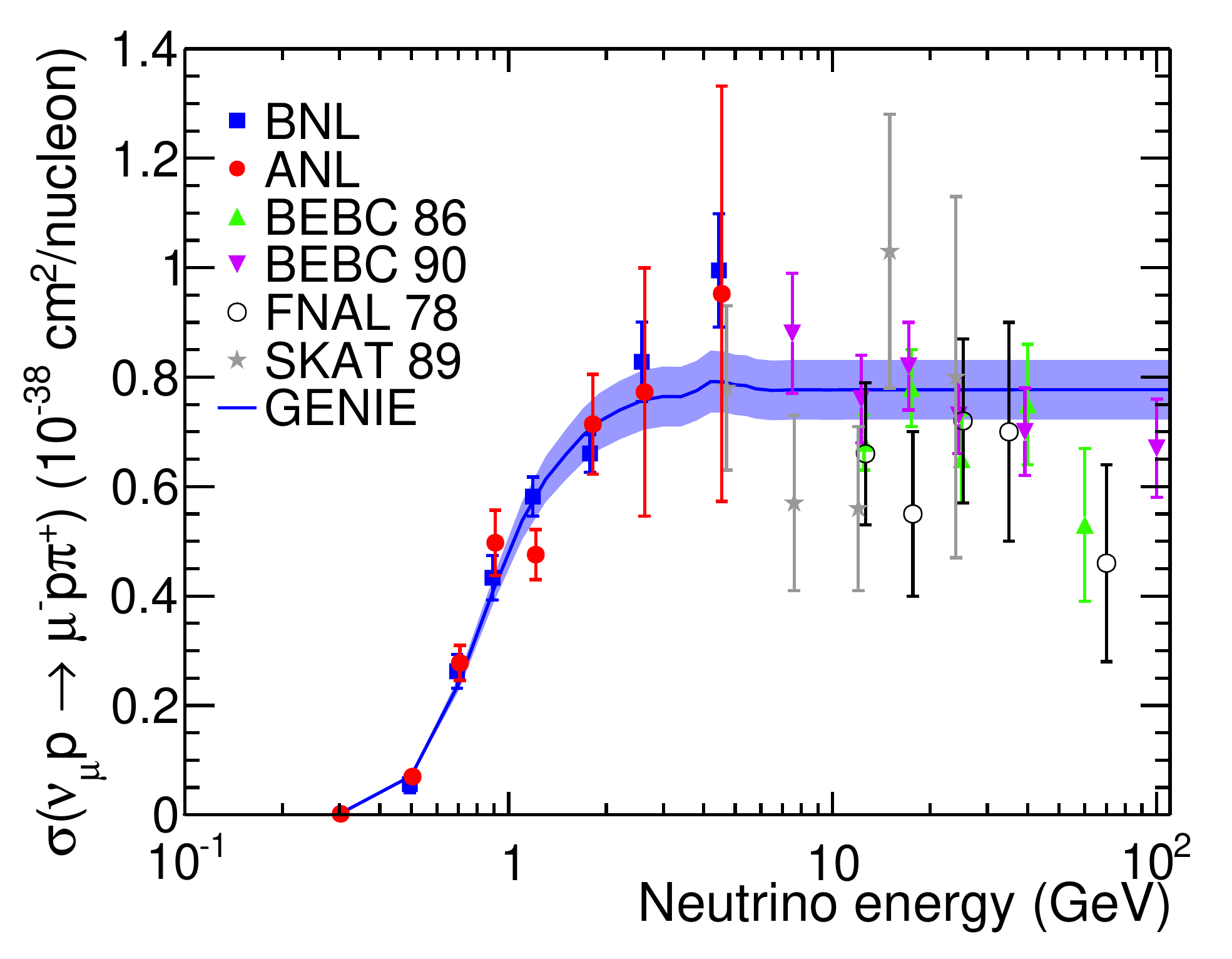}
    \caption{\ccppiplus}
  \end{subfigure}
  \begin{subfigure}{0.9\columnwidth}
    \includegraphics[width=\textwidth]{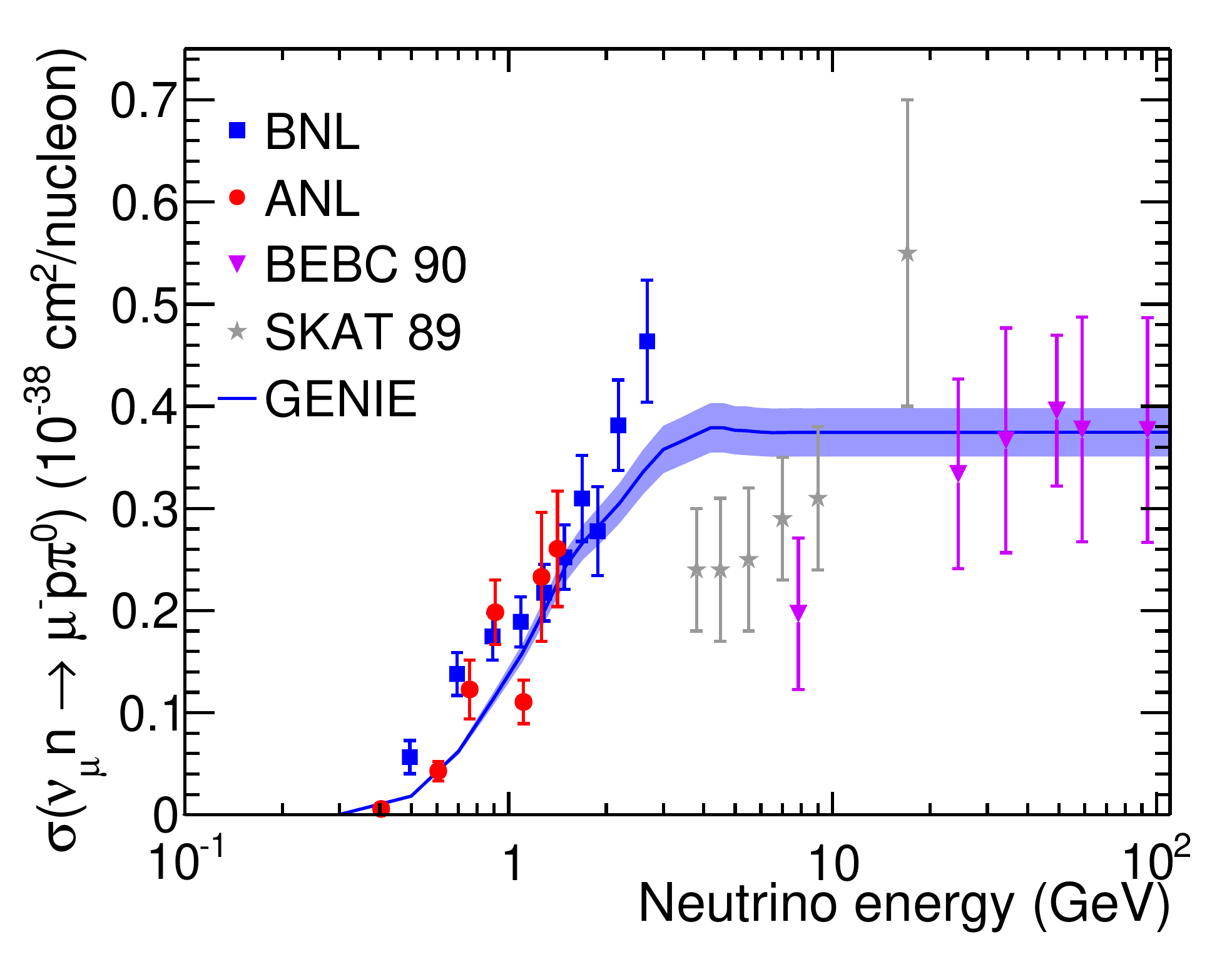}
    \caption{\ccnpizero}
  \end{subfigure}
  \begin{subfigure}{0.9\columnwidth}
    \includegraphics[width=\textwidth]{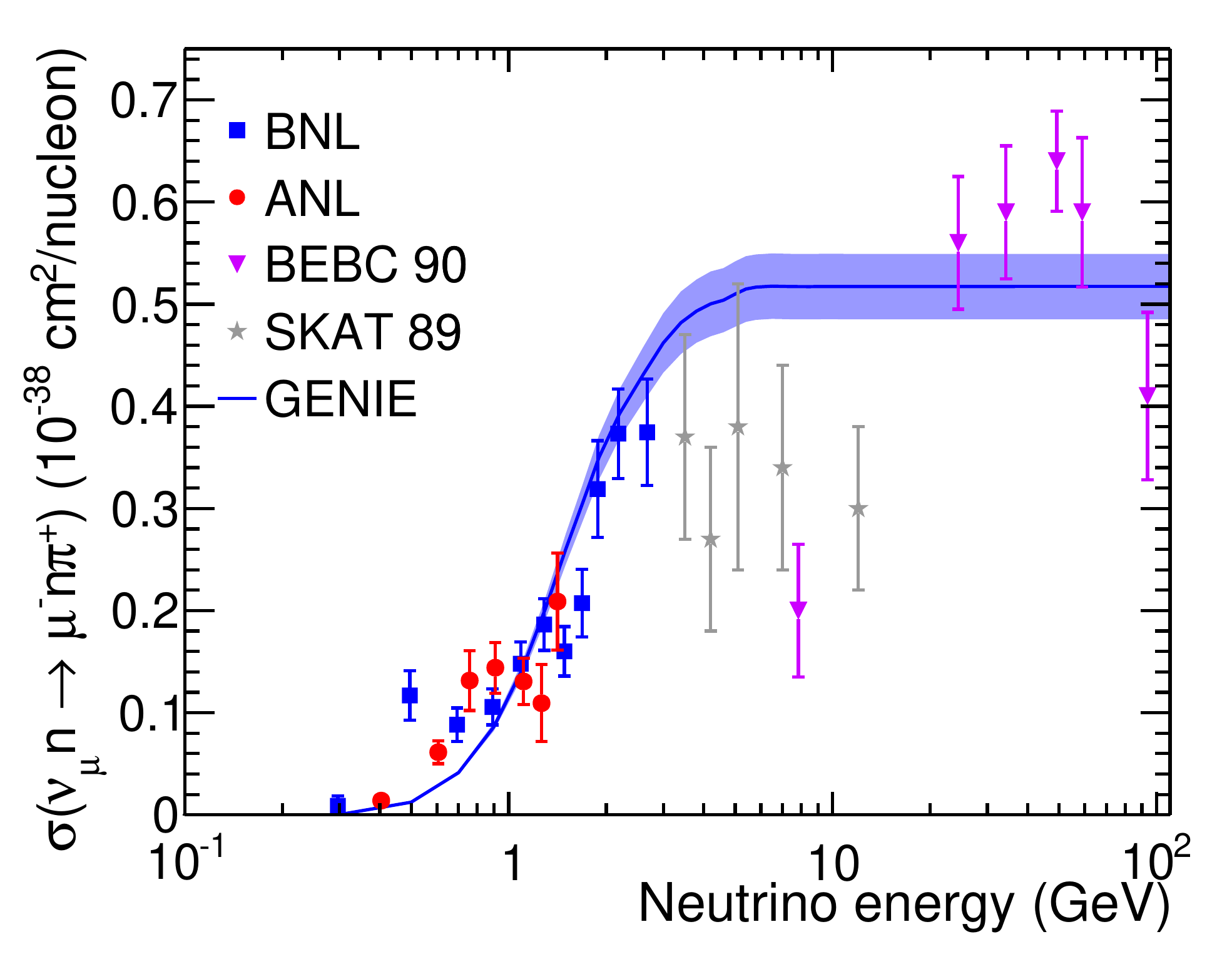}
    \caption{\ccnpiplus}
  \end{subfigure}
  \caption{The global dataset (described in Section~\ref{sec:datasets}) is compared with the best fit result and post-fit uncertainties for the GENIE v2.8.2 (RES) fit, for the three single pion production channels investigated in this work. Note that the reanalyzed ANL and BNL data shown have no invariant mass cut in the event selection, whereas the other datasets have an invariant mass cut of $W \leq \unit[2]{GeV}$ applied unless otherwise mentioned. An invariant mass cut of  $W \leq \unit[2]{GeV}$ has been applied to the GENIE prediction for this comparison.}\label{fig:global_fit_comparison}
\end{figure}

We find that the uncertainty on variable model parameters can be significantly reduced with respect to the nominal GENIE parameter uncertainties~\cite{genie_manual}, which were necessarily large to cover the disagreement between the published ANL and BNL datasets. A similar conclusion was found in the context of a different model in Reference~\cite{Alvarez-Ruso:2015eva}. The retuned uncertainties on these parameters should be used by neutrino oscillation and interaction experiments. To obtain good agreement with the data it was necessary to significantly reduce the non-resonant background normalization from the GENIE nominal prediction. The result of the GENIE v2.8.2 (RES) fit is compared to the global $E_{\nu}$-dependent data for the three single pion production channels of interest in Figure~\ref{fig:global_fit_comparison}. Most of the higher energy datasets shown (described in Section~\ref{sec:datasets}) have an invariant mass cut of $W \leq \unit[2]{GeV}$ applied, so the same invariant mass cut has been applied to the GENIE prediction shown. Note that the reanalyzed ANL and BNL data have no invariant mass cut applied, which should be borne in mind when interpreting Figure~\ref{fig:global_fit_comparison}.

We note that the recent coherent pion cross section results from \minerva~\cite{minerva_coh_2014} found a discrepancy between data and GENIE in a single pion production-dominated background sample that required significant reductions in the prediction, which may be alleviated by a reduction in the non-resonant single pion contribution as found in the fits presented here. We also note that the recent \nova results~\cite{Adamson:2016xxw} found a discrepancy between the hadronic energy distribution observed at the near detector and their GENIE simulation. There were more events in data where the hadronic system had less recoil, and fewer with high recoil, compared to the GENIE prediction. Although the discrepancy is treated as calibration effect in the \nova analysis, it is more likely to be due to deficiencies in the GENIE cross section model. The retuned pion set of production parameters described in this work will ameliorate the \nova discrepancy because it reduces the non-resonant pion production component, which will contribute events where the recoiling hadronic system has a lot of energy.

In this work, all available neutrino-nucleon single pion production data for neutrino energies below 10 GeV has been used to constrain the pion production parameters, including the reanalyzed ANL and BNL \Ev data for the first time, which is a significant step forward towards reducing the cross section uncertainties on this channel to the level required for future neutrino oscillation experiments. Recent proposals to extract neutrino-proton pion production cross sections from experiments where the target material contains hydrogen~\cite{xianguo_2015} raise the possibility of new data which will further reduce the parameter uncertainties.

\FloatBarrier
\begin{acknowledgements}
This material is based upon work supported by the US Department of Energy under Grant DE-SC0008475, and the Swiss National Science Foundation and SERI. CW is grateful to the University of Rochester for hospitality while this work was being carried out.
\end{acknowledgements}

\appendix

\section{Reanalysis of ANL and BNL \texorpdfstring{\ccnpiplus}{nu\_mu + n-->mu^- + n + pi^+} and \texorpdfstring{\ccnpizero}{nu\_mu + n-->mu^- + p + pi^0} cross section results}\label{sec:reanalysis}
In Reference~\cite{anl_bnl_reanalysis} we presented a method for removing flux uncertainties from the ANL and BNL bubble chamber datasets, which was applied to both the CC-inclusive and \ccppiplus cross sections from ANL and BNL. For the fitting work discussed in this work, it is desirable to extend this analysis to the subdominant pion production cross sections \ccnpiplus and \ccnpizero.

We note that for these subdominant channels, where one of the particles produced at the vertex is unobservable in a bubble chamber, we have to rely more heavily on the ANL and BNL reconstruction and particle identification methods than with the dominant \ccppiplus interaction where all interaction products can generally be observed\footnote{Although there is a threshold of around $p \gtrsim 150$ MeV for detecting the outgoing protons.}. It is not possible to accurately assess systematic errors for these selections, so we only quote statistical errors, which are likely to be dominant for all channels.

\subsection{Obtaining corrected cross sections}
A full description of the method can be found in Reference~\cite{anl_bnl_reanalysis}. In brief, we take the event rates from ANL and BNL for the exclusive pion production channels \ccppiplus, \ccnpizero and \ccnpiplus and CCQE as a function of neutrino energy, without any invariant mass cuts, which we correct for detector effects using the recommendations given in the original papers. Then we take the ratio of each exclusive pion production channel to the CCQE event rate (taking the ratio cancels the flux) to get a ratio of the cross sections, and then multiply by the relatively well known CCQE cross section to obtain the cross section for each single pion production channel. Essentially, we replace the flux uncertainty in the published single pion production results with the uncertainty on the CCQE cross section, at the cost of the additional statistical uncertainty on the CCQE event rate. 

\begin{figure}[htbp]
  \centering
  \begin{subfigure}{0.9\columnwidth}
    \includegraphics[width=\textwidth]{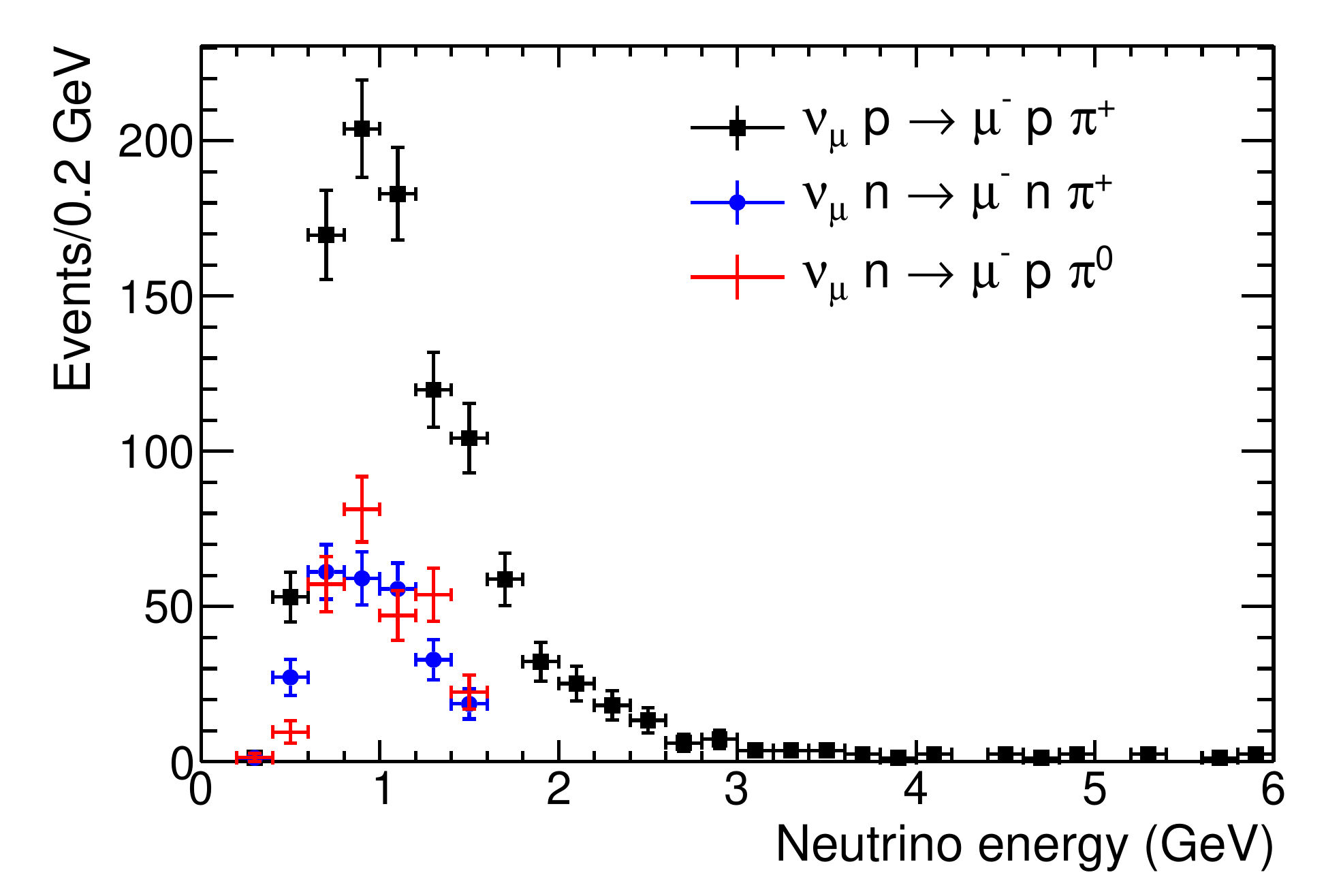}
    \caption{ANL}
    \label{subfig:ANL_rates}
    \end{subfigure}
  \begin{subfigure}{0.9\columnwidth}
    \includegraphics[width=\textwidth]{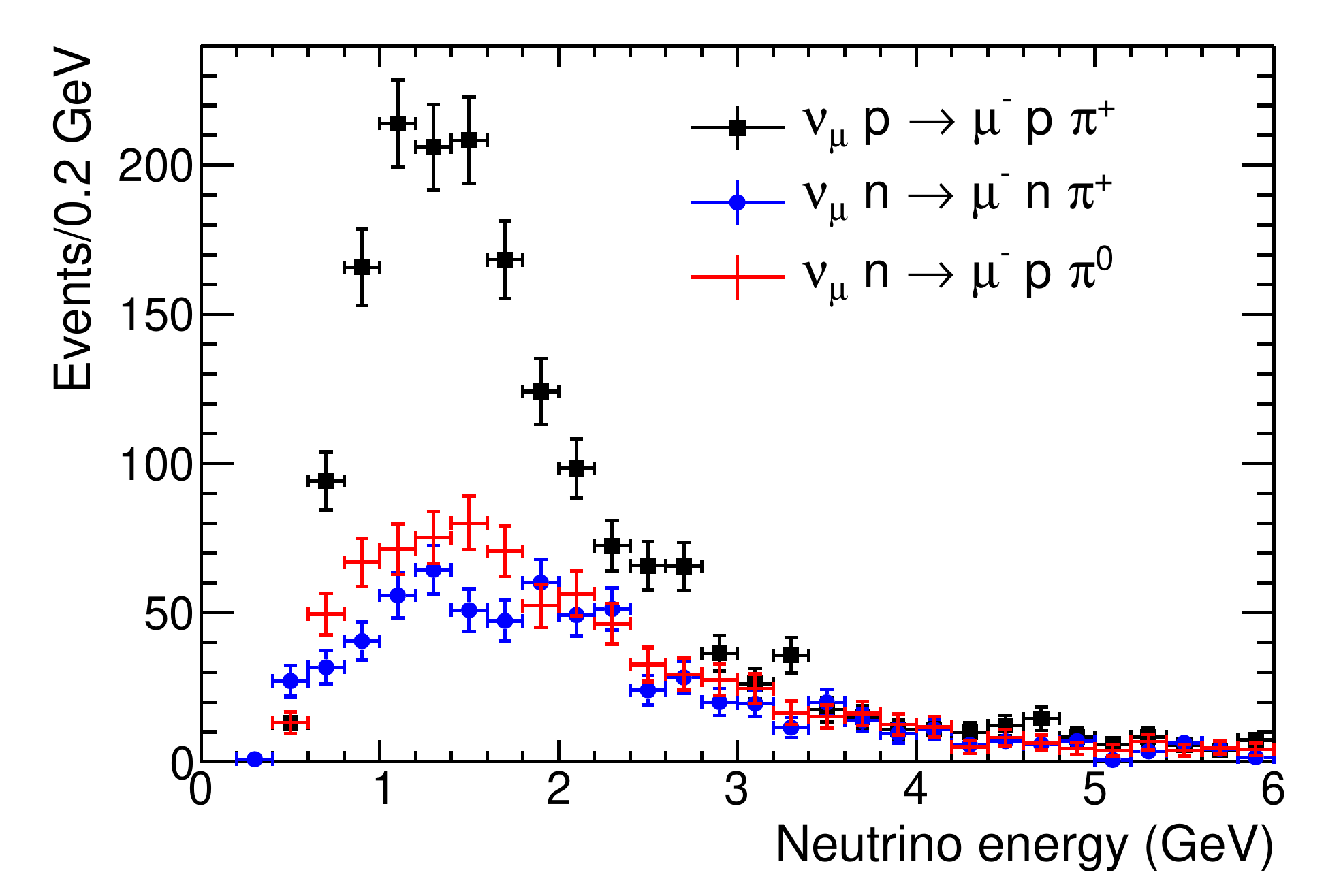}
    \caption{BNL}
    \label{subfig:BNL_rates}
    \end{subfigure}
  \caption{The digitized event rates on deuterium for the three interaction channels \ccppiplus, \ccnpizero and \ccnpiplus, as a function of the reconstructed neutrino energy \Ev. The errors are statistical only. Both ANL and BNL event rates and errors have been scaled when necessary to the statistics of their full deuterium samples.}\label{fig:event_rates}
\end{figure}

For ANL, the raw event rates are digitized from References~\cite{ANL_Barish_1979} (partial dataset) and~\cite{ANL_Radecky_1982} (full dataset) and are summarized in Table~\ref{tab:ANL_log_validation}. The CCQE event rates are only given using a partial dataset using $\sim$30\% of the final ANL exposure, whereas the single pion production event rates use the full ANL dataset. The dominant pion production channel \ccppiplus was also given in Reference~\cite{ANL_Barish_1979} using the partial exposure, so the ratio of partial to full events in this channel can be used to scale the CCQE event rate to the full statistics. The final fully corrected ANL event rates for the single production channels are shown in Figure~\ref{subfig:ANL_rates}. Note that the event rates given for the partial dataset are already corrected for detector effects and backgrounds. For the full dataset, the distributions are given without detector corrections applied, but the total corrected event rate is given, so the corrected event rate can simply be obtained by scaling the raw distribution (detector corrections as a function of \Ev were not considered in the ANL analysis). Note also that the ANL data is mostly from a deuterium fill of the detector, but data is also included from an initial hydrogen fill of the detector, which makes up approximately 2\% (6\%) of the full (partial) dataset. This issue only affects the \ccppiplus channel (as all other channels here are on a neutron), and is discussed in Reference~\cite{anl_bnl_reanalysis}.
\begin{table*}[htb]
  \centering
  {\renewcommand{\arraystretch}{1.2}
    \begin{tabular}{ccccc}
      \hline
      Dataset & Channel & Digitized & Published & Corrected \\
      \hline
      \multirow{2}{*}{Partial} & $\nu n \rightarrow \mu^{-}p$  & 834.6     & 833   & -- \\
      & \ccppiplus & 395.9     & 398  & -- \\
      \hline
      \multirow{3}{*}{Full} & \ccppiplus & 843.2 & 871 & 1115.0 \\
      & \ccnpizero & 200.3 & 202.2 & 272.8 \\
      & \ccnpiplus & 203.3 & 206.2 & 255.8 \\
      \hline
  \end{tabular}}
  \caption{Numbers of events for each of the ANL samples as published by ANL and digitized for this work.}
  \label{tab:ANL_log_validation}
\end{table*}

For BNL, the raw event rates for the single pion production datasets are digitized from Reference~\cite{BNL_Kitagaki_1986}, and for CCQE from Reference~\cite{BNL_Kitagaki_1990}, and are summarized in Table~\ref{tab:BNL_event_rates}. Detector and background corrections are applied as calculated by BNL, which are only given without any \Ev dependence as used in the original BNL analysis. The final fully corrected BNL event rates for the single pion production channels are shown in Figure~\ref{subfig:BNL_rates}.

\begin{table*}[htb]
  \centering
  {\renewcommand{\arraystretch}{1.2}
    \begin{tabular}{cccc}
      \hline
      Channel  & Digitized & Published & Det. correction \\
      \hline
      $\nu n \rightarrow \mu^{-}p$ & 2693.3 & 2684 & 1.11 $\pm$ 0.04 \\
      \ccppiplus & 1534.7 & 1610  & 1.12 $\pm$ 0.07 \\
      \ccnpizero &  808.4 & 853.5 & 1.05 $\pm$ 0.14 \\
      \ccnpiplus &  802.0 & 822.5 & 0.89 $\pm$ 0.10 \\
      \hline
  \end{tabular}}
  \caption{Numbers of observed (uncorrected) events for each of the BNL samples as published by BNL and as digitized for this work. All samples shown here use the full BNL dataset.}
  \label{tab:BNL_event_rates}
\end{table*}

\begin{figure}[htbp]
  \centering
  \begin{subfigure}{0.9\columnwidth}
    \includegraphics[width=\textwidth]{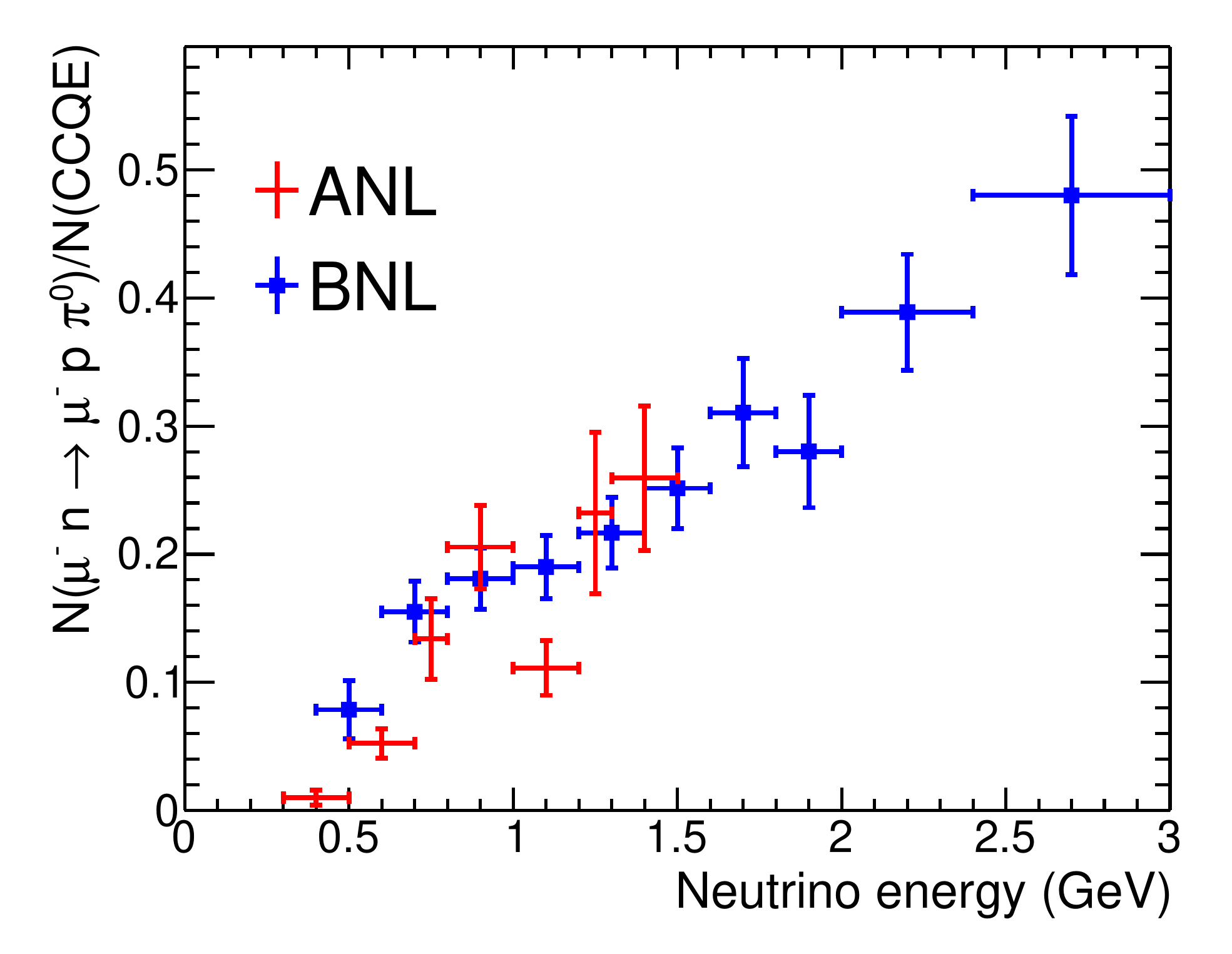}
    \caption{\ccnpizero}
    \end{subfigure}
  \begin{subfigure}{0.9\columnwidth}
    \includegraphics[width=\textwidth]{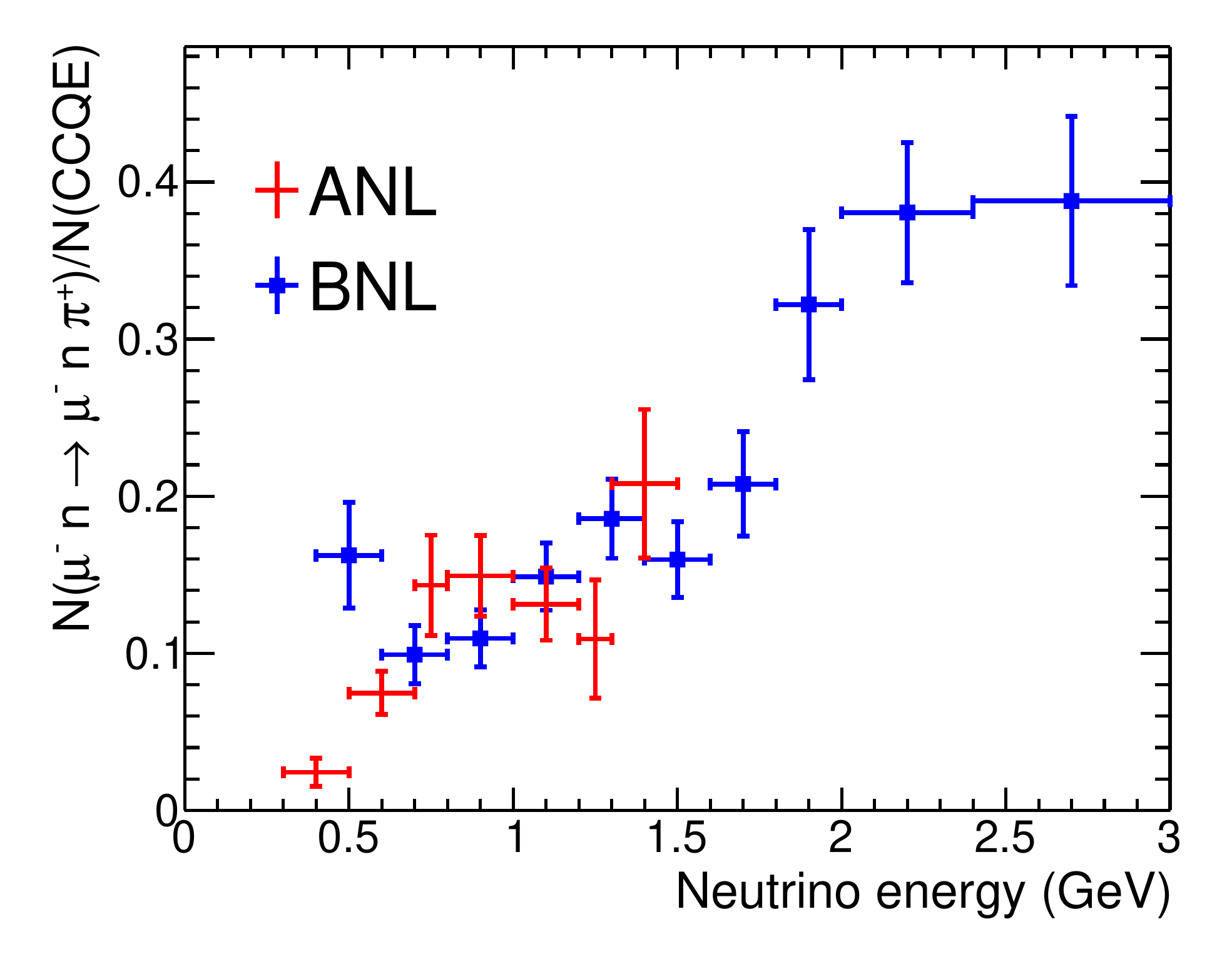}
    \caption{\ccnpiplus}
    \end{subfigure}
  \caption{The ratio of \ccnpizero and \ccnpiplus events to CCQE events as a function of \Ev for both ANL and BNL.}\label{fig:subdominant_ratios}
\end{figure}

\begin{figure}[htbp]
  \centering
  \begin{subfigure}{0.9\columnwidth}
    \includegraphics[width=\textwidth]{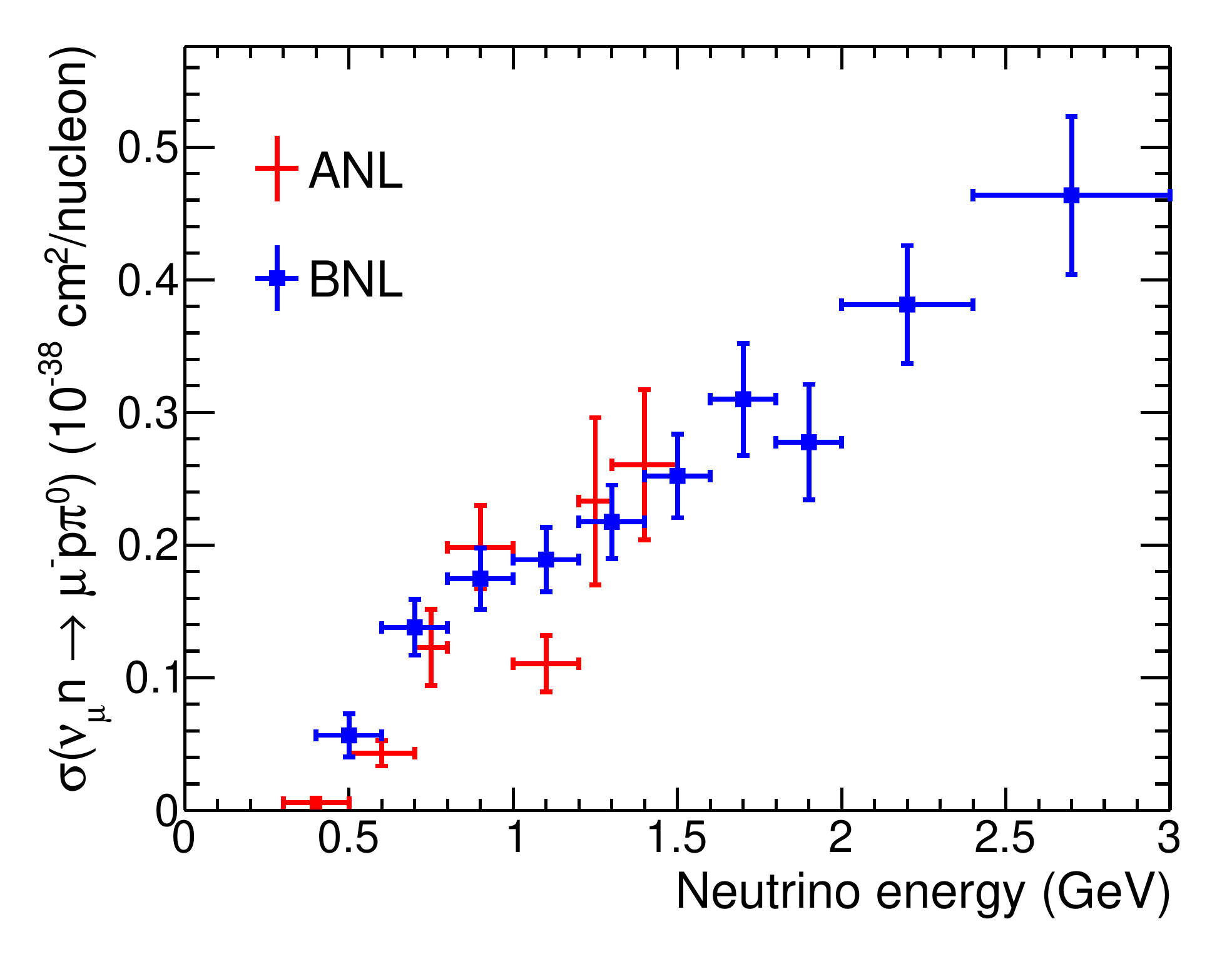}
    \caption{\ccnpizero}
    \end{subfigure}
  \begin{subfigure}{0.9\columnwidth}
    \includegraphics[width=\textwidth]{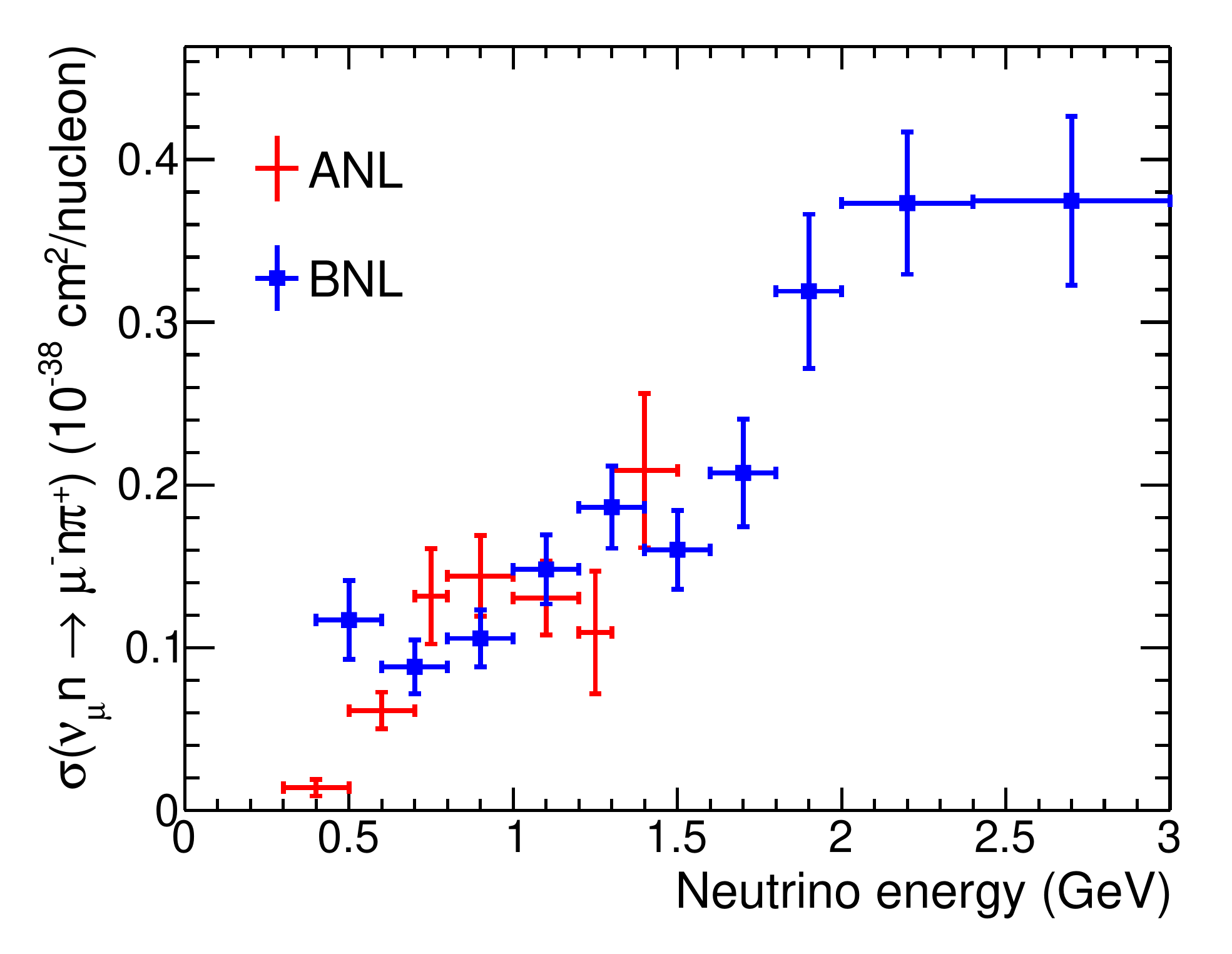}
    \caption{\ccnpiplus}
    \end{subfigure}
  \caption{Comparison of the \ccnpizero and \ccnpiplus cross sections obtained by multiplying the ratio with CCQE (shown in Figure~\ref{fig:subdominant_ratios}) by the GENIE CCQE cross section prediction for $\nu_{\mu}$--D$_{2}$ interactions.}\label{fig:subdominant_cross_sections}
\end{figure}

Using the corrected event rates in Figure~\ref{fig:event_rates}, and the corresponding distribution for CCQE (shown in Reference~\cite{anl_bnl_reanalysis}), it is possible to form ratios of \ccnpizero and \ccnpiplus over CCQE, as shown in Figure~\ref{fig:subdominant_ratios}. Finally, corrected cross sections for these subdominant pion production channels can be obtained by multiplying the ratio by the known CCQE cross section, to produce the final cross sections given in Figure~\ref{fig:subdominant_cross_sections} and used in this work. This procedure has already been applied to the \ccppiplus channel in Reference~\cite{anl_bnl_reanalysis}, so is not shown here. Details of the GENIE CCQE cross section for $\nu_{\mu}$--D$_{2}$ interactions used to produce Figure~\ref{fig:subdominant_cross_sections} are also given in Reference~\cite{anl_bnl_reanalysis}.

\subsection{Error analysis}
For all of the digitized datasets used in this work (summarized in Tables~\ref{tab:ANL_log_validation} and~\ref{tab:BNL_event_rates}), the agreement between the total digitized event rate and published event rate agrees within 1\%. We assume that the effect of digitization on the shape of the event rate distributions is small, and therefore neglect digitization uncertainties in this work, as in Reference~\cite{anl_bnl_reanalysis}. 

Only statistical errors are shown for the reanalyzed datasets, which are the dominant source of uncertainty for low-statistics bubble chamber data. Flux normalization uncertainties are the second largest source of uncertainty in the original ANL and BNL analyses, at around 15--20\%. These uncertainties are not considered here because they cancel (by construction) when taking ratios. However, we note that we have replaced the flux uncertainty with the uncertainty in the $\nu_{\mu}$--D$_{2}$ CCQE cross section, where the dominant uncertainty is the axial mass, $M_{\mathrm{A}}$, which can be considered to be $\sim$2\% normalization error on the \Ev distributions~\cite{bba03, kuzmin_2008}.

There is an uncertainty on the reconstructed neutrino energy for all channels which is estimated for BNL to be $\frac{\Delta E_{\nu}}{E_{\nu}}\sim$\unit[2]{\%} for CCQE and \ccppiplus events~\cite{BNL_Baker_1981}, and $\sim$\unit[5]{\%} for other charged current production channels which are not kinematically overconstrained (\ccnpizero and \ccnpiplus). ANL also quote an uncertainty of $\frac{\Delta E_{\nu}}{E_{\nu}}\leq$ \unit[5]{\%} for the subdominant channels, but do not quote an uncertainty on kinematically overconstrained channels~\cite{ANL_Radecky_1982}. This uncertainty is therefore more significant for the subdominant channels \ccnpizero and \ccnpiplus than the dominant \ccppiplus channel, but for all cases, the energy smearing is $\frac{\Delta E_{\nu}}{E_{\nu}}\leq$ \unit[5]{\%}.

There are additional uncertainties for all channels which come from the detector corrections and background subtractions which are discussed for ANL in Refences~\cite{ANL_Barish_1979, ANL_Radecky_1982} and for BNL in References~\cite{BNL_Kitagaki_1986, BNL_Kitagaki_1990}. These corrections are given on the total rate only, so no information is available from either experiment on how they may distort the shape of the \Ev distributions. For the overconstrained CCQE and \ccppiplus channels, these are mostly corrections for reconstruction and scanning inefficiencies, with small background corrections. A conservative estimate on the normalization uncertainty for the overconstrained channels is $\sim$5\%. For the \ccnpizero and \ccnpiplus channels, which are not kinematically overconstrained, the normalization uncertainty from the quoted correction factors are $\sim$10--15\% for both experiments. There are significantly more backgrounds for the underconstrained channels, which makes the reanalysis of these channels more dependent on the ANL and BNL calculations than the dominant \ccppiplus channel. These backgrounds are from the misreconstructed CCQE and \ccppiplus events, multipion events with unobserved final state particles and migration between the \ccnpizero and \ccnpiplus selections. In this analysis we neglect the normalization uncertainty from detector effects for the \ccnpizero and \ccnpiplus channels for simplicity. Although the $\sim$10--15\% is no longer negligible compared with the statistical errors, applying a fully correlated normalization error of this size would not change the results of this analysis significantly, and still rests on the rather simplistic assumption that all of the detector effects and backgrounds have no \Ev dependence (although this assumption is present for the published ANL and BNL cross sections measurements for these channels). We note that the size of this neglected normalization error is smaller than the flux uncertainties which are canceled in this analysis.


\section{Reanalyzed ANL and BNL results with an invariant mass cut of $W < 1.4$ GeV}\label{sec:reanalysis_wcut}

\begin{figure}[htb]
  \centering
  \begin{subfigure}{0.9\columnwidth}
    \includegraphics[width=\textwidth]{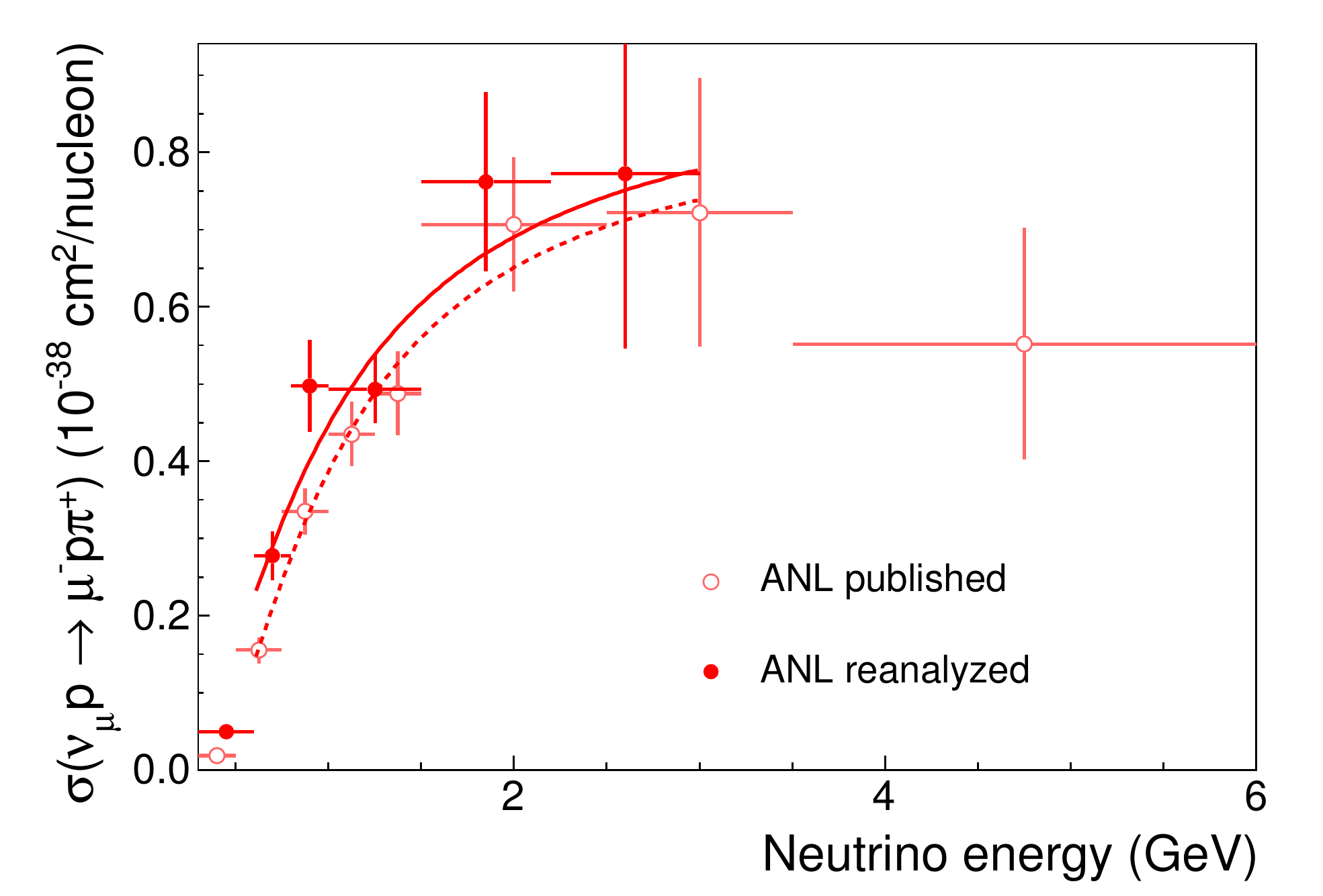}
    \caption{ANL \ccppiplus}
  \end{subfigure}
  \begin{subfigure}{0.9\columnwidth}
    \includegraphics[width=\textwidth]{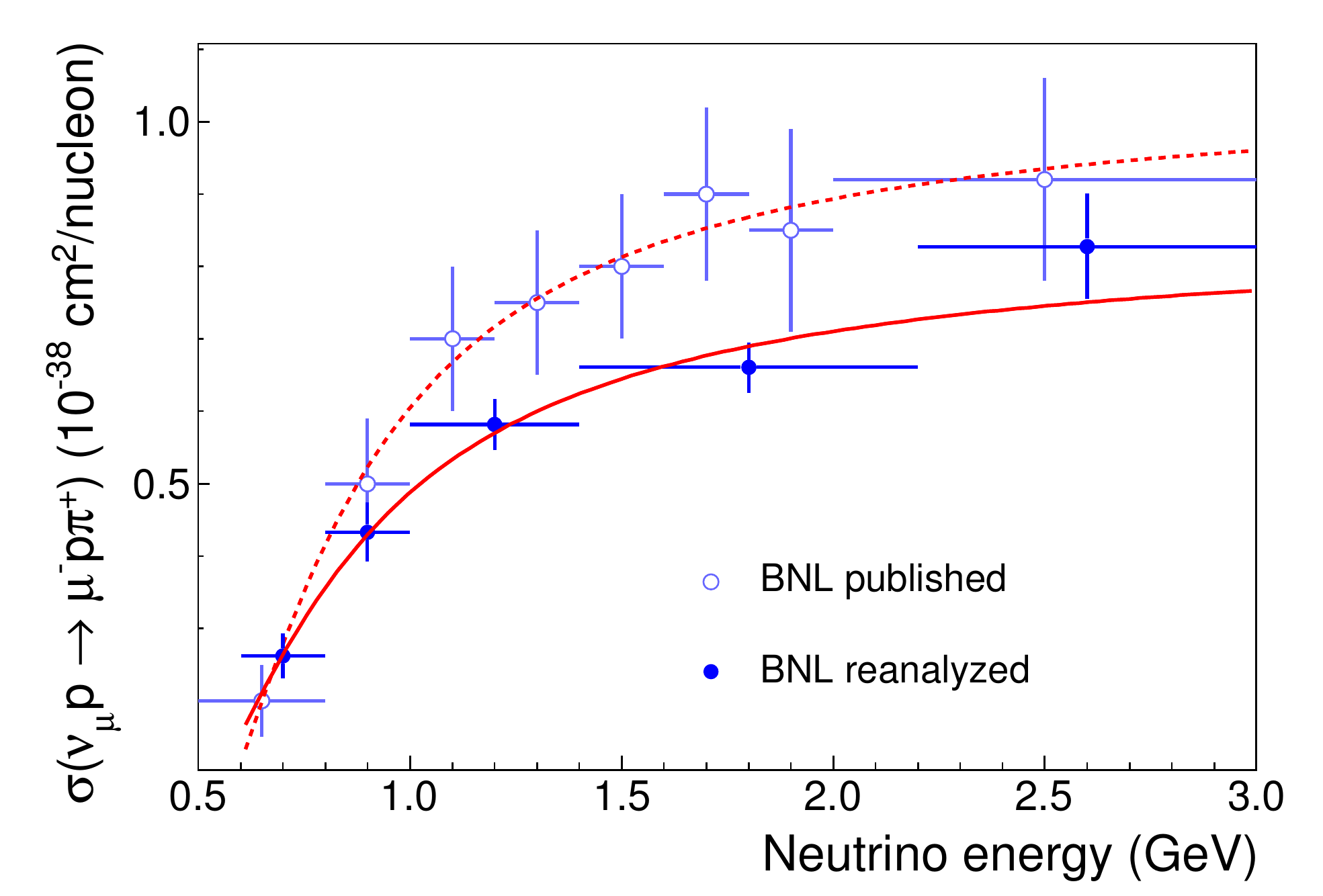}
    \caption{BNL \ccppiplus}
  \end{subfigure}
  \begin{subfigure}{0.9\columnwidth}
    \includegraphics[width=\textwidth]{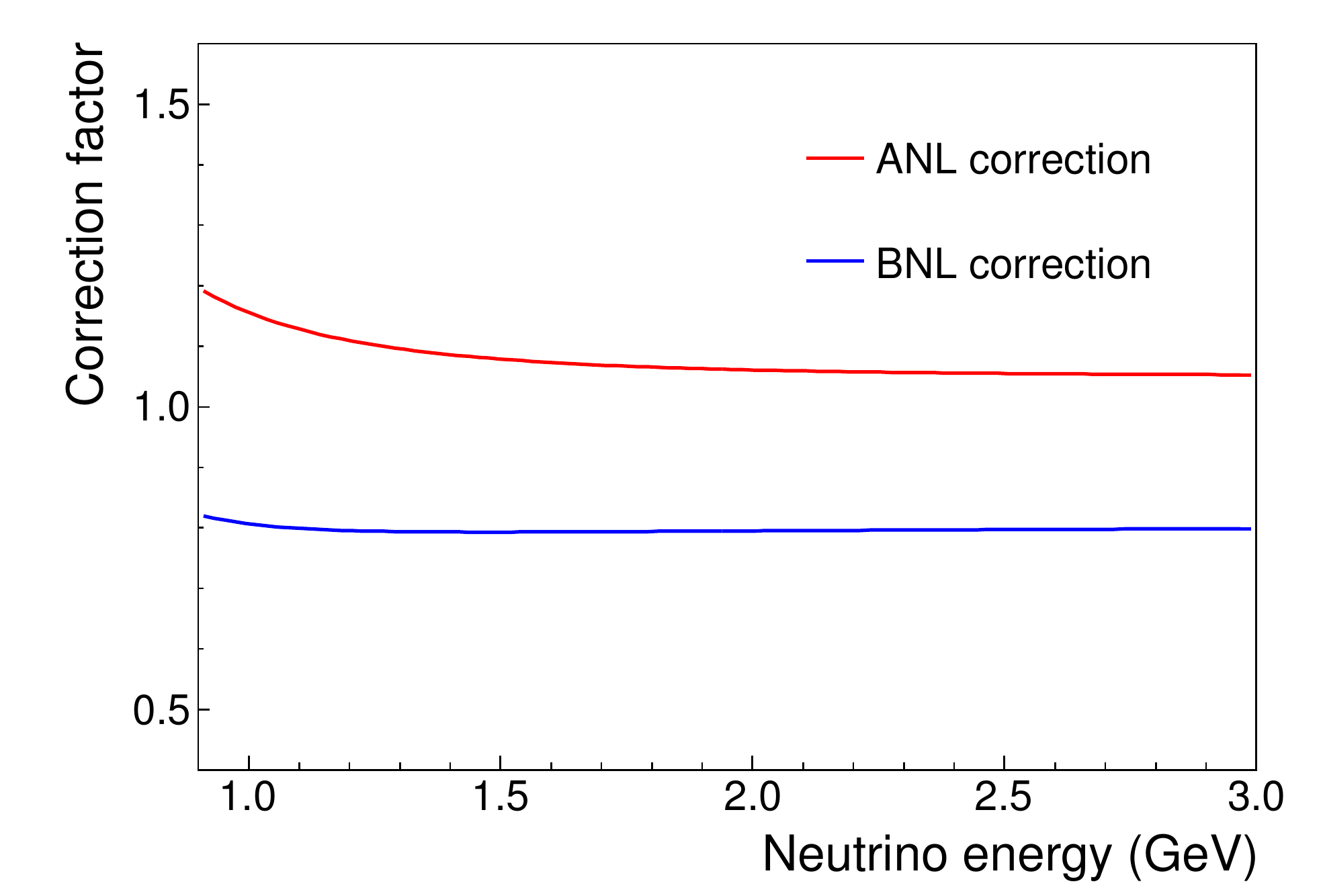}
    \caption{Correction functions}
  \end{subfigure}
  \caption{Published and reanalyzed \ccppiplus cross sections from ANL (a) and BNL (b) without invariant mass cut, with the fit from Equation~\ref{eq:lowW-fit}. (c) The ratio of the fit functions, used as a correction factor for the $W<1.4$~GeV data sets.}
\label{fig:anl-bnl-lowW}
\end{figure}

ANL and BNL also published cross sections with a cut on hadronic invariant mass $W< 1.4$~GeV, but their publications do not include event rate distributions with the same cut that would allow a similar ratio analysis as carried out in~\ref{sec:reanalysis}. Instead, we use the ratio of reanalyzed to published cross sections without an invariant mass cut as a correction factor for the $W< 1.4$~GeV cross sections. The published and reanalyzed cross sections have different binnings, so we fit a continuous function in neutrino energy $E_\nu$ to the \ccppiplus cross section:
\begin{equation}
\sigma = a_0\tan^{-1}(a_1 E_\nu + a_2)
\label{eq:lowW-fit}
\end{equation}
where $a_i$ are parameters of the fit. Figure~\ref{fig:anl-bnl-lowW}
shows the published and renalyzed data sets with their fits to
Equation~\ref{eq:lowW-fit}, and the ratio of fit functions which is
used to correct the $W<1.4$~GeV data given in References~\cite{ANL_Radecky_1982} (ANL) and~\cite{BNL_Kitagaki_1986} (BNL). For $E_\nu < 1$~GeV, the value
of the correction function at 1~GeV is used.
\begin{figure}[htb]
  \centering
  \begin{subfigure}{0.9\columnwidth}
    \includegraphics[width=\textwidth]{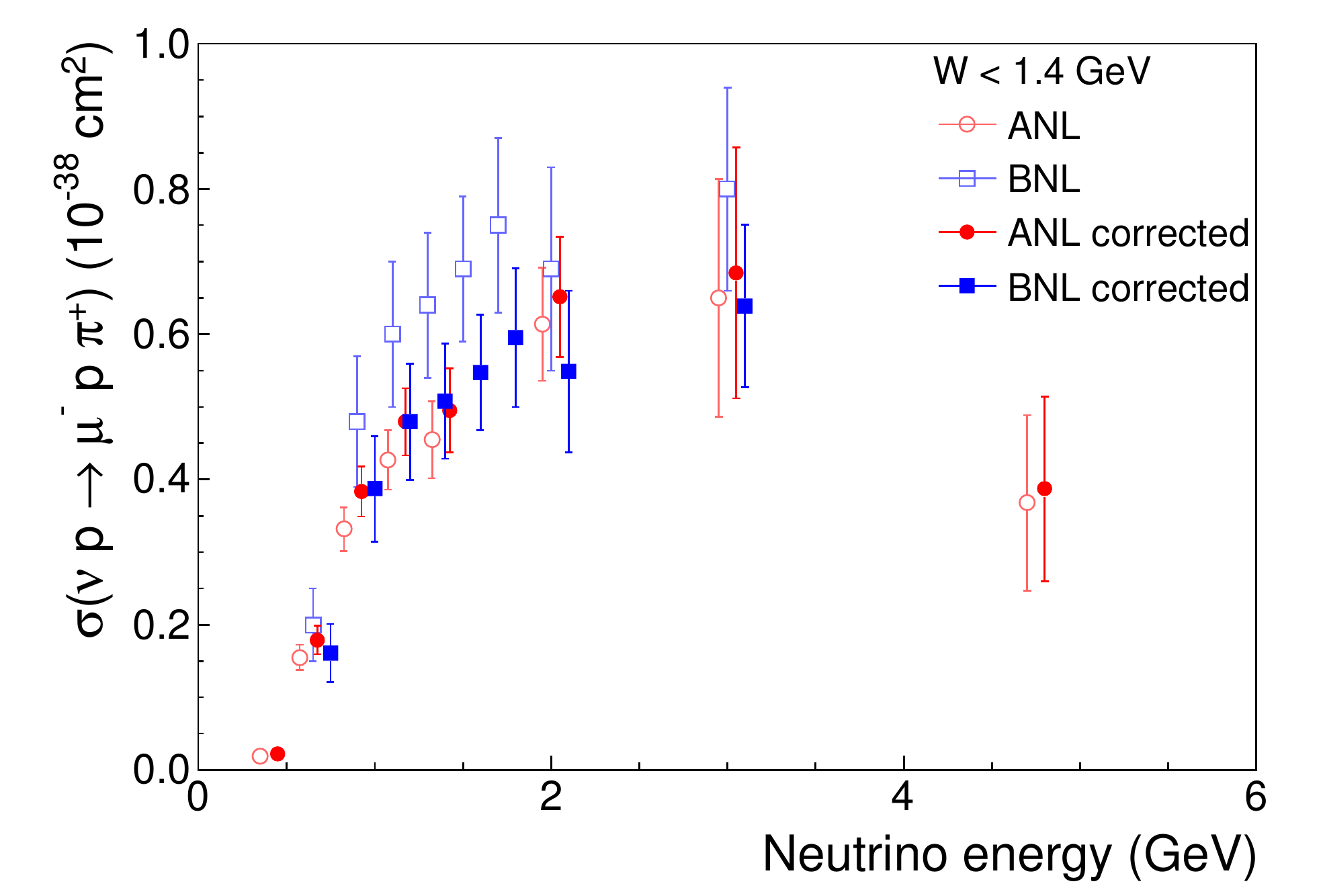}
    \caption{\ccppiplus}
  \end{subfigure}
  \begin{subfigure}{0.9\columnwidth}
    \includegraphics[width=\textwidth]{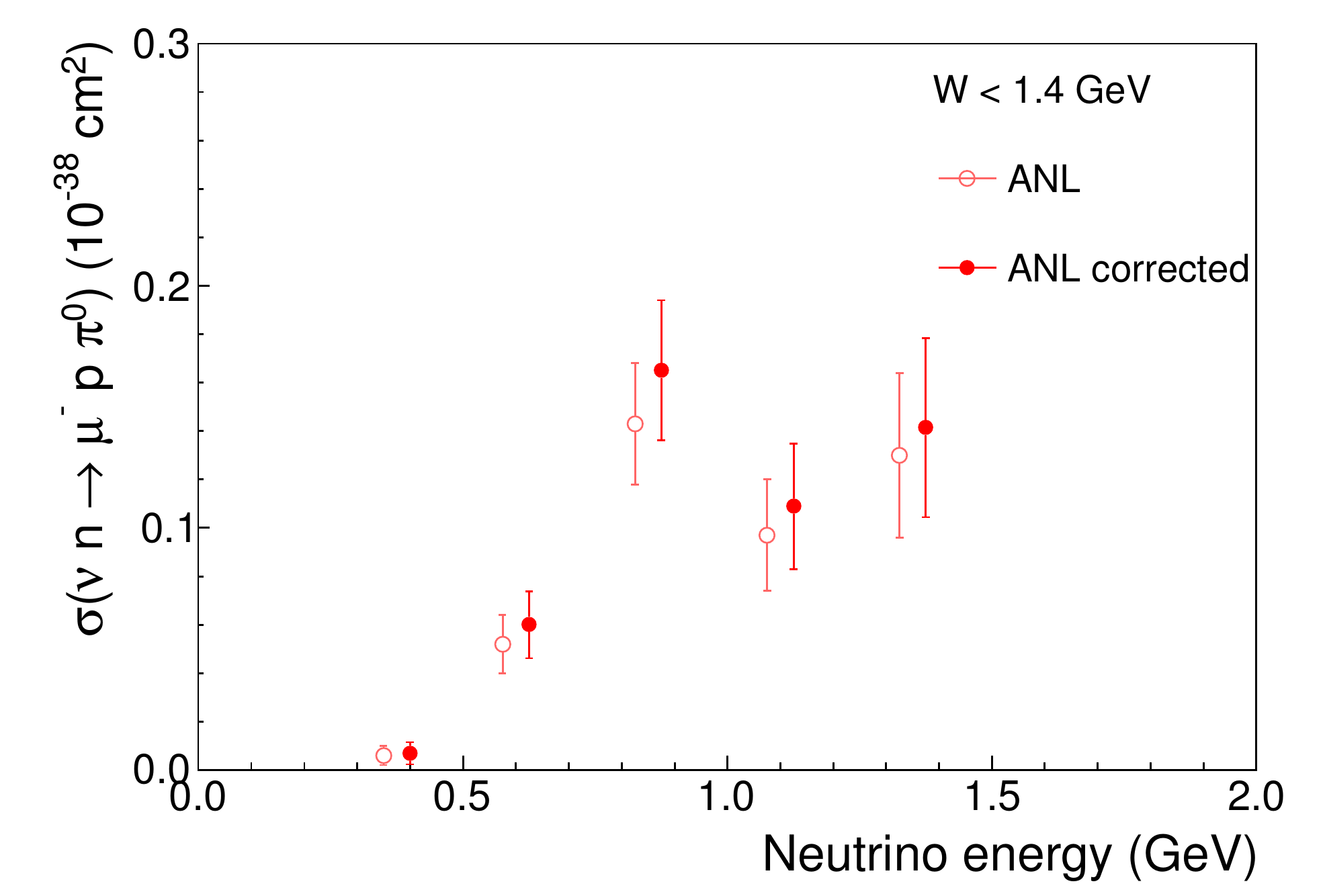}
    \caption{\ccnpizero}
  \end{subfigure}
  \begin{subfigure}{0.9\columnwidth}
    \includegraphics[width=\textwidth]{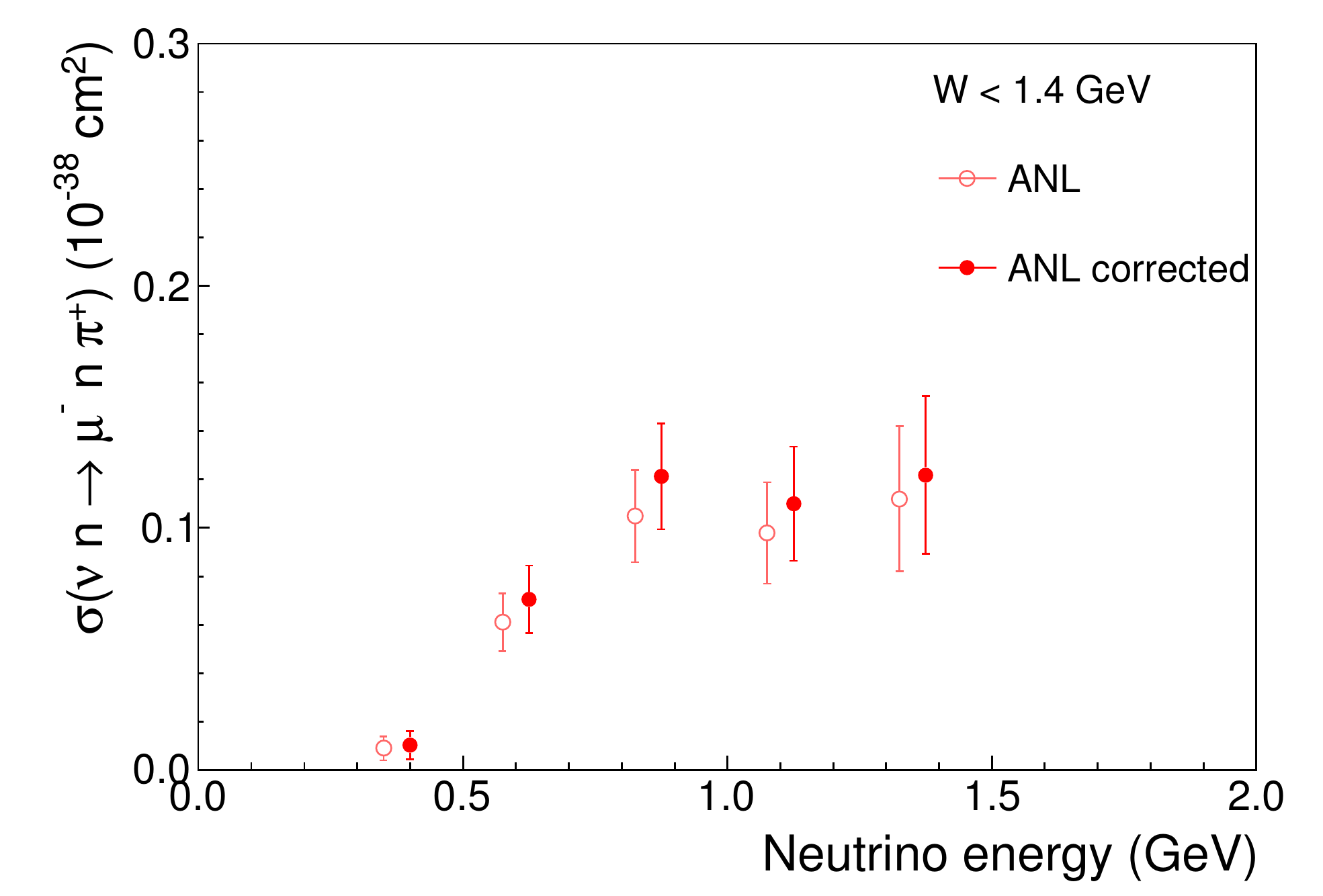}
    \caption{\ccnpiplus}
  \end{subfigure}
  \caption{Cross sections for $W<1.4$~GeV with and without the correction described in the text.}
\label{fig:anl-bnl-lowW-corrected}
\end{figure}

Figure~\ref{fig:anl-bnl-lowW-corrected} shows the cross sections for
$W<1.4$~GeV with and without the correction factor applied. In the
\ccppiplus channel, where both experiments have data, the agreement is
improved by the correction method.

\FloatBarrier
\bibliographystyle{spphys}
\bibliography{ANL_BNL_CC1pi}

\end{document}